\documentclass{aa}
\usepackage{txfonts}
\usepackage{latexsym}
\usepackage{epsfig} 
\usepackage{hangcaption} 
\usepackage{graphicx}
\usepackage{rotating}
\usepackage{natbib}
\bibpunct{(}{)}{;}{a}{}{,}
\newcommand{\NPulsars}{18~} 
\renewcommand{\deg}{^\circ}
\begin{document}
\title{Frequency dependence of orthogonal polarisation modes in
       pulsars} 
\subtitle{} 
\author{J.M. Smits \inst{1} \and B.W. Stappers \inst{2,3} \and
       R. T. Edwards \inst{4} \and J. Kuijpers \inst{1} \and
       R. Ramachandran \inst{5}} 
\offprints{J.M. Smits \email{R.Smits@\allowbreak science.\allowbreak
       ru.nl}} 
\institute{Department of Astrophysics, Radboud University Nijmegen,
  Nijmegen \and ASTRON, Dwingeloo, The Netherlands \and Astronomical
  Institute `Anton Pannekoek', Amsterdam \and Australia Telescope
  National Facility, CSIRO, Australia \and Department of Astronomy,
  University of California, Berkeley, USA}
\date{Received <date> / Accepted <date>}
\abstract{We have carried out a study of the orthogonal polarisation
  mode behaviour as a function of frequency of \NPulsars pulsars,
  using average pulsar data from the European Pulsar Network
  (EPN). Assuming that the radiation consists of two 100\% polarised
  completely orthogonal superposed modes we separated these modes,
  resulting in average pulse profiles of each mode at multiple
  frequencies for each pulsar. Furthermore, we studied the frequency
  dependence of the relative intensity of these modes. We found in
  many pulsars that the average pulse profiles of the two modes differ
  in their dependence on frequency. In particular, we found that pulse
  components that are dominated by one mode tend to increase in
  intensity with increasing frequency with respect to the rest of the
  profile.   }
\maketitle

\section{Introduction} 
Single pulse studies of the position angle (PA) of linearly polarised
radiation from a number of pulsars show that it is built up of two
modes of polarisation, which are separated in angle by 90 degrees
\citep[e.g.,][]{Manchester75, Stinebring84, Rankin88a, Gil91,
Gil92}. It is believed that these orthogonally polarised modes (OPM)
reflect the eigenmodes of the magneto-active plasma in the open
magnetic field-lines above the pulsar polar cap. Three modes of wave
propagation are allowed in this region, one of which is the ordinary
sub-luminous mode, which cannot escape the pulsar magnetosphere due to
Landau damping and is thus of no interest to the present work. The
remaining two modes are the ordinary super-luminous mode (O-mode)
which is polarised in the local plane defined by the external magnetic
field and the wave-vector, and the extraordinary mode (X-mode), which
is polarised perpendicular to this plane
\citep{Arons86,Petrova01}. According to \citet{Barnard86} refraction
can separate the X- and O-modes beams, which have different indexes of
refraction, by many beam-widths. The tendency for average profiles to
have constant widths above a critical frequency is then explained by
this separation occurring above a critical height
\citep{Sieber75}. Two conditions are required for this: (1) the radio
emission mechanism has to be broadband in frequency over a narrow
range of heights above the stellar surface and (2) the gradient in the
plasma density has a transverse component to the radial direction
\citep{McKinnon97}. The independent propagation of the two modes in
the open flux zone can produce the abrupt orthogonal transitions in
polarisation position angle that are commonly observed in studies of
individual pulse polarisation \citep{Manchester75}. The merging of the
beams might also account for depolarisation of pulse average profiles
with increasing radio frequency \citep{McKinnon97,
Hoensbroech97}. Alternatively, the OPM transitions might not be due to
refraction, but due to switching between significant and insignificant
conversion of O-mode into X-mode \citep{Petrova01}.

One of the most important questions about OPM is whether the
polarisation modes are disjoint or superposed. For disjoint modes the
polarisation of the observed radiation is given by either one of the
two modes at each point in time.  Superposed modes occur
simultaneously and the polarisation of the observed electromagnetic
radiation is given by the vectorial addition of the Stokes parameters
$Q$, $U$ and $V$ of both modes. \citet{Cordes78}, who first posed this
question based on the polarisation from \object{PSR B2020+28}, assumed the
modes to be disjoint and stated three arguments for this assumption:
(1) the degree of polarisation is fairly steady from one pulse to the
next, (2) if the modes are superposed, one might expect occasional
sign changes in the complex value $Q+iU$ when the linear polarisation
is low, due to random fluctuations in the relative strength of the two
modes and (3) the instantaneous mode of polarisation seems correlated
with the total intensity, which is consistent with disjoint occurrence
of the modes.  However, \citet{Stinebring84} found that the modes
appear to be superposed by studying sensitive polarisation
observations of a number of pulsars. It was found that average values
of the fractional linear polarisation were small at longitudes where
both polarisation modes occurred with nearly equal frequency. The same
effect was later found in the instantaneous values of the fractional
linear polarisation in single pulse observations of \object{PSR B2020+28}
\citep{McKinnon98}. This phenomenon is present on timescales of
hundreds of microseconds to hours, which is what one would expect for
superposed OPM. For disjoint modes one would expect the fractional
polarisation to be either high or low, depending on the fractional
polarisation of the active mode.

Assuming that the polarisation modes are superposed, that they are
completely polarised and completely orthogonal, the average intensity
of the individual polarisation modes as a function of longitude can be
determined from the Stokes parameters of the average pulsar
signal \citep{McKinnon00}. In this paper we make these assumptions and
determine the individual polarisation profiles of \NPulsars pulsars as
a function of both longitude and frequency (using data from the EPN
database). In section~\ref{sec:Method} we give a description of our
analysis, the results of which are shown in
section~\ref{sec:Results}. In section~\ref{sec:Discussion} we discuss
the results and finally we give our conclusions in
section~\ref{sec:Conclusions}.

\section{Method}
\label{sec:Method}
\begin{figure}[ht]
  \centering
  \includegraphics[width=0.5\textwidth]{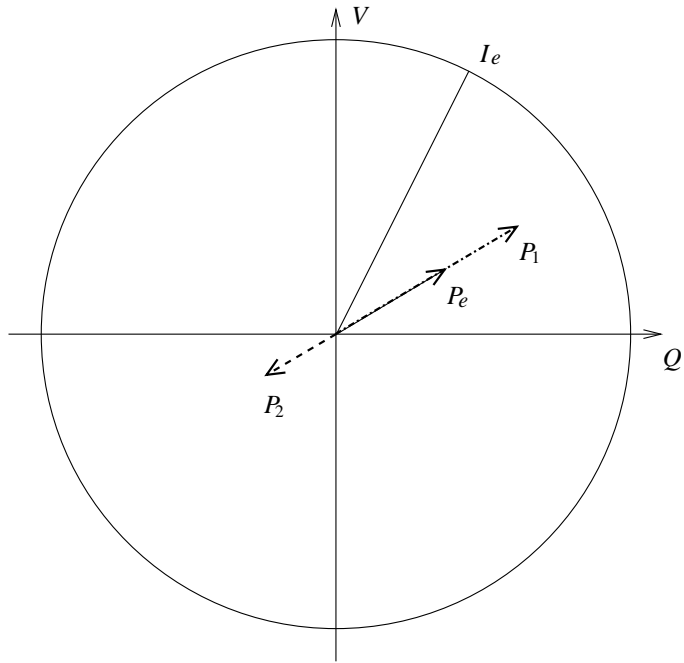}
  \caption{Poincar\'e circle that illustrates the method used to
  determine the intensity and polarisation of two orthogonally
  polarised modes from the observed radiation. $P_1$ and $P_2$
  represent the antiparallel polarisation vectors of both modes. The
  radius of the circle, $I_e$, represents the total intensity of the
  radiation, which is equal to the sum of the lengths of vectors $P_1$
  and $P_2$, because the polarisation modes are superposed
  incoherently. $P_e$ represents the polarisation of the
  electromagnetic radiation which is equal to the vectorial sum of
  $P_1$ and $P_2$. Note that the polarisation vector of the observed
  radiation has been rotated around the $V$ axis to lie in the $Q$-$V$
  plane.}
  \label{fig:poincare}  
\end{figure}

McKinnon and Stinebring have developed a method to separate the modes
of electromagnetic radiation that is built up out of two 100\%
polarised completely orthogonal modes that are superposed
\citep{McKinnon00}. This method is illustrated by means of vectors in
Fig.~\ref{fig:poincare}. Each pulse longitude will have a value
for Stokes $Q$, $U$ and $V$, which can be expressed by a polarisation
vector in $Q$-$U$-$V$-space (Poincar\'e space). The intensity
corresponds to the radius of the sphere. Polarisation vectors lying on
the sphere indicate full polarisation, whereas vectors lying within
the sphere indicate partial polarisation. The fraction of
polarisation is given by the length of the polarisation vector divided
by the radius of the sphere (the intensity). Furthermore, two vectors
with orthogonal polarisation will be antiparallel. The method first
fixes the orientation of the polarisation vectors to lie in the
$Q$-$V$ plane, shown in Fig.~\ref{fig:poincare}, by rotating the
vector around the V axis. In this figure, the average intensity of the
observed radiation is given by the circle with radius $I_e$, the
observed polarisation vector is given by $P_e$ and the underlying
polarisation of the orthogonal modes are given by $P_1$ and
$P_2$. There are now two relationships between the observed intensity
and polarisation vector ($I_e$ and $P_e$) and the polarisation vectors
of the orthogonal modes ($P_1$ and $P_2$). First, $I_e$ is equal to
the sum of the lengths of $P_1$ and $P_2$ and second, $P_e$ is equal
to the vectorial sum of $P_1$ and $P_2$. With these relationships the
two orthogonal polarisation vectors can easily be determined from the
Stokes parameters.

We apply this method to \NPulsars pulsars for which polarimetric
average pulse profiles are available in the EPN database. Given
the assumptions that the radiation is built out of two superposed
orthogonal modes that are 100\% polarised, the separation is
completely correct. Because the superposition is linear, separation of
the average polarisation profile at each longitude will result in the
means of the individual modes at that longitude. Because of this, we
do not see the polarimetric behavior of single pulses, nor any
polarimetric behavior on small timescales. If, for example, the
separation from the average polarisation profile at a fixed longitude
would reveal both modes to be of equal strength, then one solution
might be that at this longitude all single pulses are built out of the
superposition of two modes with equal strength. Yet it might also be
the case that at this longitude one half of the single pulses would be
built out of just one of the orthogonal modes and the rest of the
single pulses would be built out of just the other mode. The data
from the EPN database consists of time-averaged data at multiple
frequencies. Of these pulsars, 11 are known from previous studies to
show OPM. It should be noted that this method gives separated
intensities corresponding to both modes at each longitude, but there
is no direct knowledge as to determine which value belongs to which
mode. We therefore have to be cautious when making a longitude plot of
the intensity of both modes as there are always two possibilities when
connecting intensity points from one longitude to the next. Generally,
it is safe to assign values with the highest intensity to one mode and
values with the lowest intensity to the other. When these values lie
very close to each other, the assignment of value to modes is made on
the basis of visual inspection for changes in modal dominance as
evidenced by sharp jumps of nearly 90$\deg$ in the PA. In some cases
the intensity points are connected by finding a smooth line through
the values. We have defined the polarisation mode with the highest
integrated intensity at the lowest available frequency as the
strongest mode. We thus produce the intensity of the separated modes,
as well as the PA and ellipticity angle
($\frac12\mathrm{atan}(V/\sqrt{Q^2+U^2}$) as a function of frequency.

Furthermore, we have calculated the ratio of the longitude-averaged
intensity of both modes at each frequency. For pulsars with two pulse
components this has been done for both components separately and for
both components added together. The error on the average intensity was
calculated as $\sigma_I = \sqrt{N}\cdot\sigma$, where N is the number
of bins of the profile and $\sigma$ is the noise level, given by
the standard deviation of the noise as estimated from data in the
off-pulse region of the profile. For 14 pulsars, the pulse profiles
at all frequencies were aligned by visual comparison of the profiles
with the corresponding plots from \citet{Kuzmin98}. When a profile at
a particular frequency had no direct comparison, the alignment was
estimated from the frequency development from the plots from
 \citet{Kuzmin98}. Pulsars B0144+59, B1039$-$19, B1800$-$21 and B1944+17
were not listed in this paper and were aligned by comparing pulse
components and characteristics in the polarisation from frequency to
frequency. We can therefore not guarantee that these pulsars are
aligned correctly.

\section{Results}
\label{sec:Results}
\begin{figure}[ht]
  \centering
  \includegraphics[width=0.5\textwidth]{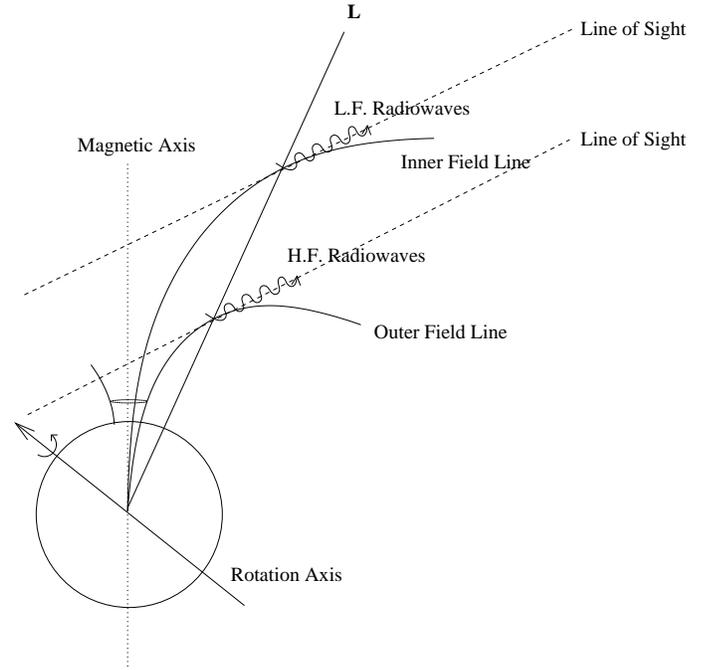}
  \caption{Schematics of two field lines from which radio waves of
  two different frequencies are observed. L is the line that connects
  the locations where the dipolar field lines are directed towards the
  observer. Due to radius-to-frequency mapping, low frequency radio
  waves are emitted higher in the magnetosphere than high frequency
  radio waves. As the image shows, the low frequency radio waves that
  can be observed are emitted on field lines that intersect the
  surface closer to the magnetic axis than the field lines that emit
  the observed high frequency radio waves.}
  \label{fig:fieldlines}  
\end{figure}

If we assume that emission altitude is inversely related to the
frequency of emission (radius to frequency mapping) and that the
magnetic field-lines are dipolar shaped at the emission height,
observing at different frequencies has two implications: (1) radiation
at high frequency arises lower in the pulsar magnetosphere than
radiation at low frequency and (2) radiation observed at high
frequency is emitted at field-lines that intersect the surface further
away from the magnetic axis than radiation observed at the same pulsar
phase at low frequency (see Fig.~\ref{fig:fieldlines}). When studying
the average profiles at multiple frequencies, this should be kept in
mind.

Figures~\ref{fig:0144} through~\ref{fig:2020} show the results of the
mode separation at all frequencies for all pulsars. The left columns
show the average total intensity (solid line) and the average
intensity of both modes (dotted and dashed lines), the middle columns
show the PA of the average Stokes parameters and the right
columns show the ellipticity angle at all frequencies. The PA and
ellipticity angle at pulse longitudes where the polarised
intensity is less than twice the RMS noise are not
shown. Figures~\ref{fig:ratio_of_modes_1}
through~\ref{fig:ratio_of_modes_3} show the frequency development of
the ratio of the average intensities of both modes, plotted on a
log-log scale, for all pulsars. The straight line through the points
was determined by the least squares fitting technique. Whenever
these points show little deviation from the straight line fit, the
related power law index is shown in
Table~\ref{tab:characteristics}. As there is no preferred mode, the
values are always taken to be positive. The point-style indicates the
reference corresponding to the source of the data in the EPN, which is
listed below Fig.~\ref{fig:ratio_of_modes_3}. The polarisation of the
1.41\,GHz data from \citet{Hoensbroech97} was frequently found to
differ from other 1.41\,GHz EPN data and from the development of the
polarisation as a function of frequency. Thus these data were
not included in the analysis.

Let us now consider each pulsar in detail. For some pulsars we will
refer to the classification by \citet{Hoensbroech98}, who have
classified a number of pulsars into four groups, named after one of
the pulsars in that group. The first group of pulsars is
exemplified by type 0355+54, for which one component is highly
polarised, whereas the rest of the profile has little
polarisation. The component that is highly polarised has a flatter
spectral behaviour than the rest of the profile and therefore dominates
at higher frequency. The second group consists of 0525+21-like
pulsars, which, in the classification system
from \citet{Rankin83}, are all examples of conal doubles. They have
two moderately linearly polarised components, the PA follows the
rotating vector model (RVM) of \citet{Radhakrishnan69} nearly
perfectly and there is usually a core-type bridge between the
components. The third group consists of 1800$-$21-like pulsars,
which are young, have a very high spin down power, $\dot{E}$, a
relatively flat radio spectrum behaviour, detectable
X-ray emission and radio emission over a wide fraction of the pulse
period. They're also nearly fully polarised, linearly and circularly
and do not show the commonly observed effect of depolarisation towards
high frequencies. The final group consists of 0144+59-like
pulsars, which show an unusual polarisation behaviour. Their degree of
circular polarisation increases strongly with frequency.

\subsection*{B0144+59}
Pulsar B0144+59 defines the class of 0144+59-like pulsars, as
described above.  Figure~\ref{fig:0144} shows that at all frequencies
except 4.85\,GHz there is a mode-switch around $-2$$\deg$, which is
associated with a jump in the PA.  At 0.4\,GHz the first component can
be associated with one mode, whereas the second and third components
seem to be associated with the other mode. Although the
intensity fluctuations around 2$\deg$, which suggest a second
and third component, are not significant and might be just one
component. The association of the modes with components is also valid
at higher frequencies, even when the first component becomes
weaker and almost disappears at 4.85\,GHz. At this frequency the
remaining component is almost completely dominated by one mode. Also,
the slope of the position angle traverse increases. This pulsar shows
an example of a component (left component) being dominated by a single
mode and slowly disappearing with increasing frequency. It would be
interesting to see whether this trend continues towards even lower
frequencies. Unfortunately, such data was not at our disposal. The
frequency development of these modes is nicely fit by a power law.

\subsection*{B0355+54}
Pulsar B0355+54 defines the class of 0355+54-like pulsars, as
mentioned above. Figure~\ref{fig:0355} shows that the leading
component consists of mainly one mode (the stronger mode) and the
second component is dominated by the other mode. At all frequencies,
except perhaps for 10.55\,GHz, the PA jumps and the ellipticity angle
switches sign at the same longitude where the modes change
dominance. At low frequency the dominance of the weaker mode is only a
few degrees in longitude, but becomes broader with increasing
frequency. This mode can be associated with the second component, but
the stronger mode is present there as well. The first component rises
in intensity at higher frequencies, relative to the intensity of the
first component, which contains only one mode. This behavior has
already been noted by \citet{Hoensbroech98}. The PA sweep of the left
component is straight and has a small slope, which slightly increases
with increasing frequency. The right component shows the 90$\deg$
jump, after which the sweep remains straight for a few degrees (the
number of degrees depending on frequency) and then has a smaller jump,
followed by a curved sweep. The second jump in the PA can be
associated with a second change in dominance of the two
modes. However, at high frequencies this is not clear. It is also not
clear whether the curve at the trailing edge of the PA is the natural
PA sweep of one of the modes, or that the curve is the result of a
(non-orthogonal) superposition of both modes. The frequency
development of the modes is nicely fit by a power law, except for a
jump at 1.642\,GHz.

\subsection*{B0450+55}
Pulsar B0450+55 falls into the same group of pulsars as B0355+54
 \citep{Hoensbroech98}. Figure~\ref{fig:0450} shows that at a
 frequency of 0.910\,GHz the PA jumps 90$\deg$ twice around 0$\deg$
 longitude, indicating a brief mode-change which is visible in the
 average profile of the modes. The first jump shows a gradual
 change. The ellipticity angle switches sign at the same longitude in
 accordance with the mode-change. At lower frequencies the PA also
 jumps twice at the same longitudes, but this does not show up in the
 average profile of the modes. These jumps are less than 90$\deg$ and
 also show a gradual change. It is interesting that at 0.404\,GHz the
 first jump is upwards while at 0.610\,GHz the first jump is
 downwards, while the rest of the PA sweep is very alike. The first
 component of the average intensity is strongly dominated by one mode
 and the intensity of this component increases with frequency relative
 to the second component, in which both modes are present and nearly
 equally strong. Because it is dominated by one mode,
 Fig.~\ref{fig:ratio_of_modes_1} does not show the frequency
 development of the modes in the left component. For both the right
 component and the entire profile, the frequency development of the
 modes do not seem to be fit by a power law.

\subsection*{B0525+21}
Pulsar B0525+21 defines the class of B0525+21-like pulsars, as
described above. Figure~\ref{fig:0525} shows that both modes are
active in both components. At high frequency, there is strong
depolarisation in both components, which is represented by the weaker
mode becoming stronger at higher frequency. The modes in both
components behave very similarly, although the ratio between integrated
intensity of the stronger and weaker mode is higher in the right
component. The intensity of this component increases with frequency,
relative to the left component. The PA sweep at 4.85\,GHz shows a
90$\deg$ phase jump, suggesting a mode-switch somewhere in the bridge
between the two components. However, the frequency development of the PA
sweep clarifies that this jump is actually the result of the normal
RVM. The ellipticity angle gives little information about mode-switching,
because of the low circular polarisation. This leaves us with no
indication of mode-switching in this pulsar.  The frequency development
of the modes are nicely fit by a power law in both components.

\subsection*{B0809+74}
The polarisation of pulsar B0809+74 has been well studied in the
past. Recently, \citet{Rankin04} has shown that the two orthogonal
modes of polarisation are the primary reason for a variation in the
longitudinal separation between the drifting subpulses of pulsar
B0809+74. Furthermore, \citet{Edwards04} has confirmed that the
polarisation fluctuations are the result of two orthogonally polarised
modes, although an apparantly randomly polarised component needs to be
superposed. The pulsar has two components. The first component arises
with increasing frequency and becomes visible around 0.925\,GHz and
higher and is almost completely dominated by the stronger mode (see
Fig.~\ref{fig:0809}). In the second component both modes are almost
equally strong. At frequencies of 0.606 and 1.408\,GHz, the PA jumps
by approximately 60$\deg$ and 80$\deg$, respectively. This does not
seem to be associated with a mode-change.  There is no clear frequency
development of the modes.

\subsection*{B0823+26}
Pulsar B0823+26 has only one component, in which both modes are active
(see Fig.~\ref{fig:0823}). Both the profile and the relative
intensity of the modes do not change very much over frequency. Around
the middle of the profile there is a small jump in the PA sweep, which
increases in size with increasing frequency. At the same longitude as this
jump the modes are almost of equal strength. Also, the weaker mode
becomes stronger with respect to the stronger mode at higher
frequency. The frequency development of the modes are nicely fit by a
power law.

\subsection*{B0834+06}
Pulsar B0834+06 consists of two components with a high bridge
connecting them. From studies of individual-pulse polarisation
observations it has been concluded that both modes are equally strong
in the two components, but only one mode dominates across the bridge
 \citep{Stinebring84}. However, Fig.~\ref{fig:0834} shows that the
difference in average intensity between the modes is small across the
bridge as well. The PA sweep at 0.408\,GHz and 1.408\, GHz are very
similar, however, the PA sweep at 0.610\,GHz has a drop in the center
of the profile. Except for one point with a large error, the frequency
development of the modes can be fit by a power law. But since
there are only four points in total, this is not very significant.

\subsection*{B0919+06}
The average profile of pulsar B0919+06 changes greatly over frequency.
At low frequency there seem to be two components visible, as well as a
slope at the leading edge. With increasing frequency the two
components merge into one and the slope becomes smaller and even
disappears at 4.85\,GHz. The profile is dominated by one mode, with
the exception of the slope and part of the left component, in which
both modes are almost equally strong (see Fig.~\ref{fig:0919}). The
weaker mode moves to the right with respect to the stronger mode, at
rising frequency. This results in the extra component at low frequency
(0.408\,GHz), associated with the weaker mode, to disappear towards
higher frequencies. There is no indication of mode-switching. The
frequency development of the modes seem to be fit well by a power law,
with the exception of the datapoint at 4.85\,GHz.

\subsection*{B0950+08}
Pulsar B0950+08 is remarkably similar to pulsar B0919+06. They both
have one dominant component a weak second component at low frequencies
and a slope at the leading edge. However, the polarisation modes of
pulsar B0950+08 are almost equally strong over the entire frequency
range and do not show any drift (see Fig.~\ref{fig:0950}). There is
no evidence of a mode-switch in the average profile. The PA does rise
and fall where the modes have nearly equal power, but only by
$70\deg$. The rise in the PA becomes broader with increasing frequency
and at 1.642 and 4.850\,GHz, the PA has changed significantly. At
1.642\,GHz it is clear that the PA drops instead of rises and by a
smaller amount than at lower frequency. The ellipticity angle falls
and rises at the same longitude at which the PA changes. This dip also
becomes broader towards higher frequency. There does not seem to be a
clear frequency development of the modes.

\subsection*{B1039$-$19}
Pulsar B1039$-$19 is B0525+21-like. Figure~\ref{fig:1039} shows that
both modes are active in both components and that the PA does indeed
follow the RVM. At higher frequency, the weaker mode becomes
stronger. At low frequency the PA sweep shows a typical RVM curve, but
towards higher frequency the sweep starts with a rise and then drops
in a straight line.  For the first components, the frequency development
of the modes seems to be fit by a power law. For the second
component the fit is not so good, but the same trend is visible.

\subsection*{B1133+16}
This pulsar has two components. In the left component both modes are
almost equally strong (see Fig.~\ref{fig:1133}). The right component
is dominated by one mode at low frequencies, but the modes become
almost equally strong at higher frequencies. This behavior is the
opposite of what would be expected by comparing the results on the
polarisation of B1133+16 from \citet{Backer80} (hereafter BR) with
the results from \citet{Stinebring84} (hereafter SCRWB). It can be
seen there, that at low frequency the right component shows OPM
(Fig. 9 in BR), which becomes less visible at higher frequency
(Fig. 17 in SCRWB). However, OPM are expected to occur when both
modes are about equal in strength which, according to our results, is
at high frequency rather than at low frequency. This could indicate
that the method for separating the modes is not succesful in this
case. However, we must keep in mind that we are only showing average
profiles. If the (refracted) path of one of the polarisation modes in the
magnetoactive plasma can vary in time, then the strength of the
polarisation modes can briefly change, possibly making OPM visible in
single pulse studies. At 1.408\,GHz the PA jumps upward by 90$\deg$ and jumps
downward again one degree later, which results in a mode-change, as
can be seen in the plot of the average intensity of the modes. At
1.642\,GHz this jump is still present, but has become more gradual, is
less than 90$\deg$ and does not result in a mode-change. At 1.71 and
4.85\,GHz these jumps are also present, but are first downward, then
upward. At 4.85\,GHz the average intensity profile of the modes show a
brief mode-change. The modes in the left component seems to have a
clear frequency development, whereas the modes in the right component do
not.

\subsection*{B1237+25}
Pulsar B1237+25 is known for its profile ``mode-swit\-ching'' (which
means that it has two different average profiles, called ``modes'',
which are not to be confused with the orthogonal polarisation
modes) \citep{Bartel82}. By comparing the EPN profiles of pulsar
B1237+25 with those in \citep{Bartel82}, we conclude that the abnormal
mode can only be faintly present in the average pulse profiles.  There
are many components in this pulsar, which are all dominated by the
stronger mode (see Fig.~\ref{fig:1237}). While the weak component at
a longitude of -3$\deg$ can be associated almost completely with the
stronger mode, the weak component at a longitude of 1$\fdg$5 can be
associated with the weaker mode. These associations are strongest at
low frequencies. At higher frequency, the weaker mode becomes stronger
with respect to the stronger mode. According to BR, this pulsar shows
OPM in the two main components at a frequency of 0.430\,GHz, which, as
in the right component of pulsar B1133+16, is not consistent with the
result of one dominant mode. At almost all frequencies, the position
angle rises and falls along the bridge between the two dominant
components. At this same longitude, there is a lot of circular
polarisation. This discontinuity in the PA is not associated with
OPM \citep{Stinebring84}, but it does occur at a longitude where the
weaker mode becomes stronger with respect to the stronger mode. Both
the frequency development of the intensity of the modes in the left
and right half of the profile follows a power law, but the modes in
the right half have a very low power law index.

\subsection*{B1737$-$30}
Pulsar B1737$-$30 has only one component and falls into the same group
of pulsars as B0144+59 \citep{Hoensbroech98}. At a frequency of
1.408\,GHz and higher, the component is almost completely dominated by
one polarisation mode. At the trailing edge of the component the
amount of circular polarisation increases. This behaviour is intrinsic
to the modes themselves, as there is mainly one mode active throughout
the profiles. The frequency development of the intensity of the modes
seem to follow a power law. 

\subsection*{B1800$-$21}
Pulsar B1800$-$21 defines the class of 1800$-$21-like pulsars, as
described above. At low frequency this pulsar has up to 4 components,
but at high frequency this pulsar has only two components, both
dominated by the stronger mode (see Fig.~\ref{fig:1800}). The first
component arises at a frequency of 1.408 GHz. At 1.408, 1.56, 1.642
and 4.852\,GHz there is a gap in the PA due to the low signal in the
bridge between the two main components. The PA at 4.85\,GHz shows the
PA sweep over this bridge and clarifies that the stronger mode is
dominant in both components. There is, however, a brief mode-switch in
the bridge between the two main components at 1.408\,GHz and another
one at the trailing tail of the second component.  At frequencies
below 4.85\,GHz the peak of the weaker mode is separated by
approximately 15$\deg$ ($\approx 5.6$\,ms) from the peak of the
stronger mode. At a frequency of 4.85\,GHz, the weaker mode has almost
completely disappeared. There is no clear frequency development of the
intensity of the modes, but it is clear that the weaker mode becomes
weaker with increasing frequency.

\subsection*{B1929+10}
The profile of pulsar B1929+10 changes greatly over frequency, but at
all frequencies there is only one dominant mode (see
Fig.~\ref{fig:1929}). There is OPM at the trailing part of the main
component \citep{Stinebring84}, which coincides with the rise of the
weaker component. At 4.85\,GHz the position angle suggests that there
is a mode-switch in the leading edge of the profile. Due to the low
intensity it is difficult to confirm this with the intensity plots. It
is interesting that the weaker mode retains its shape and position,
whereas the components of the stronger mode change greatly in shape
and position. This is especially evident at 4.85\,GHz where one might
naively expect the two components in the average intensity profile to
lie at the same longitude as the second and third components of the
1.71-GHz profile, rather than to coincide with the first and second
components. However, using the alignment from \citet{Kuzmin98} the
weaker mode does not change in longitude over frequency, which
suggests the alignment to be correct. There is no clear frequency
development of the intensity of the modes, but it seems that the
weaker mode becomes stronger with increasing frequency.

\subsection*{B1944+17}
Pulsar B1944+17 switches modes at the leading and trailing edge of the
pulse at 0.610 and 1.408\,GHz (see Fig.~\ref{fig:1944}). Overall,
the weaker mode has different intensity at different frequencies. It
is roughly 1/4 of the dominant mode at frequencies of 0.410 and
0.925\,GHz and about 1/2 of the strength of the dominant mode at
frequencies of 0.61, 1.408 and 4.85\,GHz. However, there is no clear
frequency development of the intensity of the modes.

\subsection*{B2016+28}
The profile of pulsar B2016+28 has two components, which become less
distinct towards higher frequency (see
also \citep{Izvekova93}). Fig.~\ref{fig:2016} shows that these
components can be associated with the two different modes. At low
frequency, the weaker mode consists of two components and of only one
component at higher frequency. The peak of the modes move towards each
other with increasing frequency and become of almost equal
strength. At frequencies higher than 1.408\,GHz, the shapes of the two
modes are almost similar. Due to this behavior, the left part of the
profile is dominated by one mode at low frequencies while it is
composed of two equally strong modes at high frequency.  This pulsar
is known for its mode-changing \citep{Stinebring84, Arons86}. This can
be seen at a frequency of 0.408\,GHz, where the modes switch
dominance. At the same longitude the PA jumps by approximately
90$\deg$, however the PA at 1.408\,GHz has a similar behavior and does
not seem to be associated with a mode-change. This suggests that the
jump in the PA at 0.408\,GHz and 1.408\,GHz are due to the RVM and
respresent the PA sweep of the dominant mode. Moreover, at 0.610 and
0.925\,GHz the PA jumps downwards and does not seem to be associated
with a mode-change. It therefore remains unclear what the true PA
sweep of this pulsar is. There does not seem to be a frequency
development of the intensity of the modes.

\subsection*{B2020+28}
\begin{figure}[ht]
  \resizebox{\hsize}{!}{\includegraphics{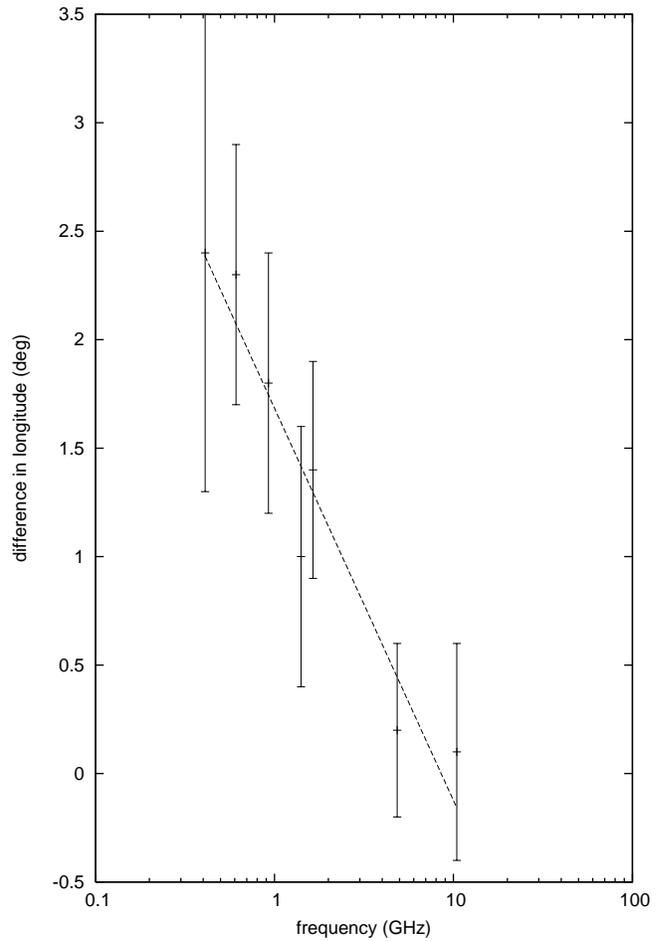}}
  \caption{Separation in longitude between the peaks of the average
  intensity of the two orthogonally polarised modes of  \object{PSR B2020+28} as
  a function of frequency. The dashed line is the best straight line
  fit through the points.}
  \label{fig:drift}
\end{figure}

Pulsar B2020+28 is known for its abrupt mode-change \citep{Cordes78,
Stinebring84, Arons86}. This can be seen in the PA at all frequencies
(see Fig.~\ref{fig:2020}). This pulsar also shows a large drift of
the modes with respect to each other. Fig~\ref{fig:drift} shows the
difference in longitude between the peaks of the two modes as a
function of frequency. (The large errors are because of the
uncertainty in the position of the peak of each mode.) The line
through the points is a fit of the function $a\log(f)+b$ of the
points, where $f$ is the frequency in GHz and $a$ and $b$ are fit
parameters. The values of these fit parameters are $a=0\fdg56 \pm
0\fdg11$ and $b=8\fdg54 \pm 0\fdg13$. There does not seem to be a
frequency development of the intensity of the modes.

\begin{table*}
  \caption{List of characteristics of the orthogonally polarised modes of analysed pulsars}
  \begin{tabular}{lccccccccc}
  \hline\hline
  Pulsar & 0144+59 & 0355+54 & 0450+55 & 0525+21 & 0809+74 &
  0823+26 & 0834+06 & 0919+06 & 0950+08 \\
  \hline
  Mode-changing & $\surd$ &  $\surd$ & $\surd$ &  $-$     & $-$     & $-$     & $-$ & $-$     & $-$ \\
  \hline
  Modes drift & $-$ & $-$ & $-$ & $-$ & $-$ & $-$ & $-$ & $\surd$& $-$ \\
  \hline
  Power law index  & 0.502 & 0.315 & & 0.316 & & 0.137 & 0.115 & 0.198& \\
  of ratio of modes  & $\pm$0.02 & $\pm$0.004 &  $-$ & $\pm$0.005&  $-$ & $\pm$0.003& $\pm$0.011 & $\pm$0.010 &  $-$ \\
  \hline
  Are modes associated & & & & & & & & & \\
  with components & $\surd$ & $\surd$ & $\surd$ & $-$& $\surd$& $-$ & $-$ & $\surd$ & $-$  \\
  \hline
  Jump in the PA but  & & & & & & & & & \\
  no mode-switch & $-$ & $-$ & $\surd$& $-$ & $\surd$ & $\surd$ & $-$ & $-$ & $\surd$ \\
  \hline
  Period (s) & 0.20 & 0.16 & 0.34 & 3.75 & 1.29 & 0.53 & 1.27 & 0.43  & 0.25 \\
  \hline
  Type & T& St & T  & D & Sd& St& D & T & Sd \\
  \hline
  $\alpha$ ($\deg$)  & $-$& 51 & 32 & 21 & 9 & 84 & 30 & 48 & 12 \\
  \hline
  $\beta$ ($\deg$) & $-$  & +4.4 & +3.3 & +0.6 & +4.5 & +1.9  & +3.4 & +4.8 & +8.5 \\
  \hline
  $w_{50}$ (mP)  & 35 & 25 & 24 & 48 & 32 & 13 & 18 & 23 & 42 \\
  \hline
  $w_{10}$ (mP) & 51 & 69 & 82 & 56 & 72 & 28 & 25 & 55 & 86 \\
  \hline
  $S_{400}$ (mJy) & 6  & 55 & 60 & 60 & 80 & 65 & 85 & 60 & 400 \\
  \hline
  $S_{1400}$ (mJy) & 2  & 25 & 13 & 9  & 10 & 10 & 4  & 4  & 85 \\
  \hline
  log$\dot{E}(\mathrm{ergs/s})$ & 33.1& 34.7& 33.4& 31.5& 30.5& 32.7& 32.1& 33.8& 32.7 \\
  \hline
  \hline
  Pulsar        &  1039$-$19 & 1133+16 & 1237+25 &  1737$-$30 & 1800$-$21 & 1929+10 & 1944+17 & 2016+28 & 2020+28 \\
  \hline
  Mode-changing & $-$ & $\surd$ & $-$ & $-$ & $\surd$ & $-$ & $\surd$ & $\surd$ &$\surd$ \\
  \hline
  Modes drift   & $-$ & $-$ & $-$ & $-$ & $-$ & $-$ & $-$ & $-$ & $\surd$ \\
  \hline
  Power law index & 0.32 & 0.234 & 0.248 & 2.82 & & & & & \\
  of ratio of modes & $\pm$0.03 &$\pm$0.002~[1]&$\pm$0.006~[1]& $\pm$0.05&$-$ &$-$ &$-$ &$-$ & $-$ \\
  \hline
  Are modes associated & & & & & & & & & \\
  with components & $-$ & $-$ & $\surd$ & $\surd$ & $\surd$ & $\surd$ & $-$  & $\surd$& $\surd$ \\
  \hline
  Jump in the PA but  & & & & & & & & & \\
  no mode-switch & $-$ & $\surd$ & $\surd$ & $-$ & $-$ & $-$ & $-$ & $-$ & $-$ \\
  \hline
  Period (s)& 1.39 & 1.19 & 1.38 & 0.61 & 0.13 & 0.23 & 0.44 & 0.56 & 0.34  \\
  \hline
  Type & M & D & M & St & M & T & Tc & Sd & T \\
  \hline
  $\alpha$ ($\deg$)  & 31 & 46 & 53 & $-$ & $-$ & 88 & 19 & 39 & 72 \\
  \hline
  $\beta$ ($\deg$) & +1.7 & +4.1 & 0  &  $-$ & $-$ & $-$  & +6.1 & +7.2 & +3.6 \\
  \hline
  $w_{50}$ (mP) & 43 & 23 & 36 & 9 & 97 & 27 & 40 & 25 & 37 \\
  \hline
  $w_{10}$ (mP) & 50 & 31 & 41 & 21 & 321& 66 & 102& 37 & 49 \\
  \hline
  $S_{400}$ (mJy) & 13 & 300& 110&16.7& 20 & 250& 35 & 320& 110\\
  \hline
  $S_{1400}$ (mJy) & 4 & 30 & 10 & 6 & 15 & 40 & 10 & 30 & 38 \\
  \hline
  log$\dot{E}(\mathrm{ergs/s})$ & 31.1& 31.9& 31.2& 34.9& 36.3& 33.6& 31.1& 31.5& 33.3 \\
  \hline
  \end{tabular}
  \footnotesize [1] Left component only
  \label{tab:characteristics}
\end{table*}

\section{Discussion}
\label{sec:Discussion}
Data from the EPN is very suitable to apply the mode-separation
technique for superposed modes, as prescribed by
\citet{McKinnon00}. This makes a frequency analysis of OPM
possible. Applying this technique to EPN data has three drawbacks (1)
the pulse-profiles at different frequencies are not aligned, (2) the
information on the observed pulsar flux is seldom available and (3)
the data only show average profiles, which might not represent the
behavior of single pulses. Still, the results described here, give a
clear picture of how the average properties of the two modes behave as
a function of frequency.

In Table~\ref{tab:characteristics} we summarise the results of the
analysis of the orthogonally polarised modes of the \NPulsars pulsars,
as well as some of the known parameters. $w_{50}$ and $w_{10}$ are the
pulse widths at 50\% and 10\% of the peak flux, respectively, $S_{400}$ and
$S_{1400}$ are the flux densities at 0.4 and 1.4\,GHz respectively and
$\dot{E}$ is the spin-down luminosity, given by $-I\Omega\dot\Omega$,
where $I$ is the pulsar's moment of inertia for which we take
$10^{45}$\,g\,cm$^2$ and $\Omega$ and $\dot\Omega$ are the rotation
frequency and the time derivative of the rotation frequency of the
pulsar, respectively. Values for the period, pulse widths and flux
densities are from \citet{Taylor93}. Type and $\alpha$ and $\beta$
parameters are from \citet{Rankin93}, except for pulsars B0144+59 and
B1800$-$21 for which we have determined the type ourselves from the
polarisation characteristics. The types are based on the
classification system from \citet{Rankin83} which assumes a core
radiation zone, surrounded by a conal radiation zone. They have the
following meaning. S$_d$ is a single component profile; the line of
sight grazes the outer cone of the polar cap. D is a double component
profile; the line of sight cuts through the outer cone twice, but
misses the core component. T is a triple component profile; the line
of sight cuts through the outer cone twice as well as the core. S$_t$
is a single component profile; in this case the outer cone is missing
and the line of sight only cuts through the core. Finally there are
two types that, in this view, can only be explained by two conal
radiation zones around the core component; they are M, which is a
multiple component profile (containing usually 5 components) and
T$_c$, which is a triple component profile, which consist of three
conal components.
 
Let us now consider some of the observed phenomena.

\subsection*{Mode-changing}
\begin{figure*}
  \centering
  \includegraphics[width=17cm]{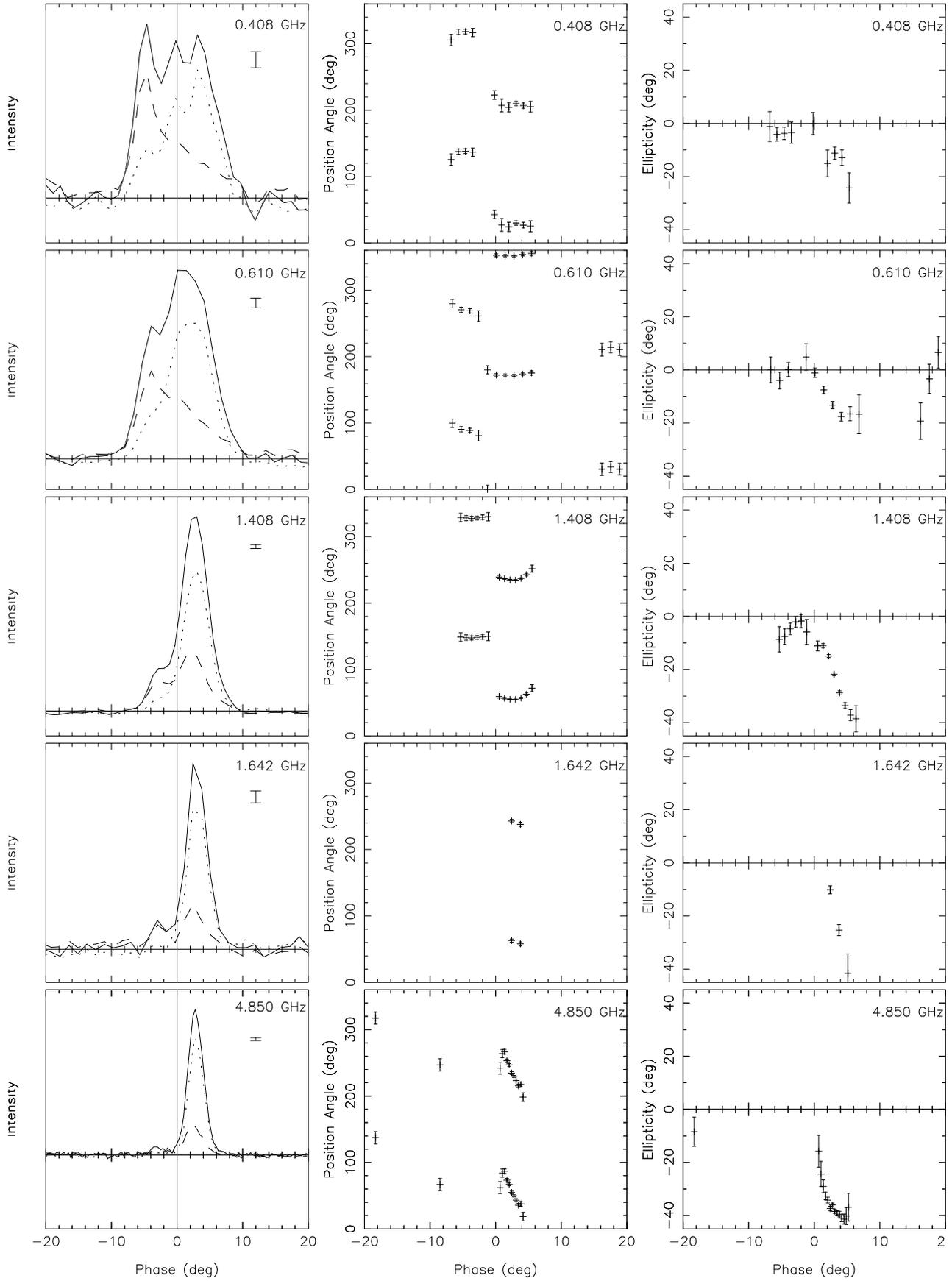}
  \caption{Total Intensity, intensity of orthogonal modes, position
  angle and ellipticity angle as a function of frequency of
  \object{PSR B0144+59}.  The error bars represent the noiselevel, given by the
  rms of the
  off-pulse region of the profile.}
  \label{fig:0144}
\end{figure*}

\begin{figure*}[ht]
  \centering
  \includegraphics[width=17cm]{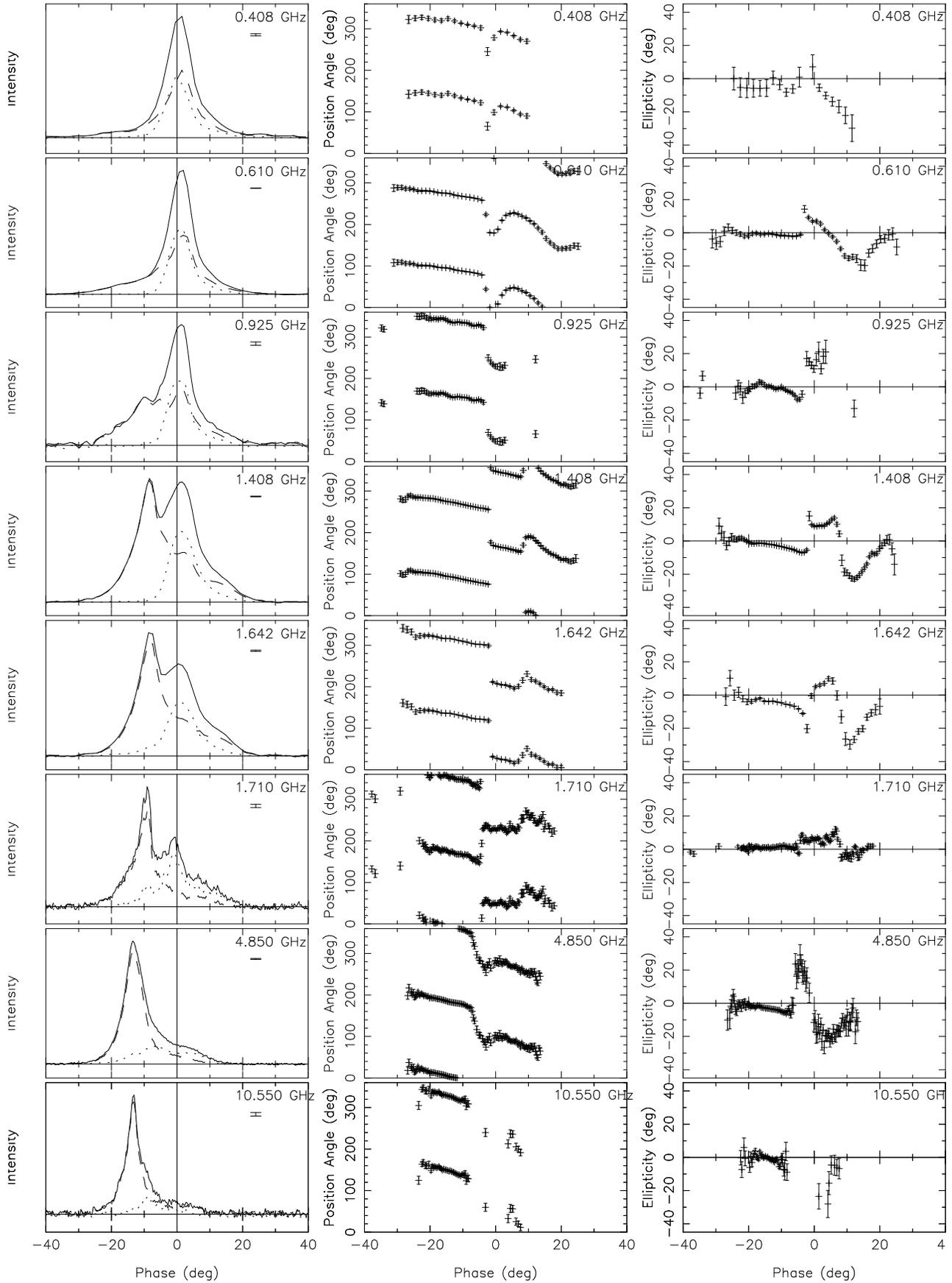}
  \caption{Total Intensity, intensity of orthogonal modes, position
  angle and ellipticity angle as a function of frequency of
  \object{PSR B0355+54}.  The error bars represent the noiselevel,
  given by the rms of the off-pulse region of the profile.}
  \label{fig:0355}
\end{figure*}

\begin{figure*}[ht]
  \centering
  \includegraphics[width=17cm]{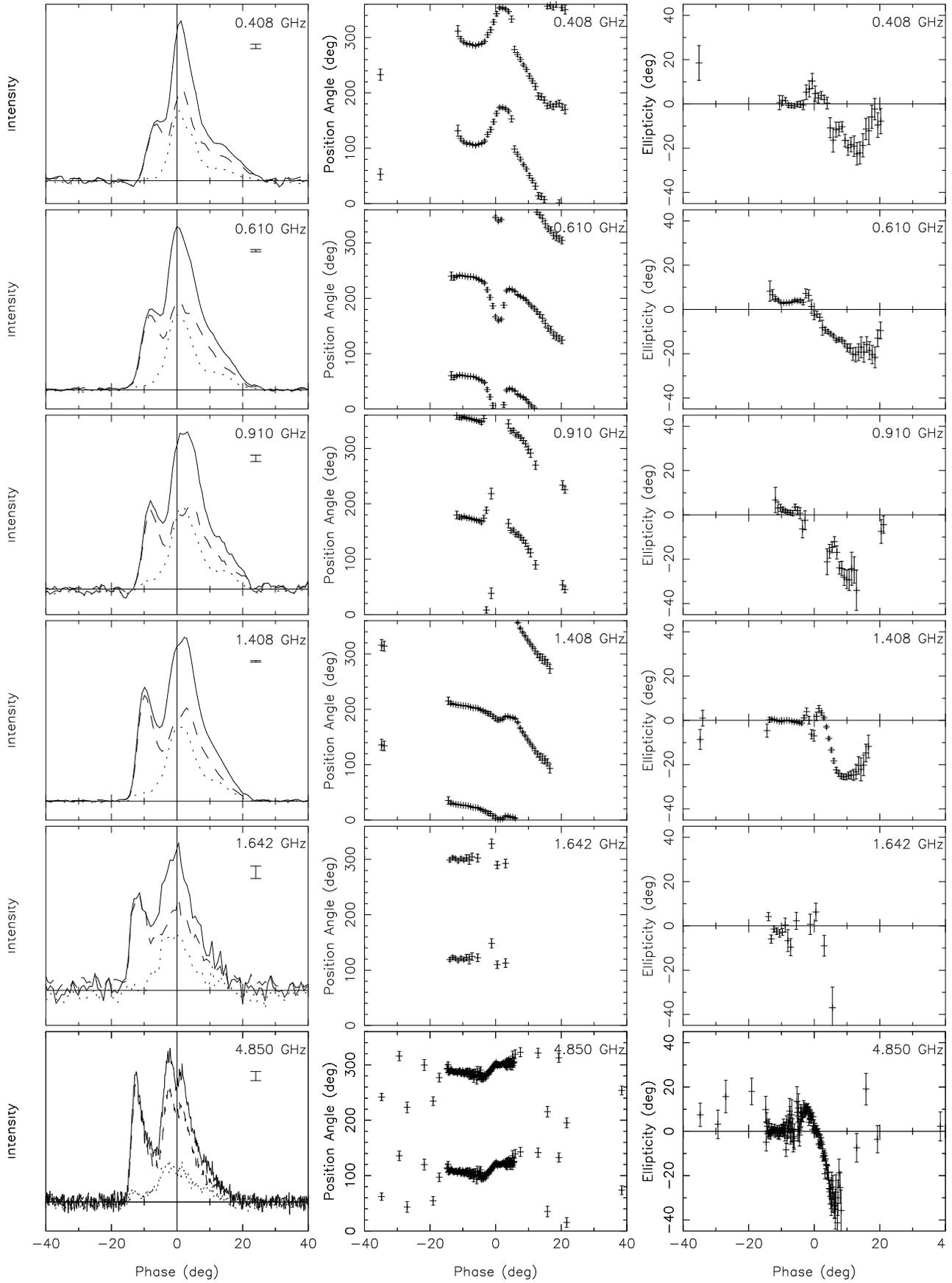}
  \caption{Total Intensity, intensity of orthogonal modes, position
  angle and ellipticity angle as a function of frequency of
  \object{PSR B0450+55}.  The error bars represent the noiselevel,
  given by the rms of the off-pulse region of the profile.}
  \label{fig:0450}  
\end{figure*}

\begin{figure*}[ht]
  \centering
  \includegraphics[width=17cm]{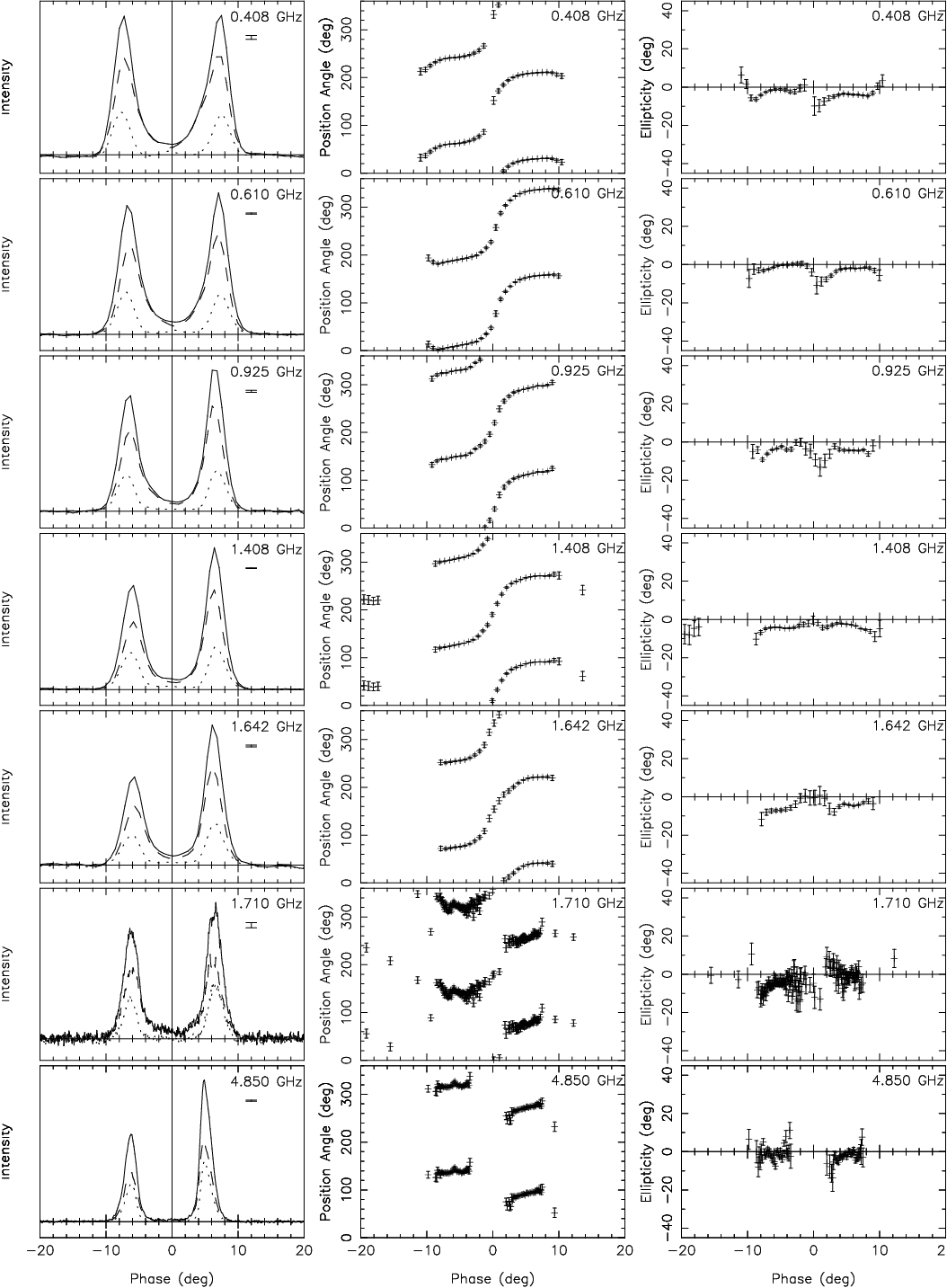}
  \caption{Total Intensity, intensity of orthogonal modes, position
  angle and ellipticity angle as a function of frequency of
  \object{PSR B0525+21}.  The error bars represent the noiselevel,
  given by the rms of the off-pulse region of the profile.}
  \label{fig:0525}  
\end{figure*}

\begin{figure*}[ht]
  \centering
  \includegraphics[width=17cm]{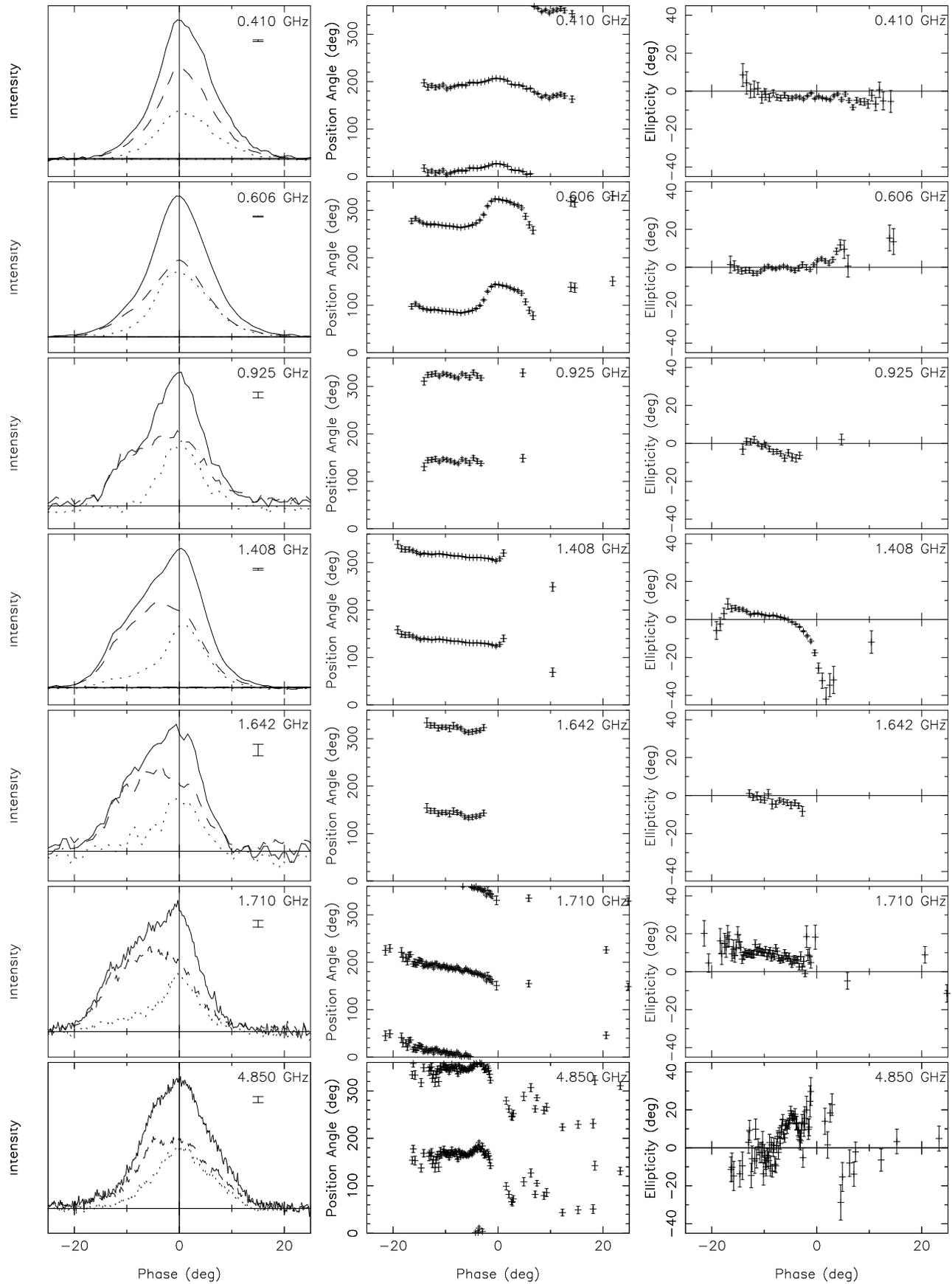}
  \caption{Total Intensity, intensity of orthogonal modes, position
  angle and ellipticity angle as a function of frequency of
  \object{PSR B0809+74}.  The error bars represent the noiselevel,
  given by the rms of the off-pulse region of the profile.}
  \label{fig:0809}  
\end{figure*}

\begin{figure*}[ht]
  \centering
  \includegraphics[width=17cm]{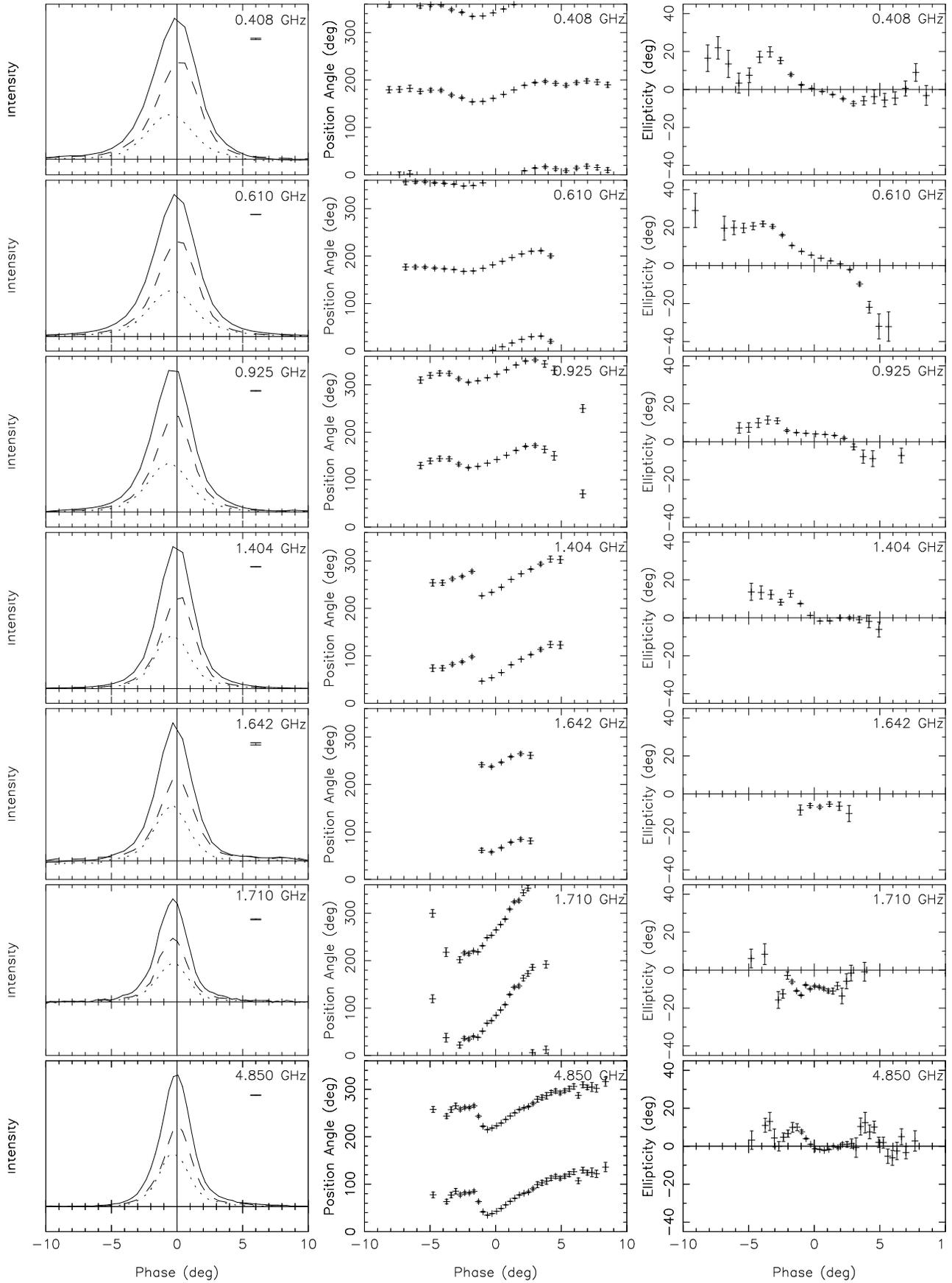}
  \caption{Total Intensity, intensity of orthogonal modes, position
  angle and ellipticity angle as a function of frequency of
  \object{PSR B0823+26}.  The error bars represent the noiselevel,
  given by the rms of the off-pulse region of the profile.}
  \label{fig:0823}  
\end{figure*}

\begin{figure*}[ht]
  \centering    
  \includegraphics[width=17cm]{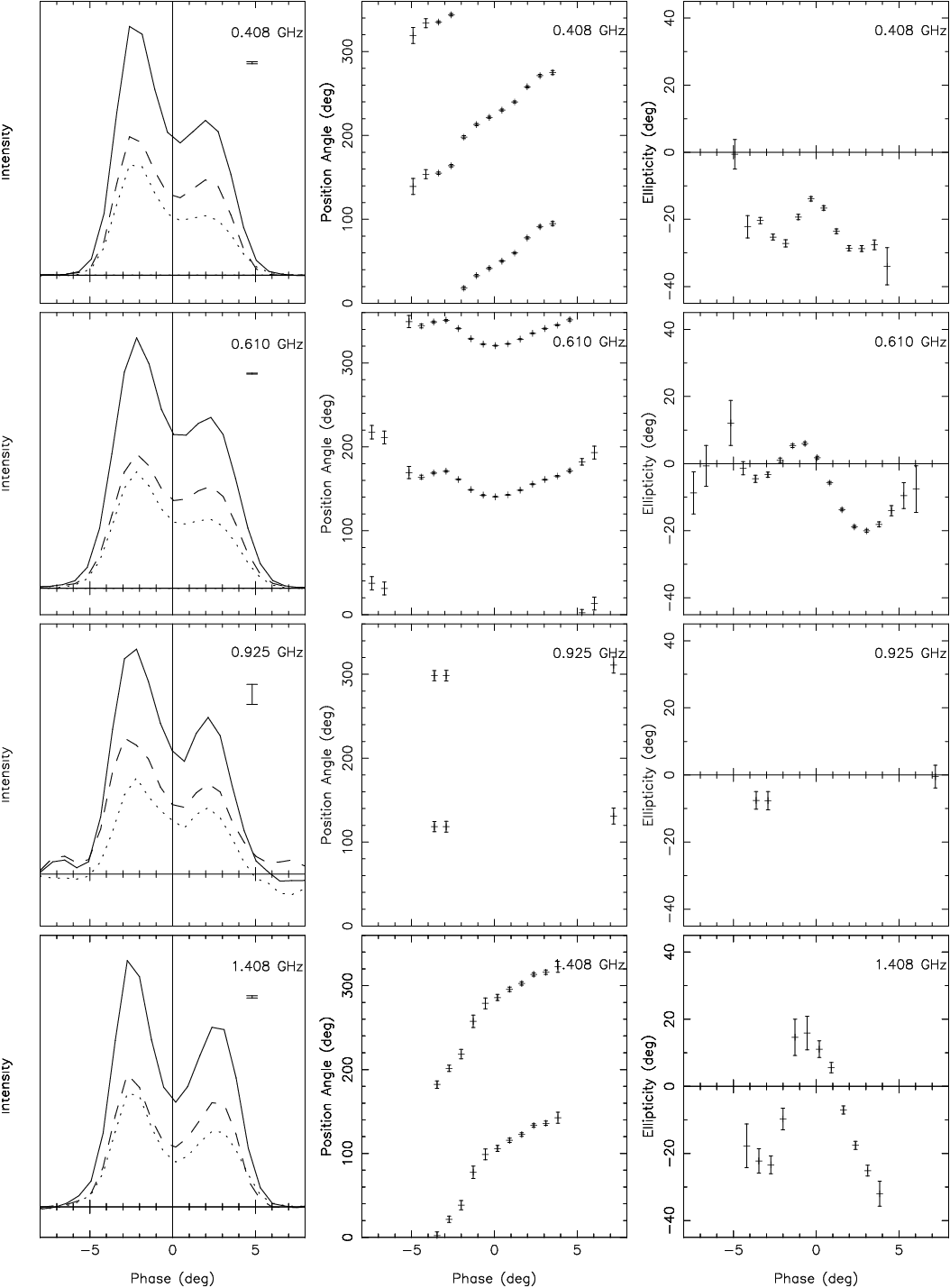}
  \caption{Total Intensity, intensity of orthogonal modes, position
  angle and ellipticity angle as a function of frequency of
  \object{PSR B0834+06}.  The error bars represent the noiselevel,
  given by the rms of the off-pulse region of the profile.}
  \label{fig:0834}  
\end{figure*}

\begin{figure*}[ht]
  \centering   
  \includegraphics[width=17cm]{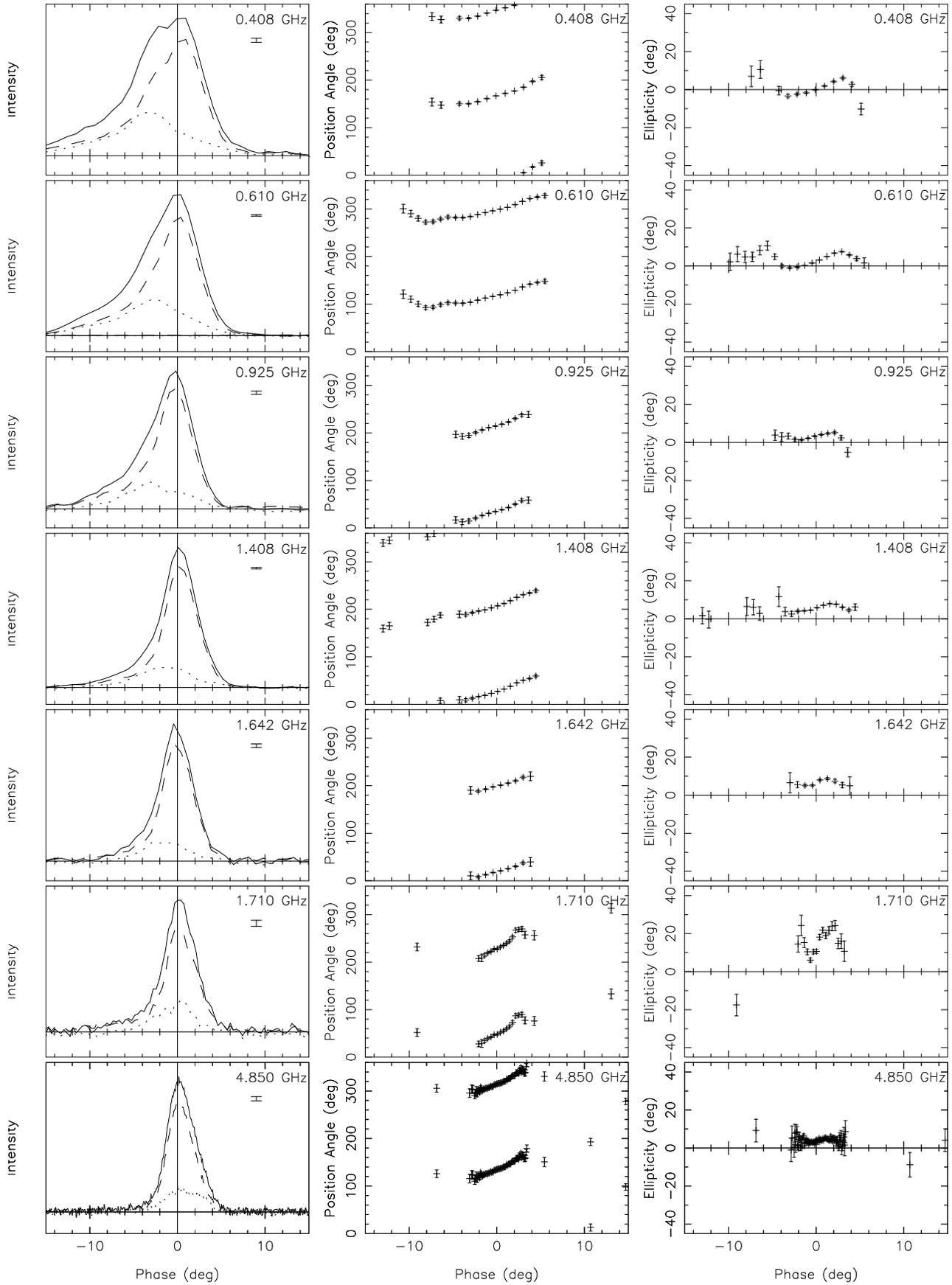}
  \caption{Total Intensity, intensity of orthogonal modes, position
  angle and ellipticity angle as a function of frequency of
  \object{PSR B0919+06}.  The error bars represent the noiselevel,
  given by the rms of the off-pulse region of the profile.}
  \label{fig:0919}  
\end{figure*}

\begin{figure*}[ht]
  \centering   
  \includegraphics[width=17cm]{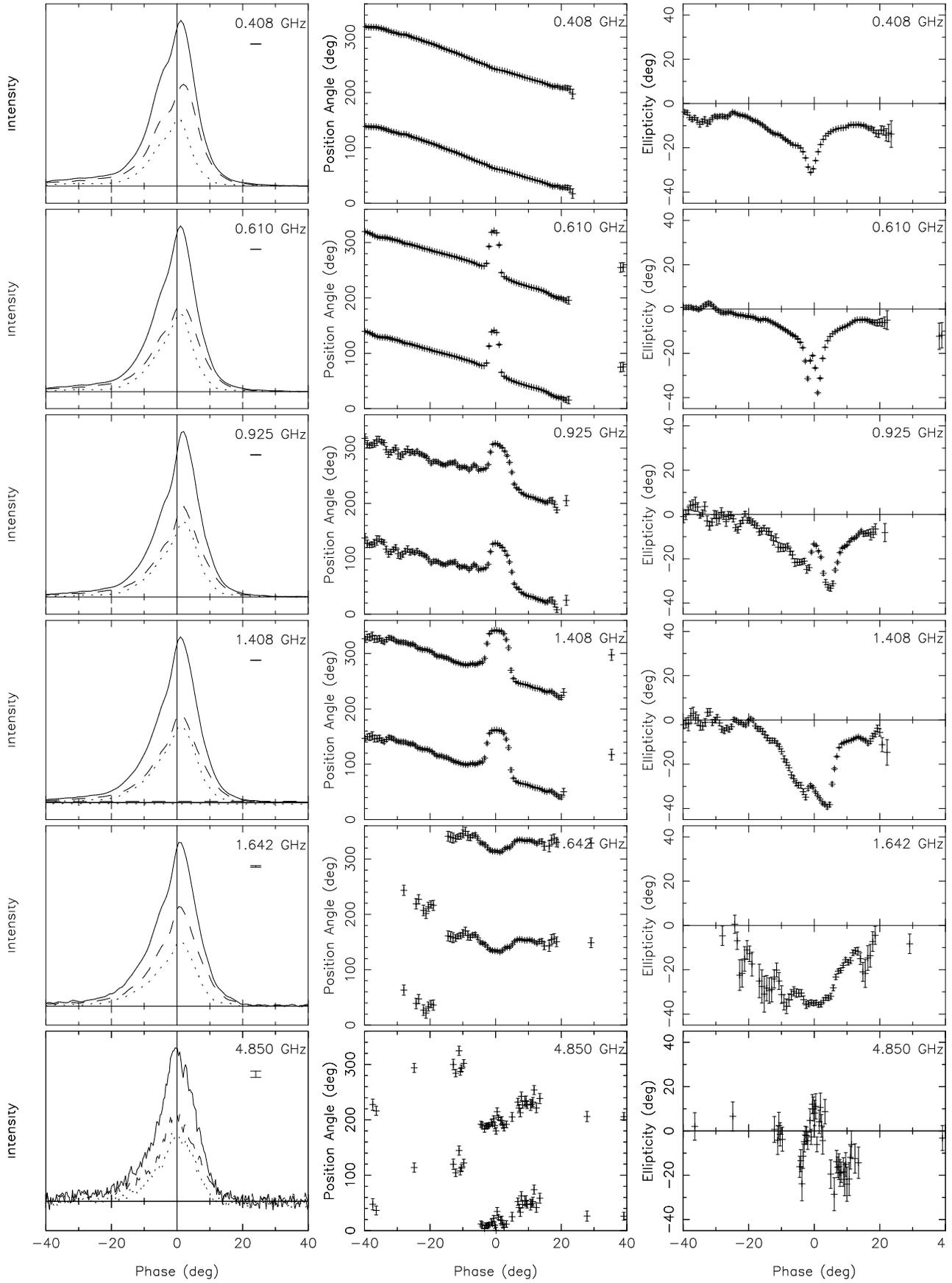}
  \caption{Total Intensity, intensity of orthogonal modes, position
  angle and ellipticity angle as a function of frequency of
  \object{PSR B0950+08}.  The error bars represent the noiselevel,
  given by the rms of the off-pulse region of the profile.}
  \label{fig:0950}  
\end{figure*}

\begin{figure*}[ht]
  \centering   
  \includegraphics[width=17cm]{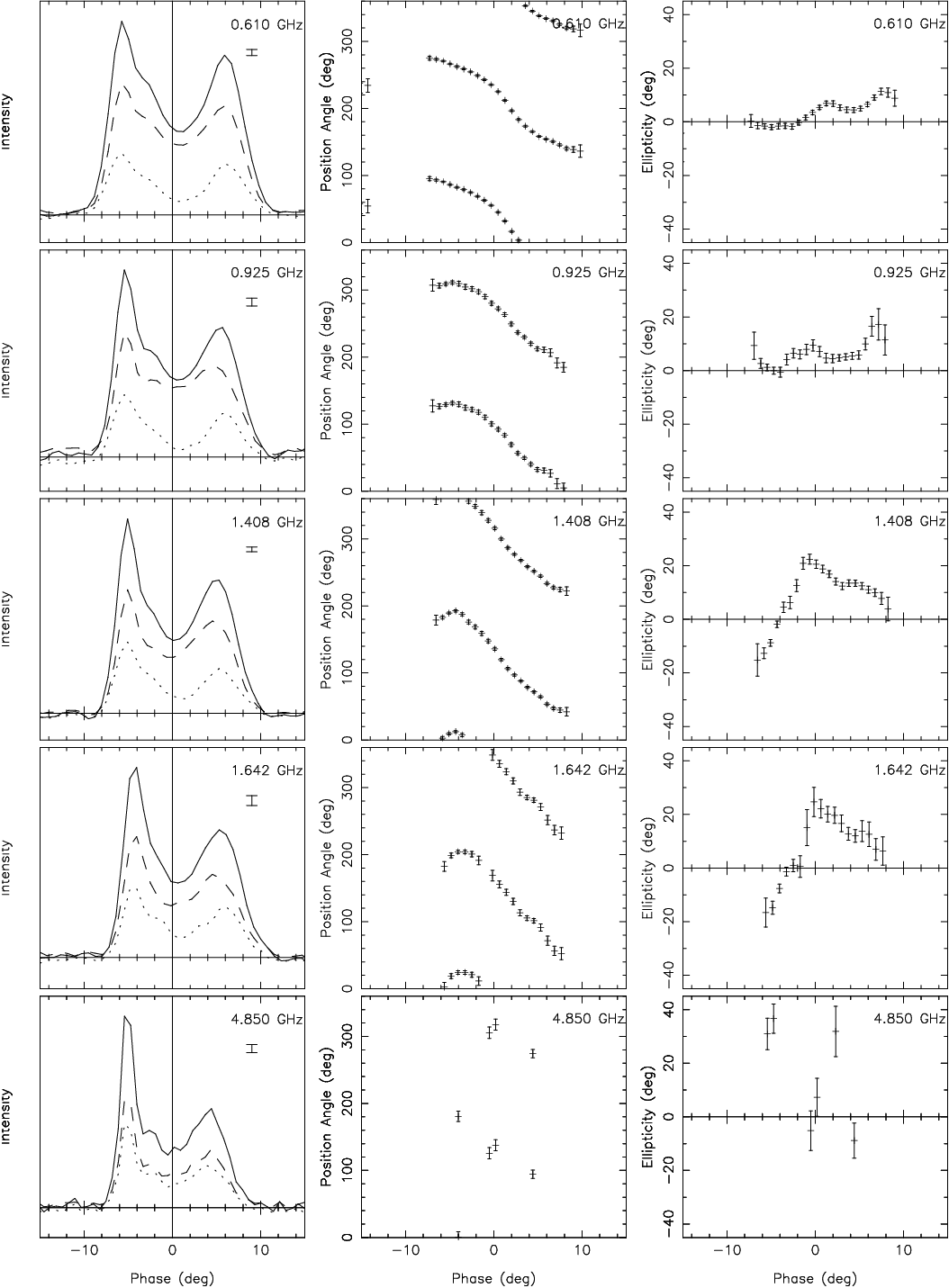}
  \caption{Total Intensity, intensity of orthogonal modes, position
  angle and ellipticity angle as a function of frequency of
  \object{PSR B1039$-$19}. The error bars represent the noiselevel,
  given by the rms of the off-pulse region of the profile.}
  \label{fig:1039} 
\end{figure*}

\begin{figure*}[ht]
  \centering   
  \includegraphics[width=17cm]{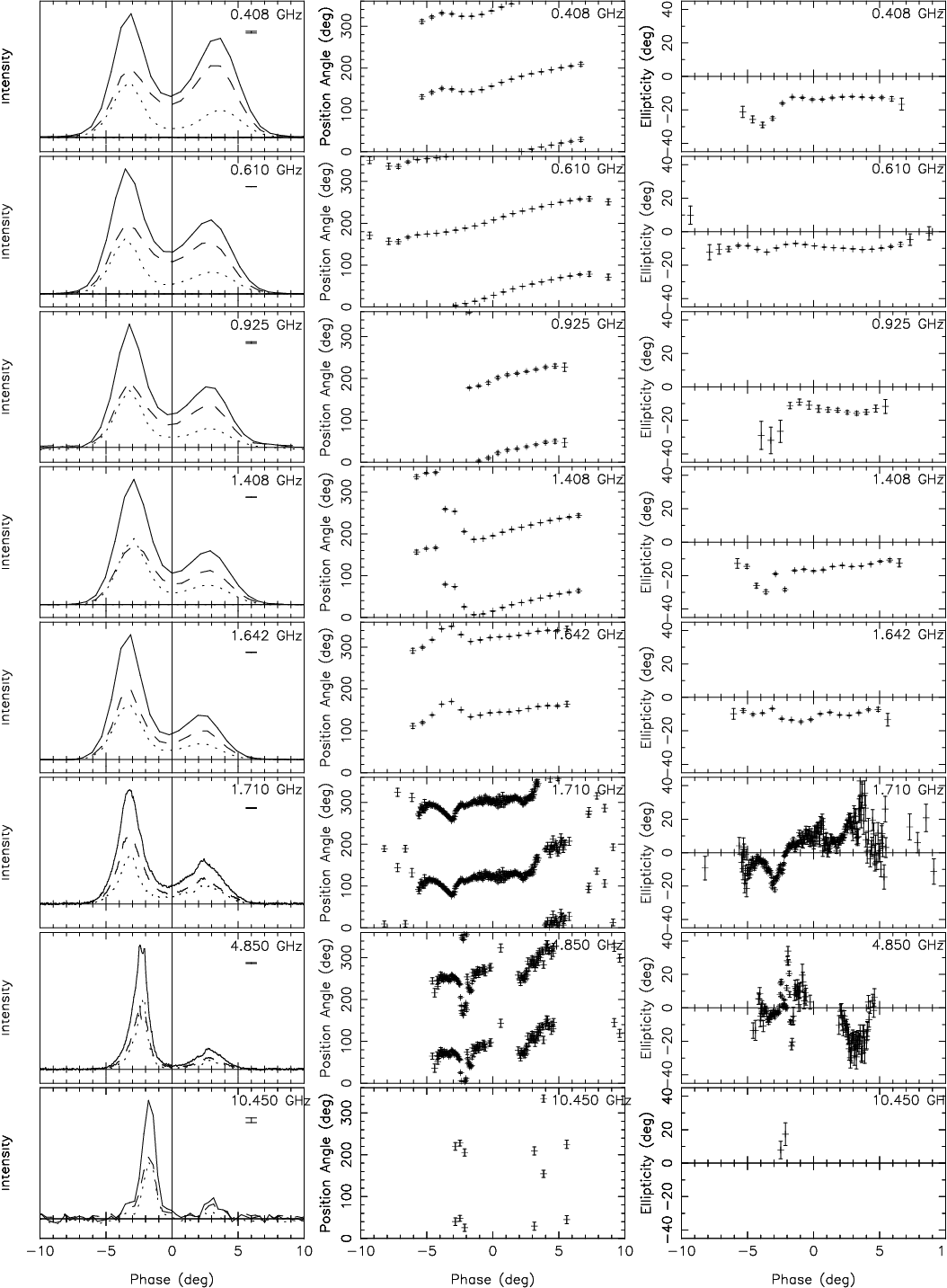}
  \caption{Total Intensity, intensity of orthogonal modes, position
  angle and ellipticity angle as a function of frequency of
  \object{PSR B1133+16}. The error bars represent the noiselevel,
  given by the rms of the off-pulse region of the profile.}
  \label{fig:1133}  
\end{figure*}

\begin{figure*}[ht]
  \centering   
  \includegraphics[width=17cm]{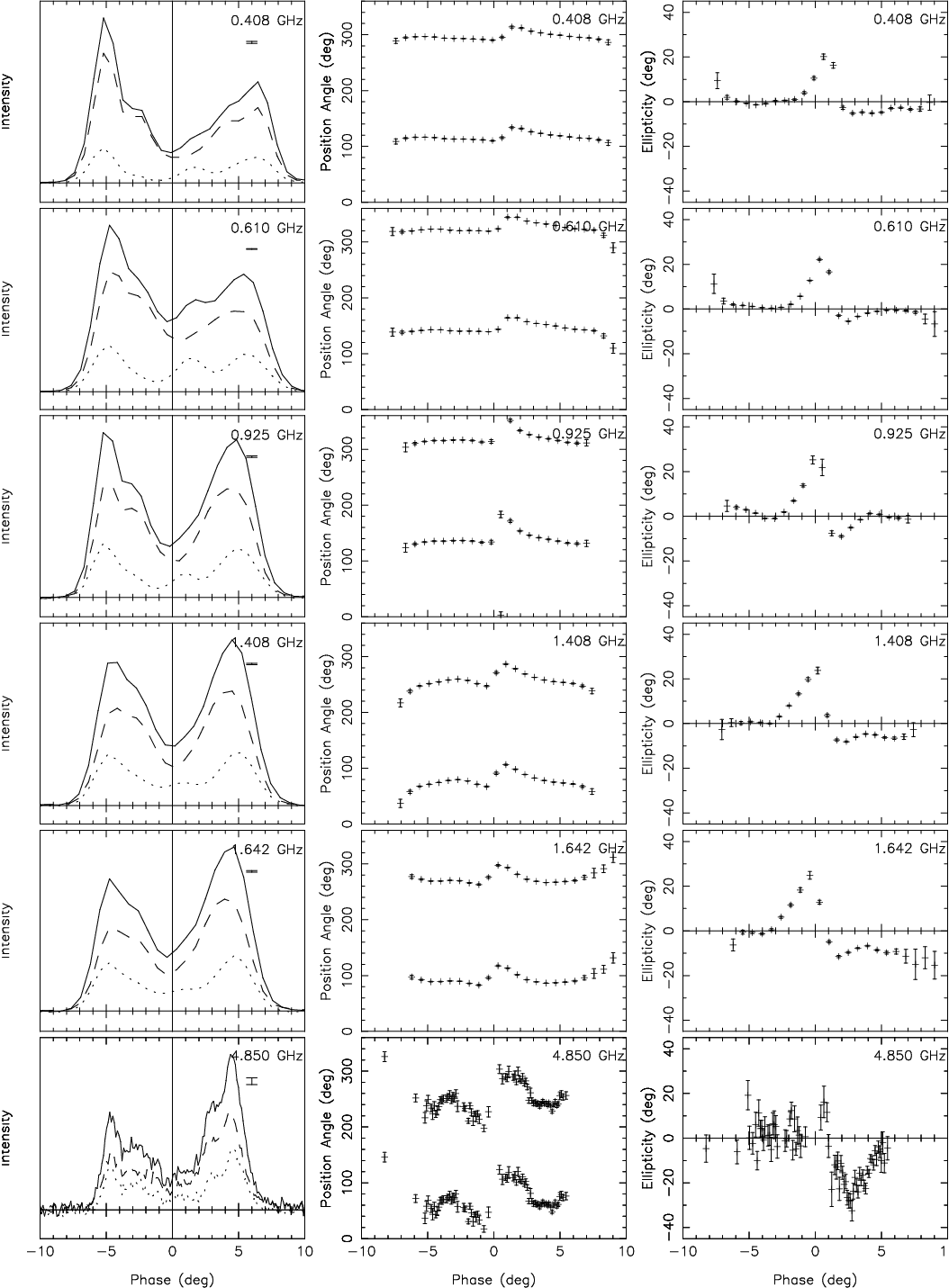}
  \caption{Total Intensity, intensity of orthogonal modes, position
  angle and ellipticity angle as a function of frequency of
  \object{PSR B1237+25}. The error bars represent the noiselevel,
  given by the rms of the off-pulse region of the profile.}
  \label{fig:1237}  
\end{figure*}

\begin{figure*}[ht]
  \centering
  \includegraphics[width=17cm]{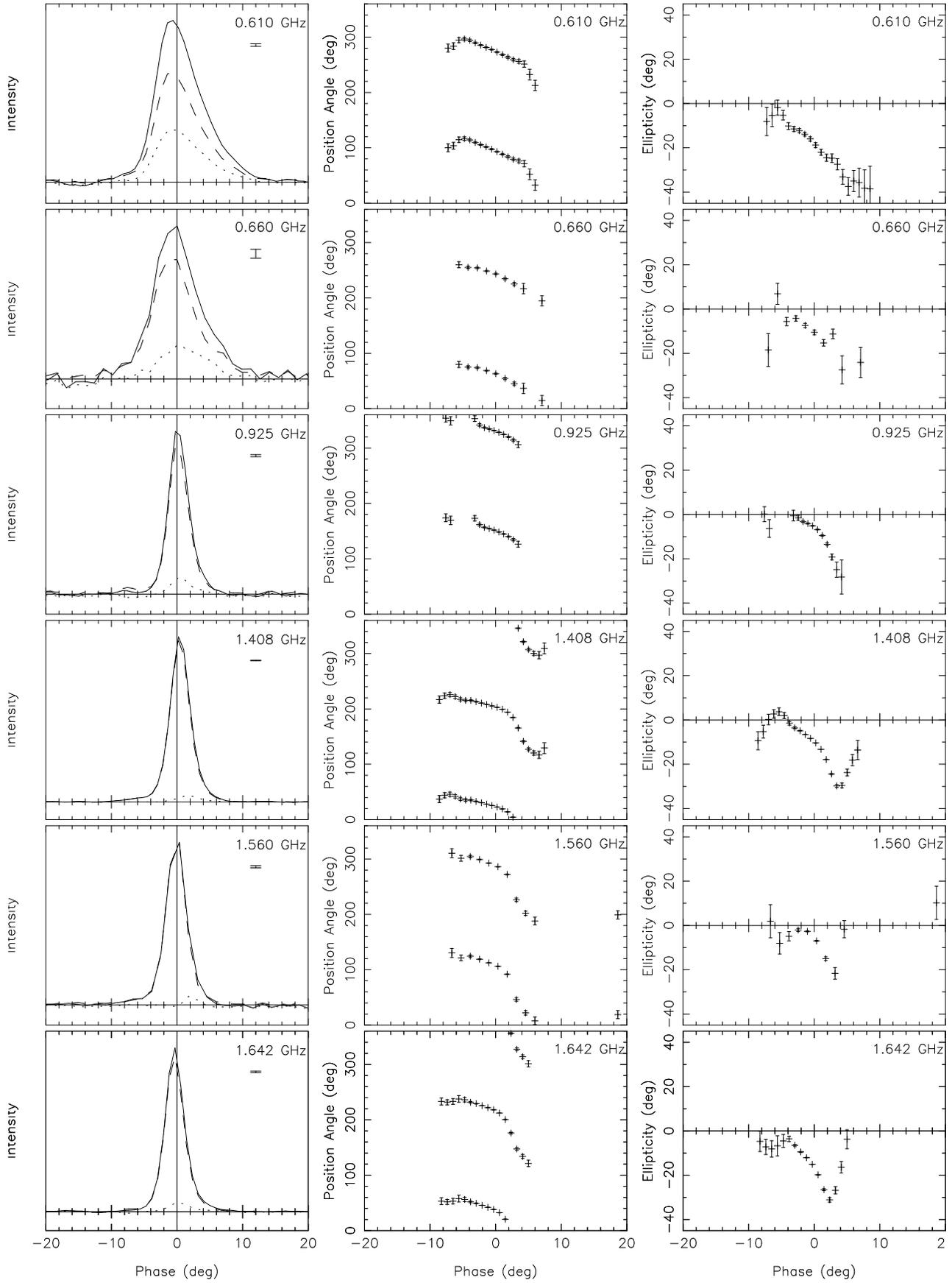}
  \caption{Total Intensity, intensity of orthogonal modes, position
  angle and ellipticity angle as a function of frequency of
  \object{PSR B1737$-$30}.  The error bars represent the noiselevel,
  given by the rms of the off-pulse region of the profile.}
  \label{fig:1737}  
\end{figure*}

\begin{figure*}[ht]
  \centering   
  \includegraphics[width=17cm]{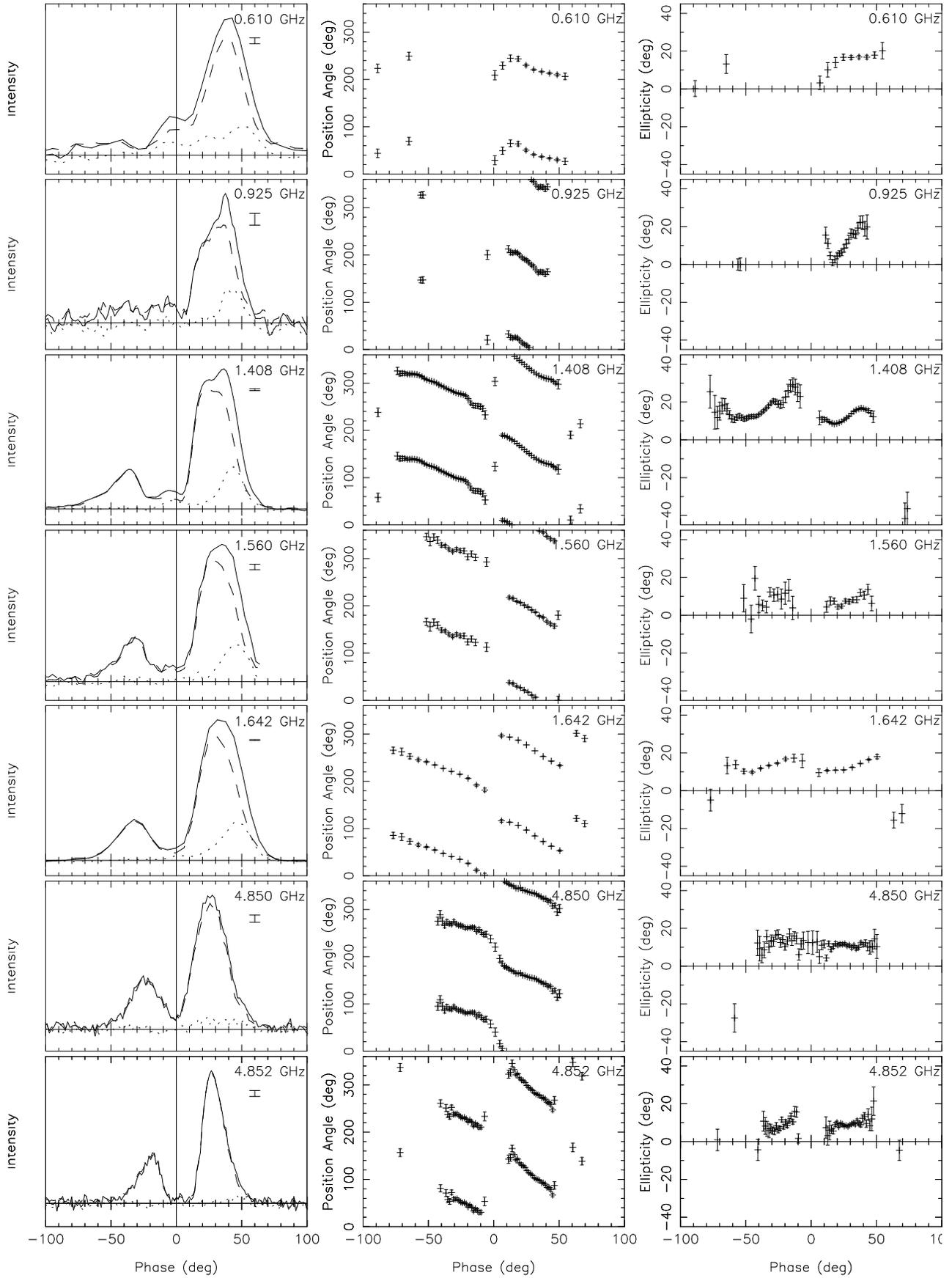}
  \caption{Total Intensity, intensity of orthogonal modes, position
  angle and ellipticity angle as a function of frequency of
  \object{PSR B1800$-$21}.  The error bars represent the noiselevel,
  given by the rms of the off-pulse region of the profile.}
  \label{fig:1800}  
\end{figure*}

\begin{figure*}[ht]
  \centering   
  \includegraphics[width=17cm]{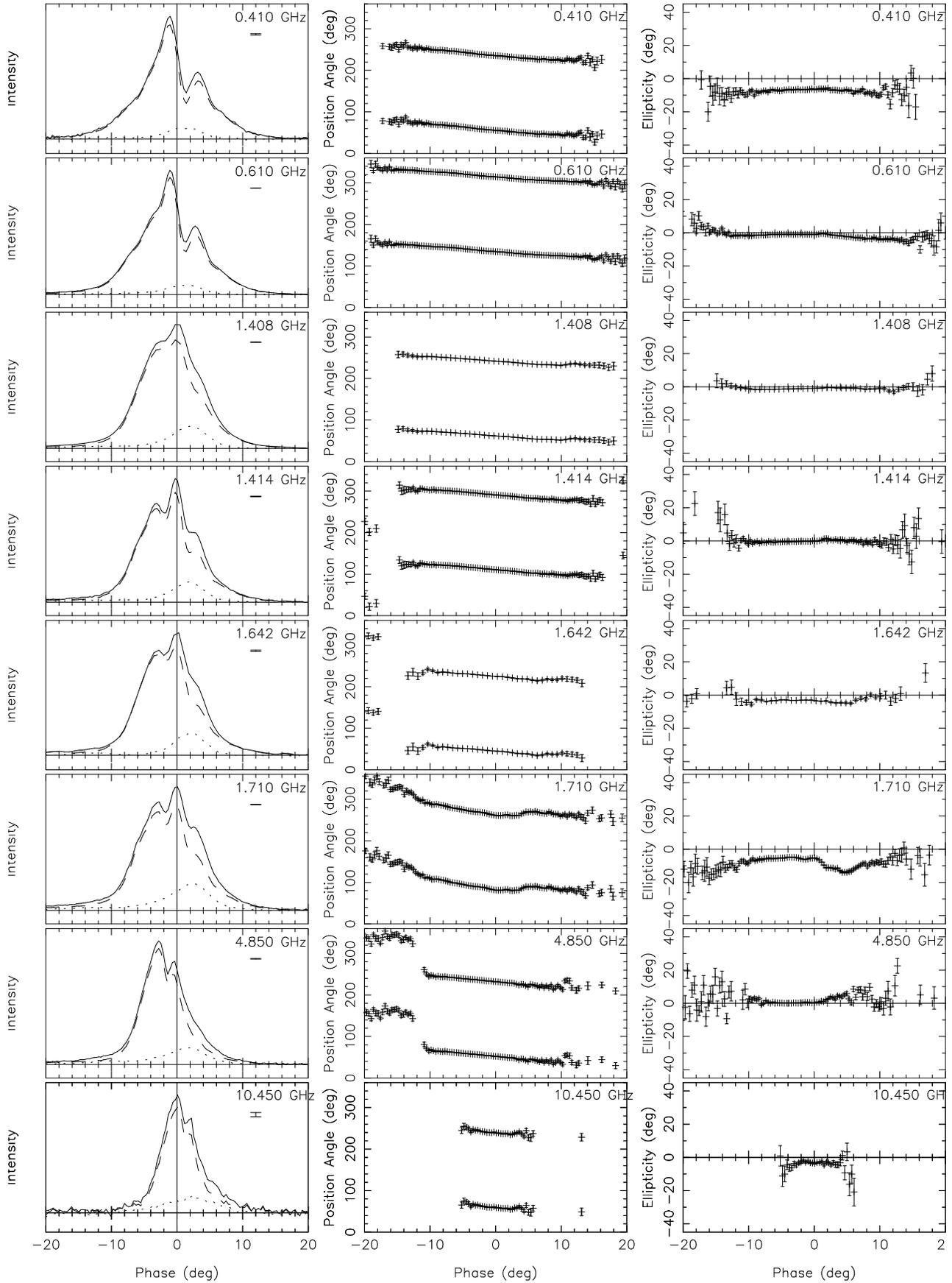}
  \caption{Total Intensity, intensity of orthogonal modes, position
  angle and ellipticity angle as a function of frequency of
  \object{PSR B1929+10}.  The error bars represent the noiselevel,
  given by the rms of the off-pulse region of the profile.}
  \label{fig:1929} 
\end{figure*}

\begin{figure*}[ht]
  \centering   
  \includegraphics[width=17cm]{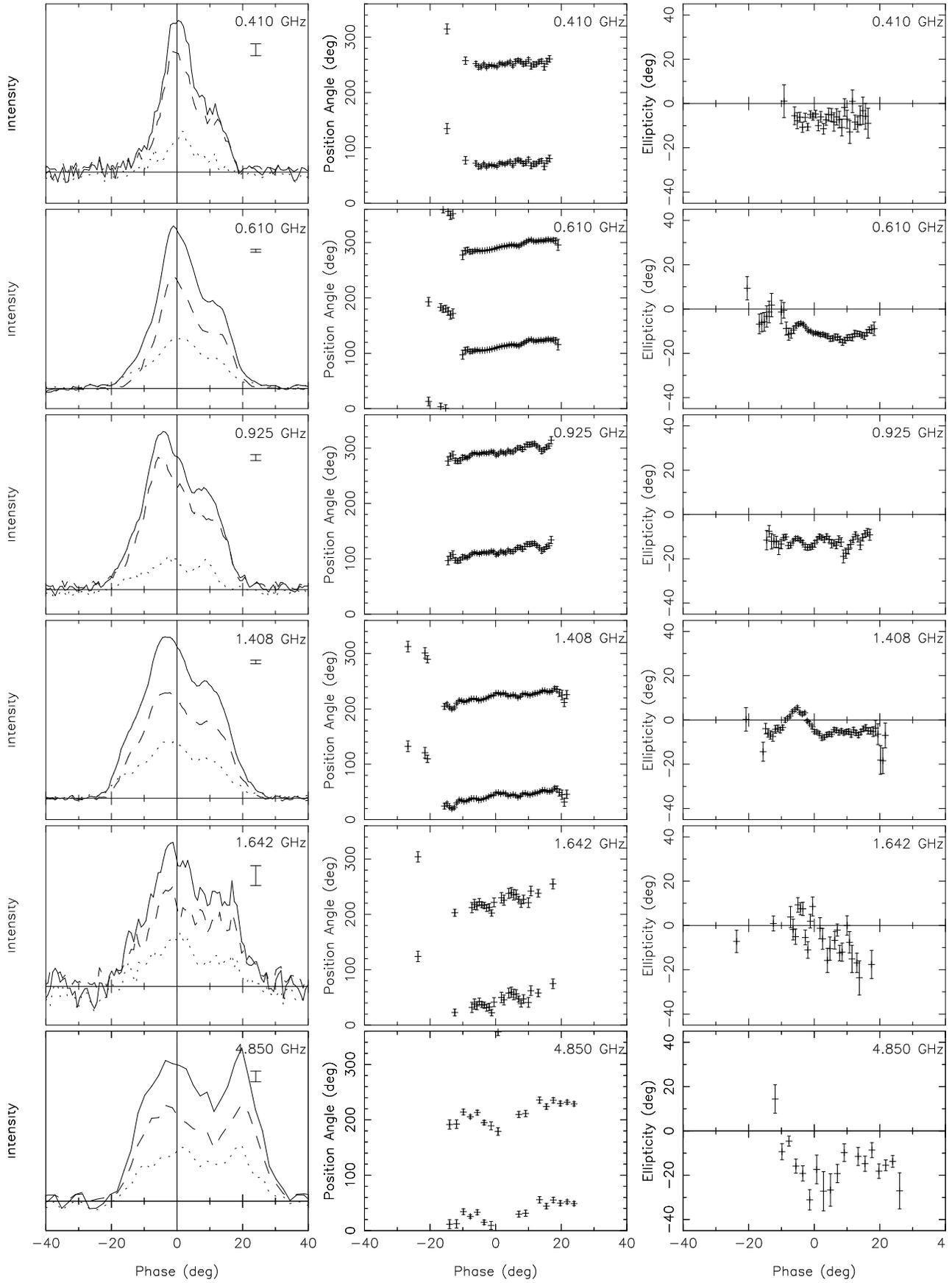}
  \caption{Total Intensity, intensity of orthogonal modes, position
  angle and ellipticity angle as a function of frequency of
  \object{PSR B1944+17}.  The error bars represent the noiselevel,
  given by the rms of the off-pulse region of the profile.}
  \label{fig:1944}  
\end{figure*}

\begin{figure*}[ht]
  \centering   
  \includegraphics[width=17cm]{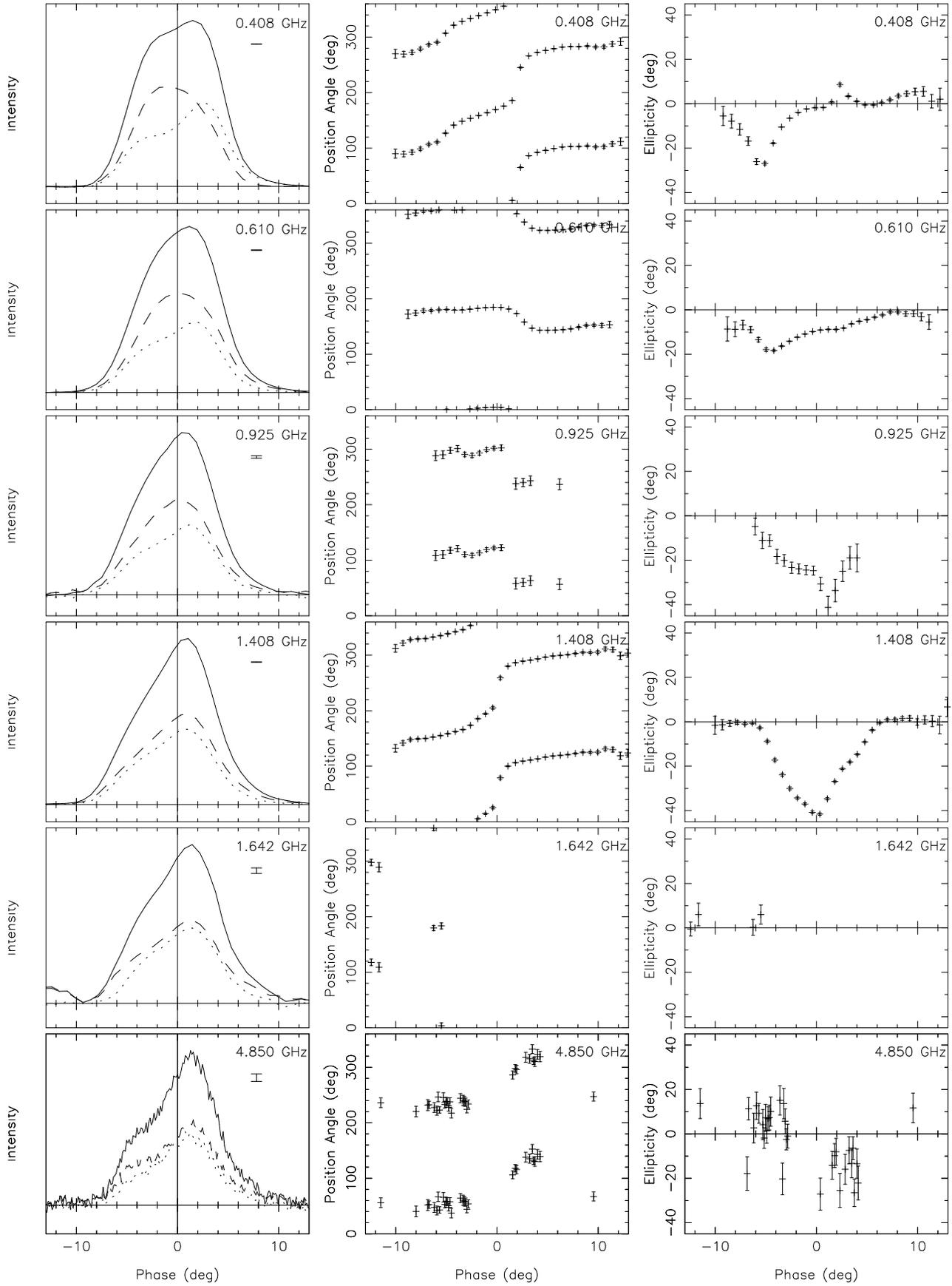}
  \caption{Total Intensity, intensity of orthogonal modes, position
  angle and ellipticity angle as a function of frequency of
  \object{PSR B2016+28}.  The error bars represent the noiselevel,
  given by the rms of the off-pulse region of the profile.}
  \label{fig:2016}  
\end{figure*}

\begin{figure*}[ht]
  \centering   
  \includegraphics[width=17cm]{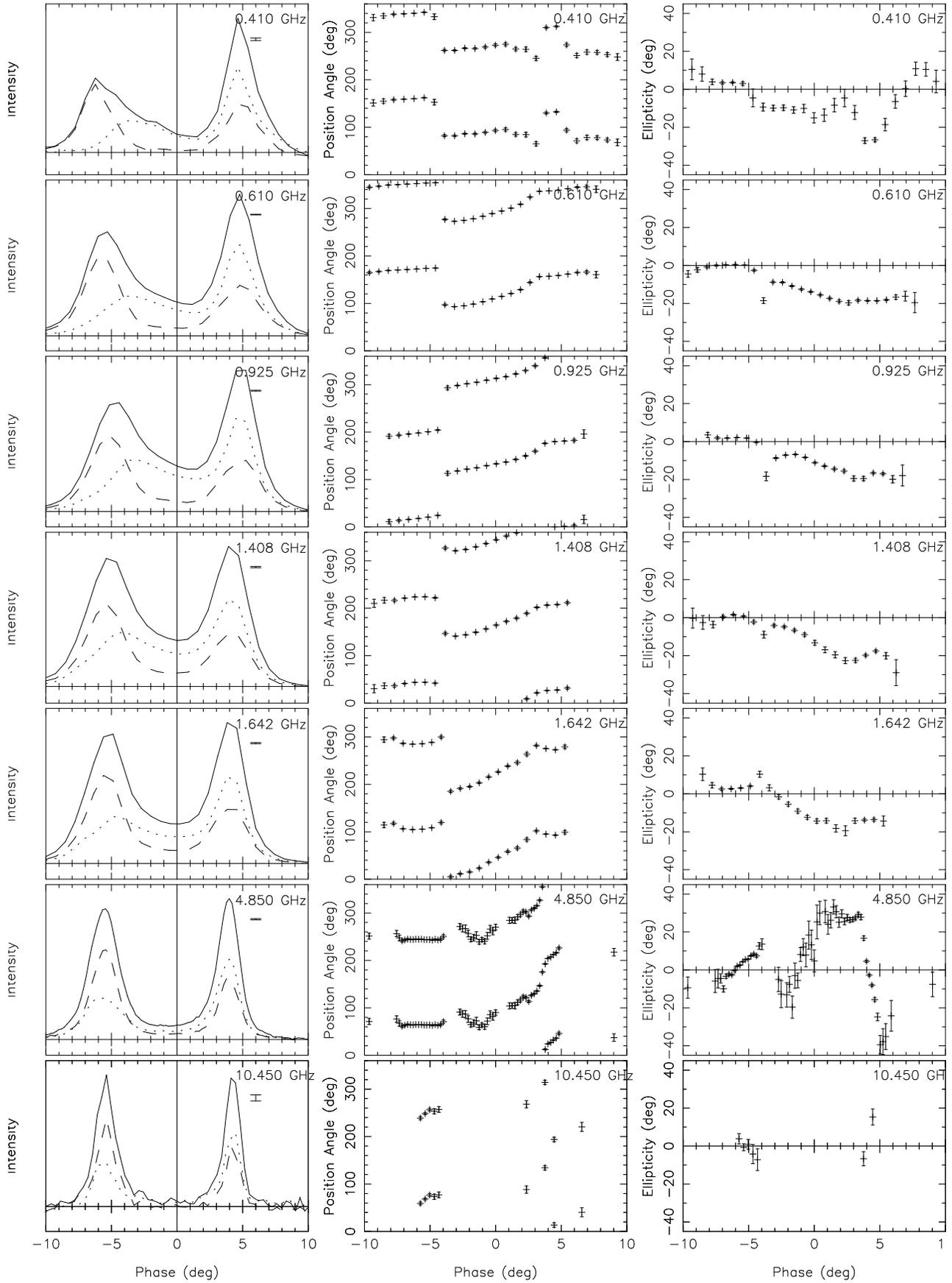}
  \caption{Total Intensity, intensity of orthogonal modes, position
  angle and ellipticity angle as a function of frequency of
  \object{PSR B2020+28}.  The error bars represent the noiselevel,
  given by the rms of the off-pulse region of the profile.}
  \label{fig:2020}  
\end{figure*}

In almost half of the pulsars, there is a jump of 90$\deg$ in the PA,
which can always be associated with a change in the dominance of the
modes. (See first row of Table~\ref{tab:characteristics}). This can be
during a fraction of the pulse period, as in pulsar B1133+16, or for
most of the remaining profile, as in pulsar B0144+59. In the
interpretation of the two orthogonal modes corresponding to the
ordinary and extraordinary modes \citep{Barnard86, Petrova01},
mode-changing is likely to be a result of the spatial separation of
these modes due to refraction. Figures~\ref{fig:0144}
through~\ref{fig:2020} show that mode-changing can occur at any
frequency. A comparison between our intensity profiles of the
polarisarion modes of pulsars B1133+16 and B1237+25 with the results
from a single pulse study \citep{Backer80} suggests that OPM can
become visible in single pulse studies even when there is one dominant
mode thoughout the average profile.  Figures~\ref{fig:0144}
through~\ref{fig:2020} show that in many pulsars there is a great
change in the average profile of the two polarisation modes over
frequency, suggesting in contrast to common belief, that refraction can
take place at all heights of emission.

\subsection*{Mode-drifting}
In two cases, the longitudinal position of one of the modes relative
to the other can be seen to change as a function of frequency (see
Table~\ref{tab:characteristics}). For pulsar B2020+28, the
longitudinal separation between the modes as a function of frequency
is shown (see Fig.~\ref{fig:drift}). Assuming radius to frequency
mapping, this drift of one of the modes can be interpreted as a change
of refraction with emission altitude. For pulsar B0919+06 the mode
drift is less pronounced and whereas the drifting mode of pulsar
B2020+28 moves away from the center with increasing frequency, the
drifting mode of pulsar B0919+06 moves towards the center with
increasing frequency. The relative drift of the two modes might be
correlated to other observable features, which have to be common
features of both pulsars. Table~\ref{tab:characteristics} shows the
common features of these two pulsars. They both have a period close to
0.4\,s, are of type T and have a similar spin-down luminosity
(log$\dot{E}$) of 33.8 and 33.3 for B0919+06 and B2020+18,
respectively.

\subsection*{Spectral behavior of the ratio of the modes}
\begin{figure*}
 \begin{tabular}{ccc}
  \includegraphics[width=0.26\textwidth]{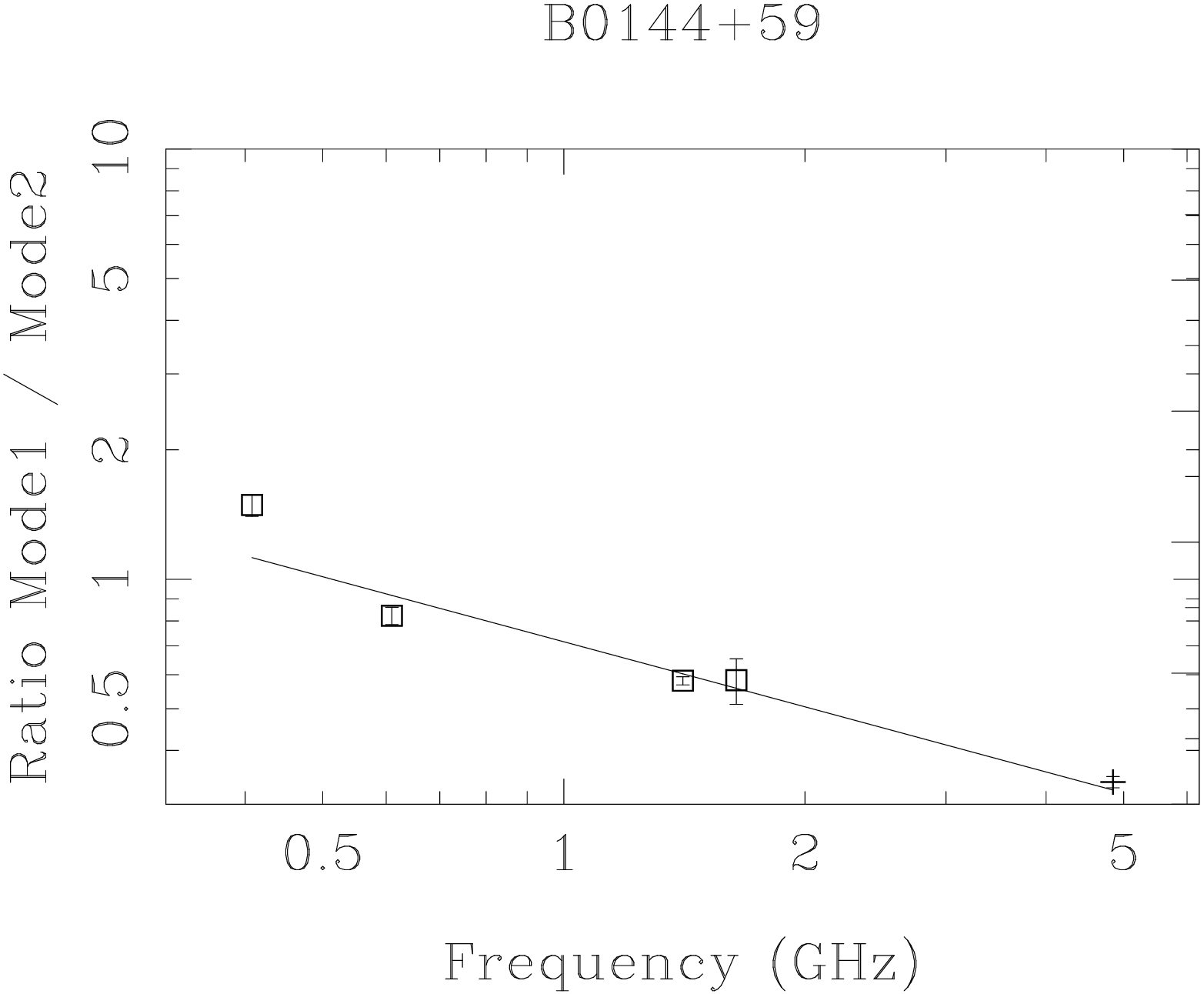}
& \hspace{1cm}
  \includegraphics[width=0.26\textwidth]{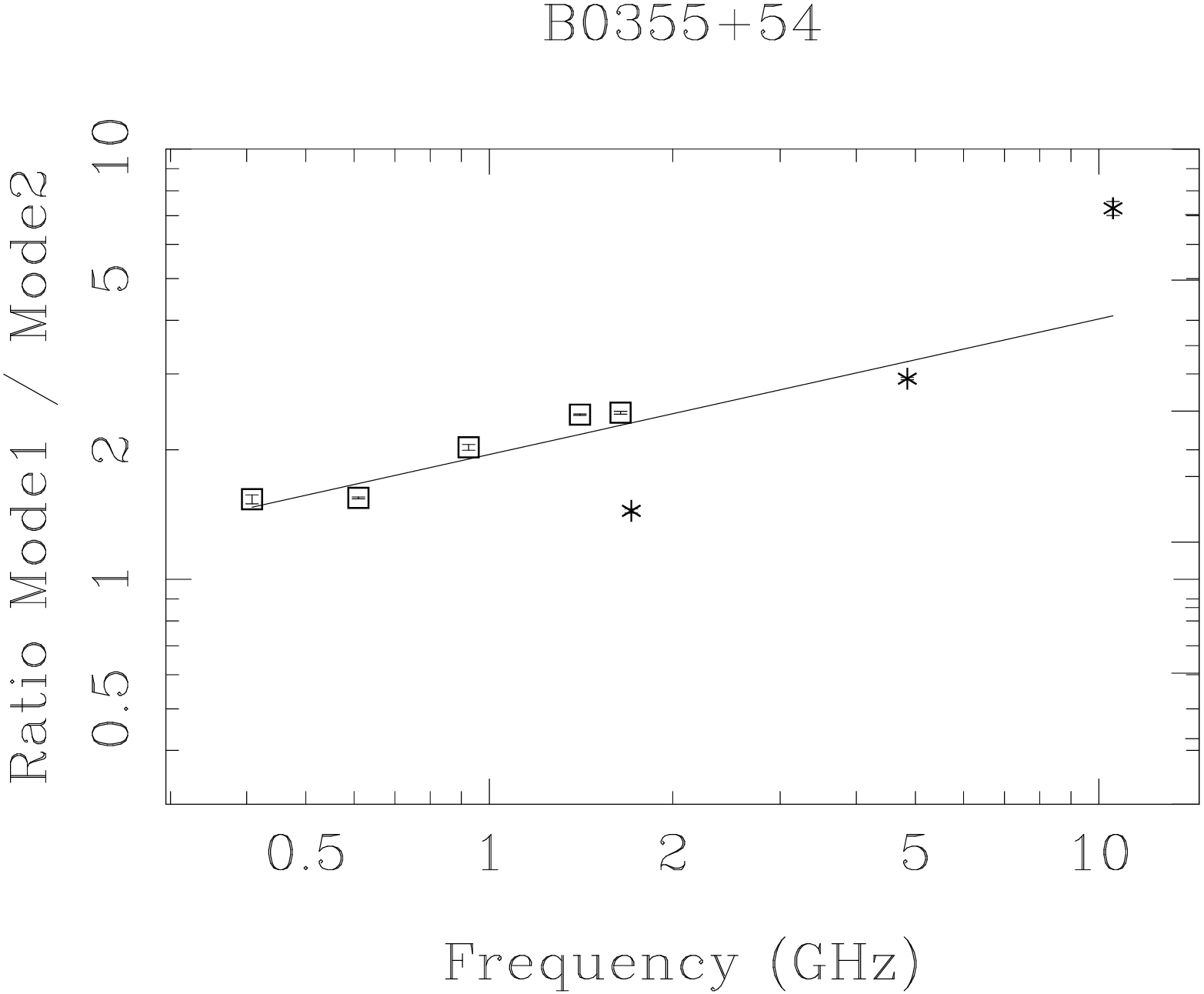}
& \hspace{1cm}
  \includegraphics[width=0.26\textwidth]{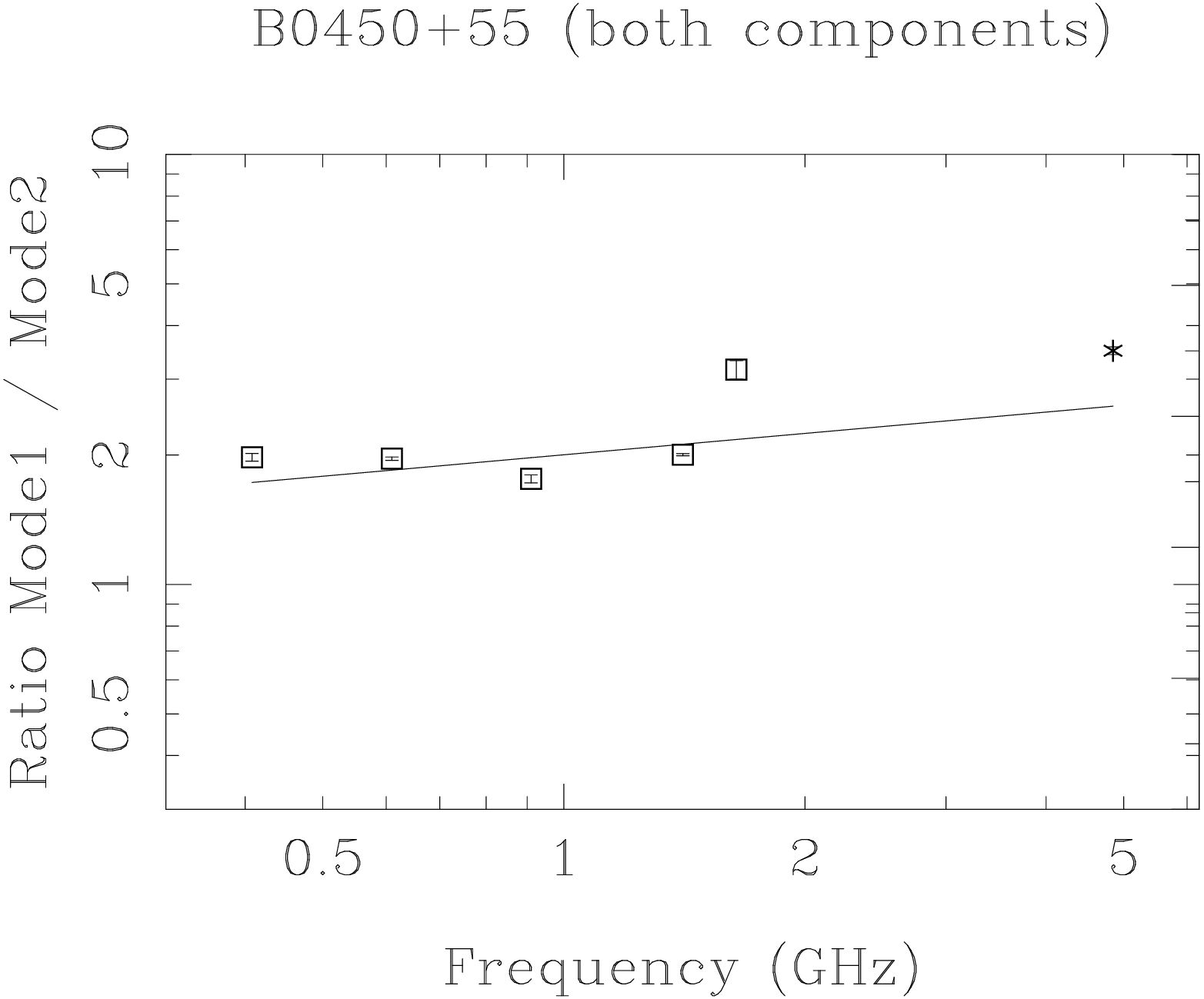}
\\
  \includegraphics[width=0.26\textwidth]{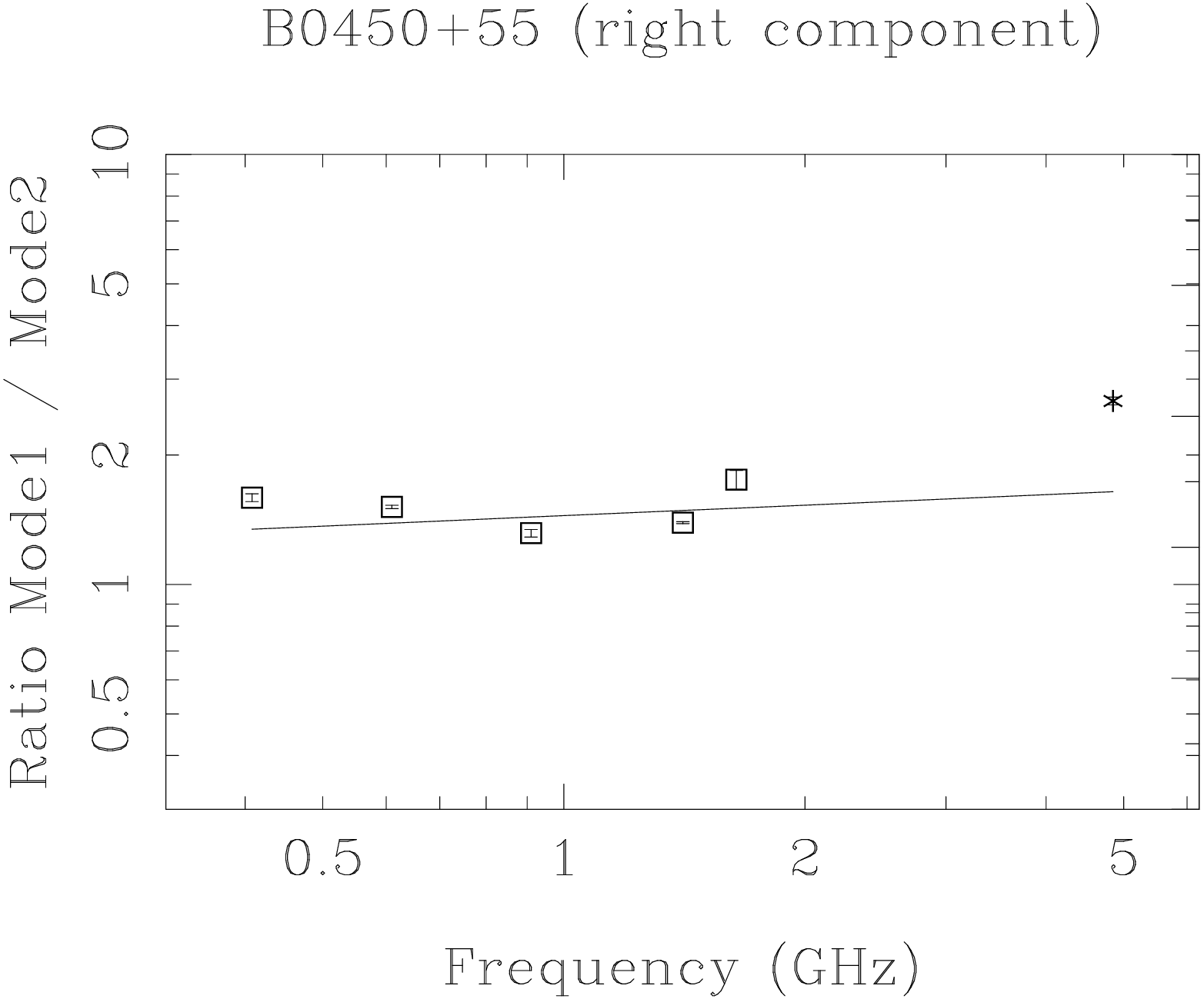}
& \hspace{1cm}
  \includegraphics[width=0.26\textwidth]{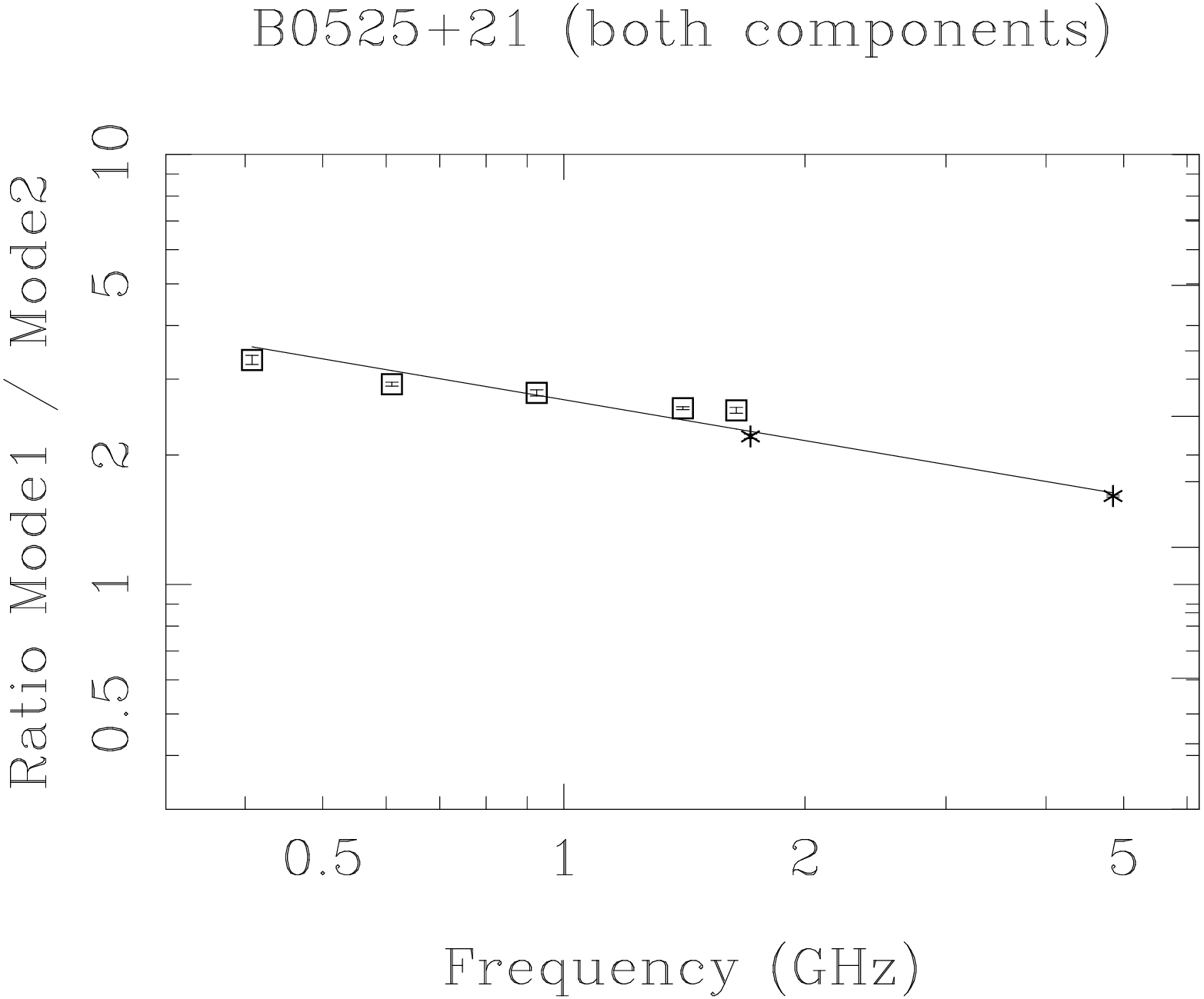}
& \hspace{1cm}
  \includegraphics[width=0.26\textwidth]{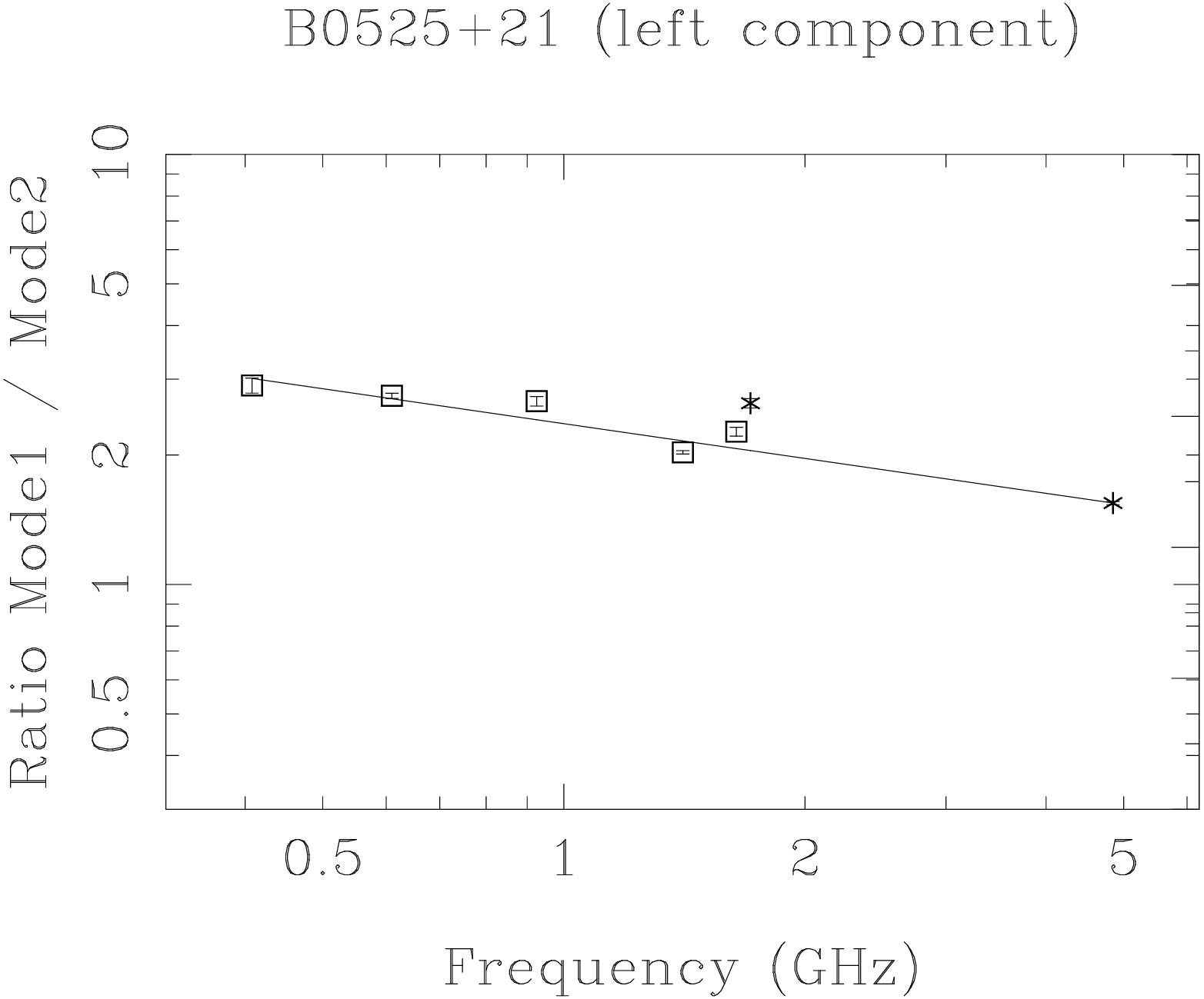}
\\
  \includegraphics[width=0.26\textwidth]{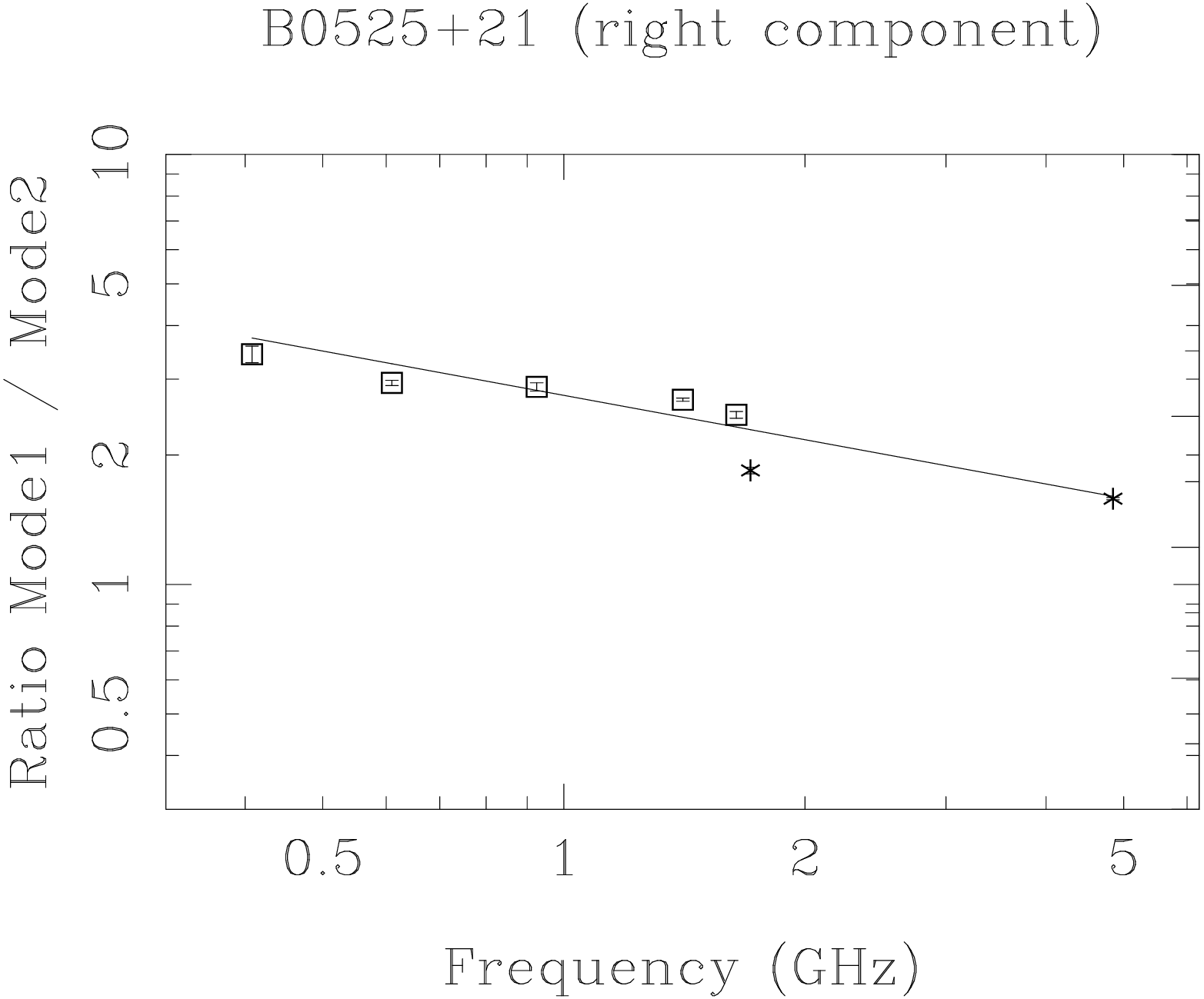}
& \hspace{1cm}
  \includegraphics[width=0.26\textwidth]{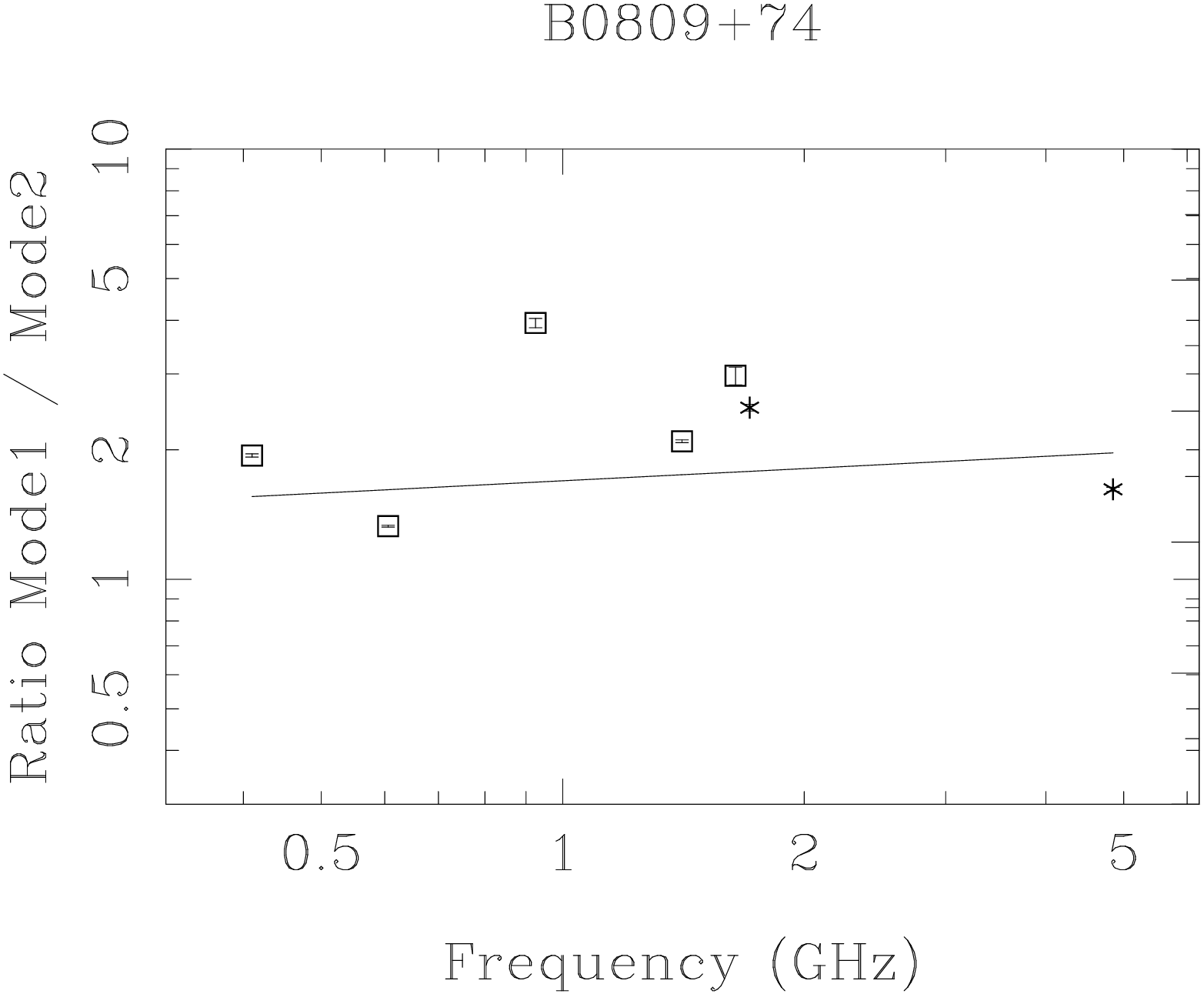}
& \hspace{1cm}
  \includegraphics[width=0.26\textwidth]{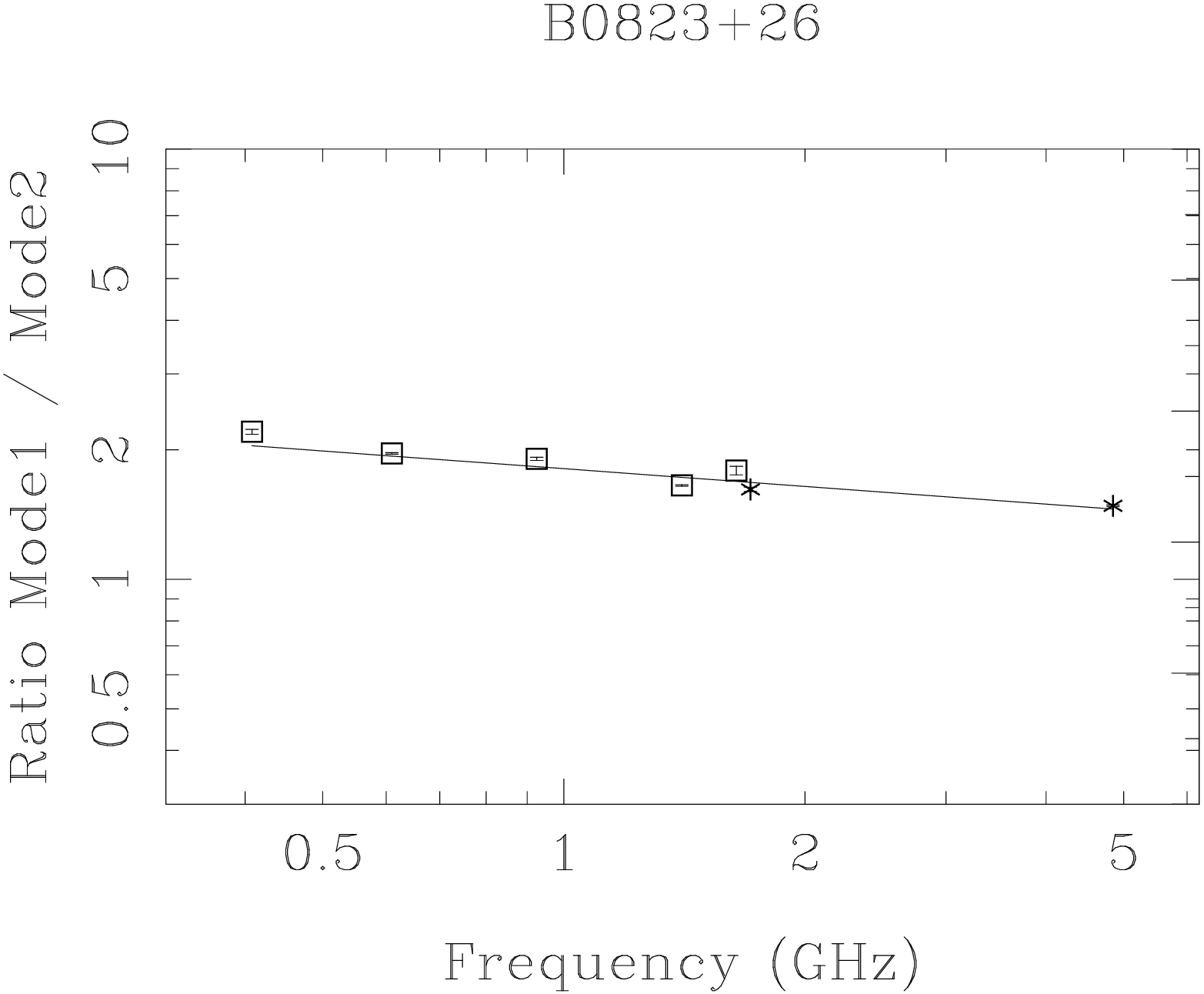}
\\
  \includegraphics[width=0.26\textwidth]{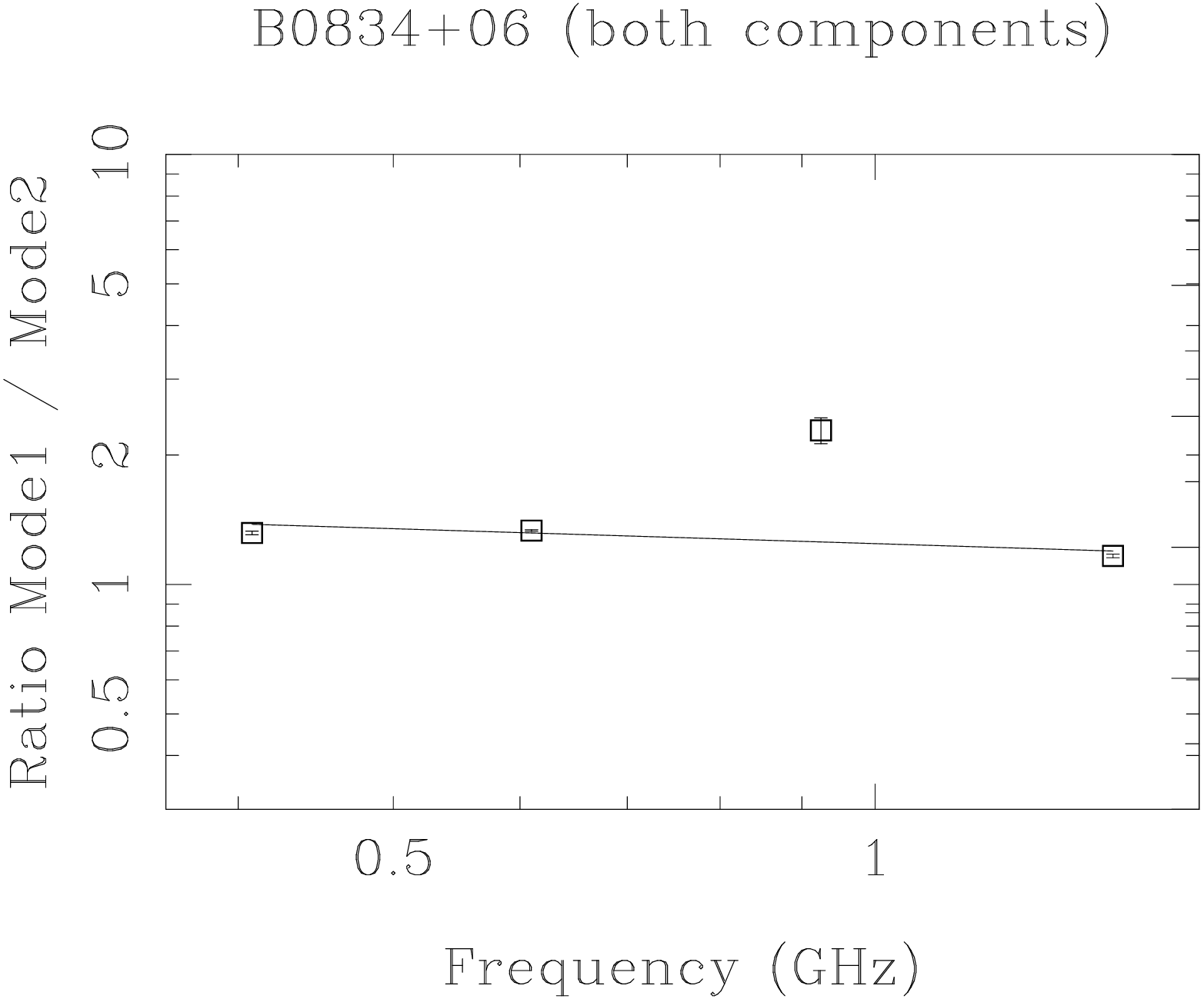}
& \hspace{1cm}
  \includegraphics[width=0.26\textwidth]{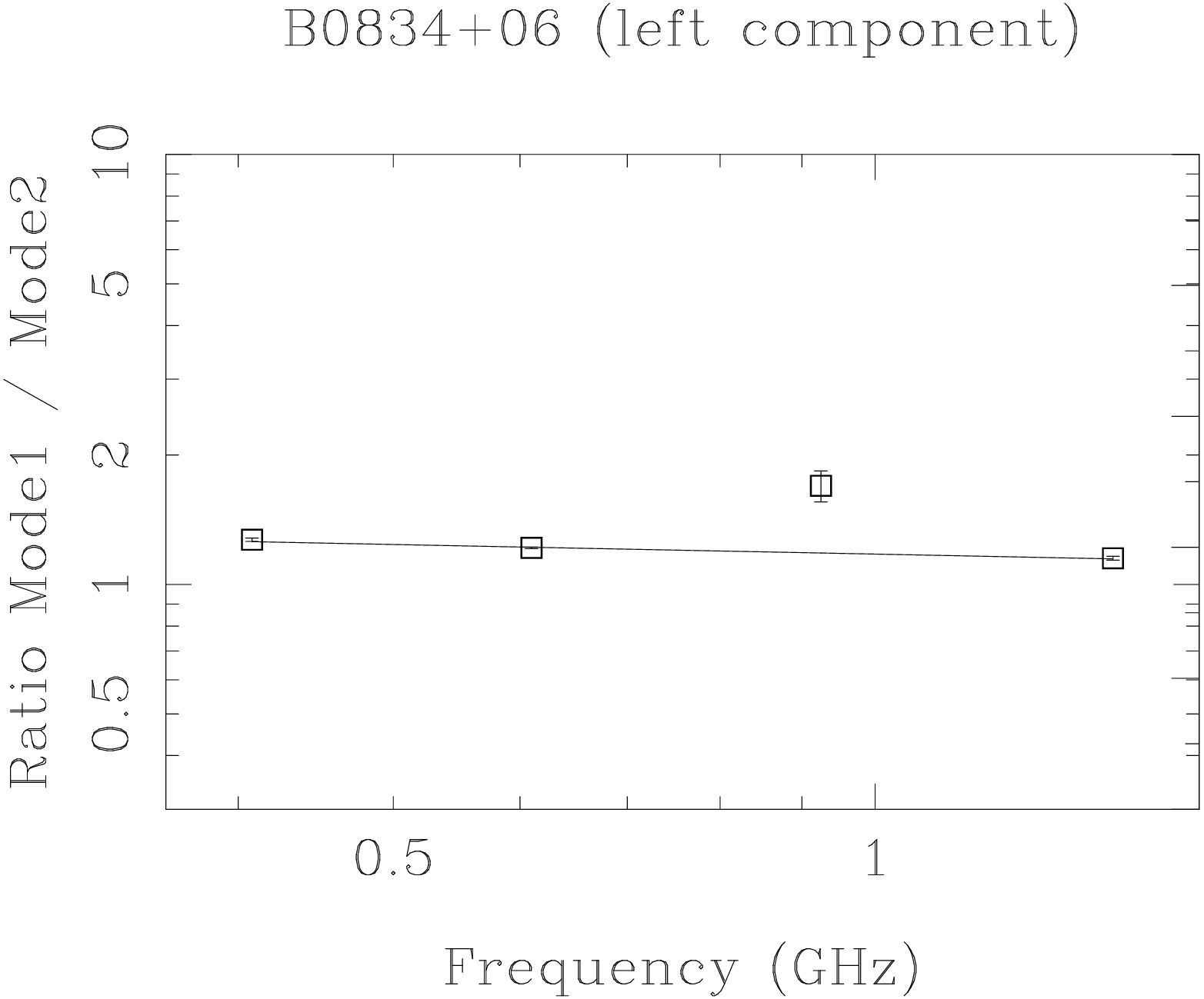}
& \hspace{1cm}
  \includegraphics[width=0.26\textwidth]{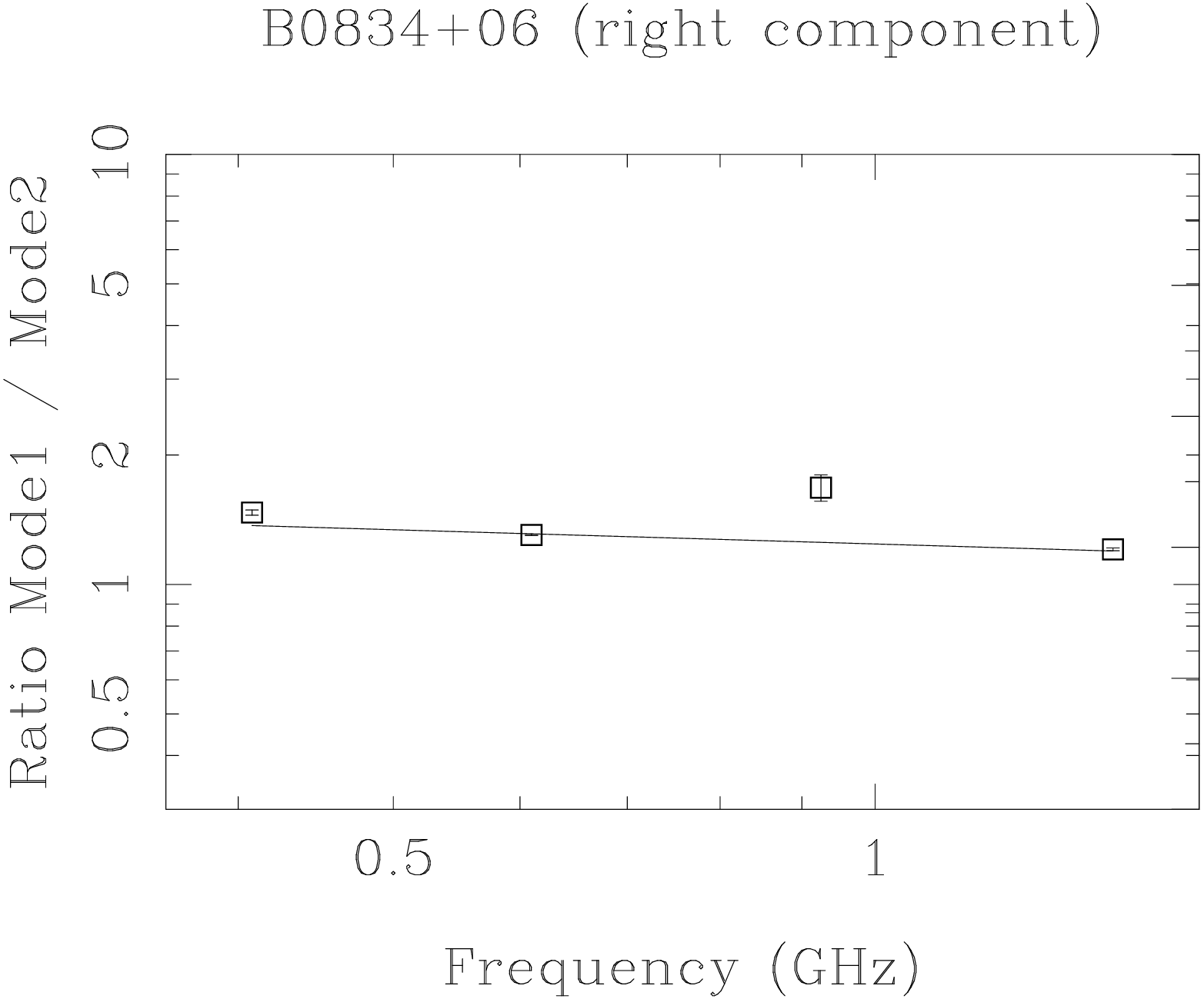}
\\
  \end{tabular}
  \caption{The ratio of the average intensity of two orthogonally
polarised modes as a function of frequency, plotted on a log-log
scale, for all pulsars. The error is calculated as $\sigma_I =
\sqrt{N}\cdot\sigma$, where N is the number of bins of the profile and
$\sigma$ is the noise level of the average intensity profile.  The
line through the points is the best straight line fit. Note that the
scales on the vertical axis are chosen to be the same as to allow for comparison
of the slopes.}
  \label{fig:ratio_of_modes_1}
\end{figure*}

\begin{figure*}
 \begin{tabular}{ccc}
  \includegraphics[width=0.26\textwidth]{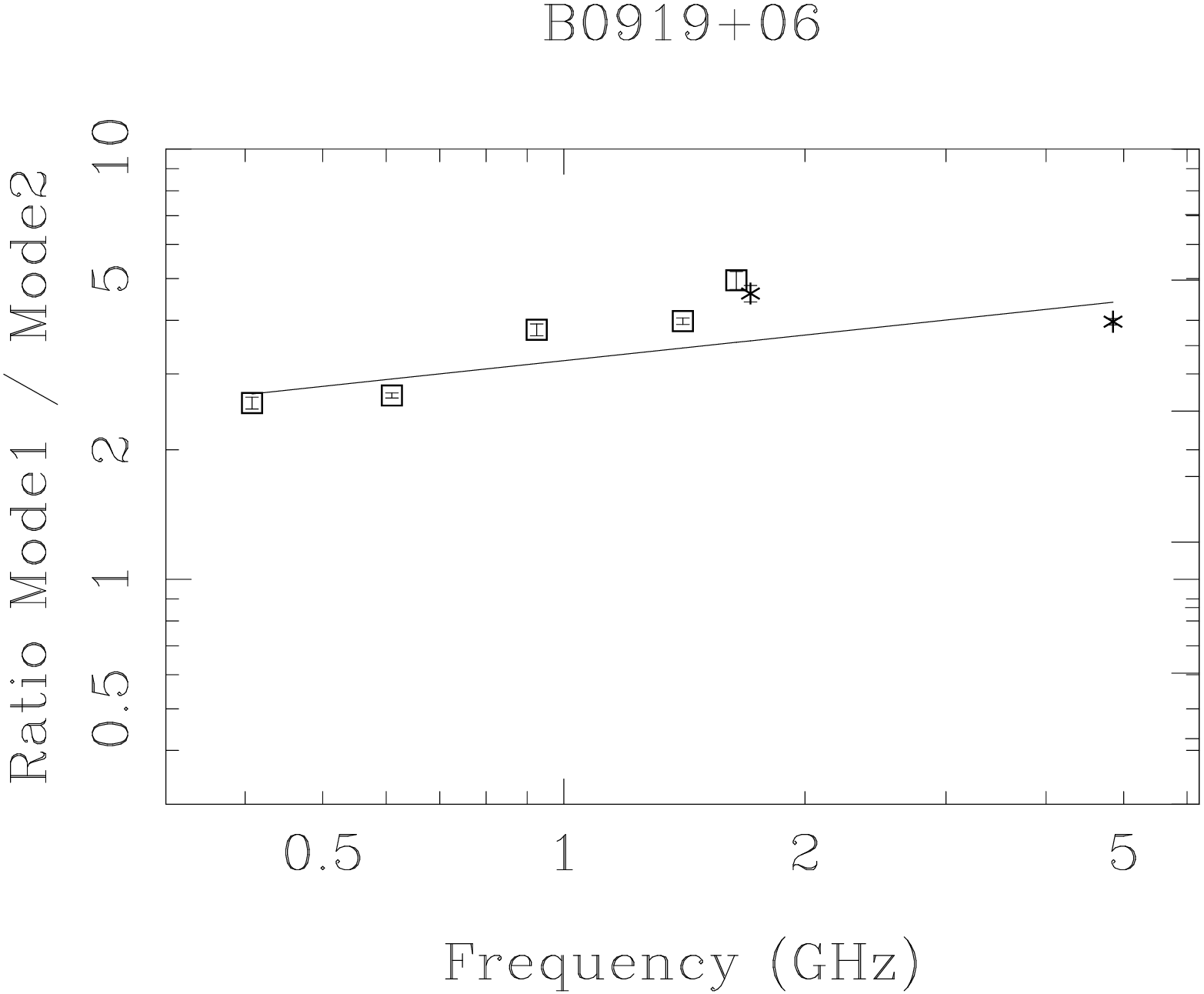}
& \hspace{1cm}
  \includegraphics[width=0.26\textwidth]{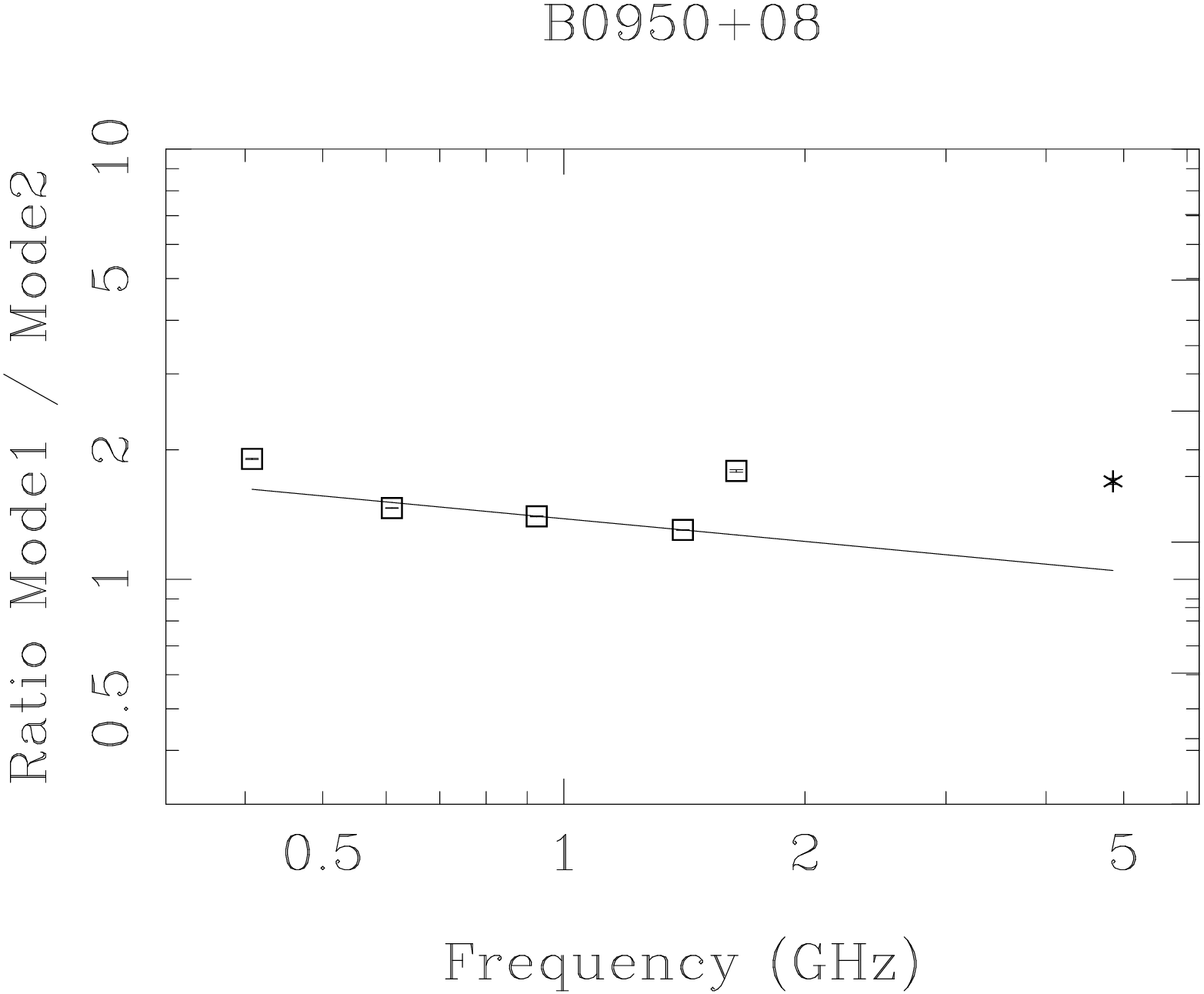}
& \hspace{1cm}
  \includegraphics[width=0.26\textwidth]{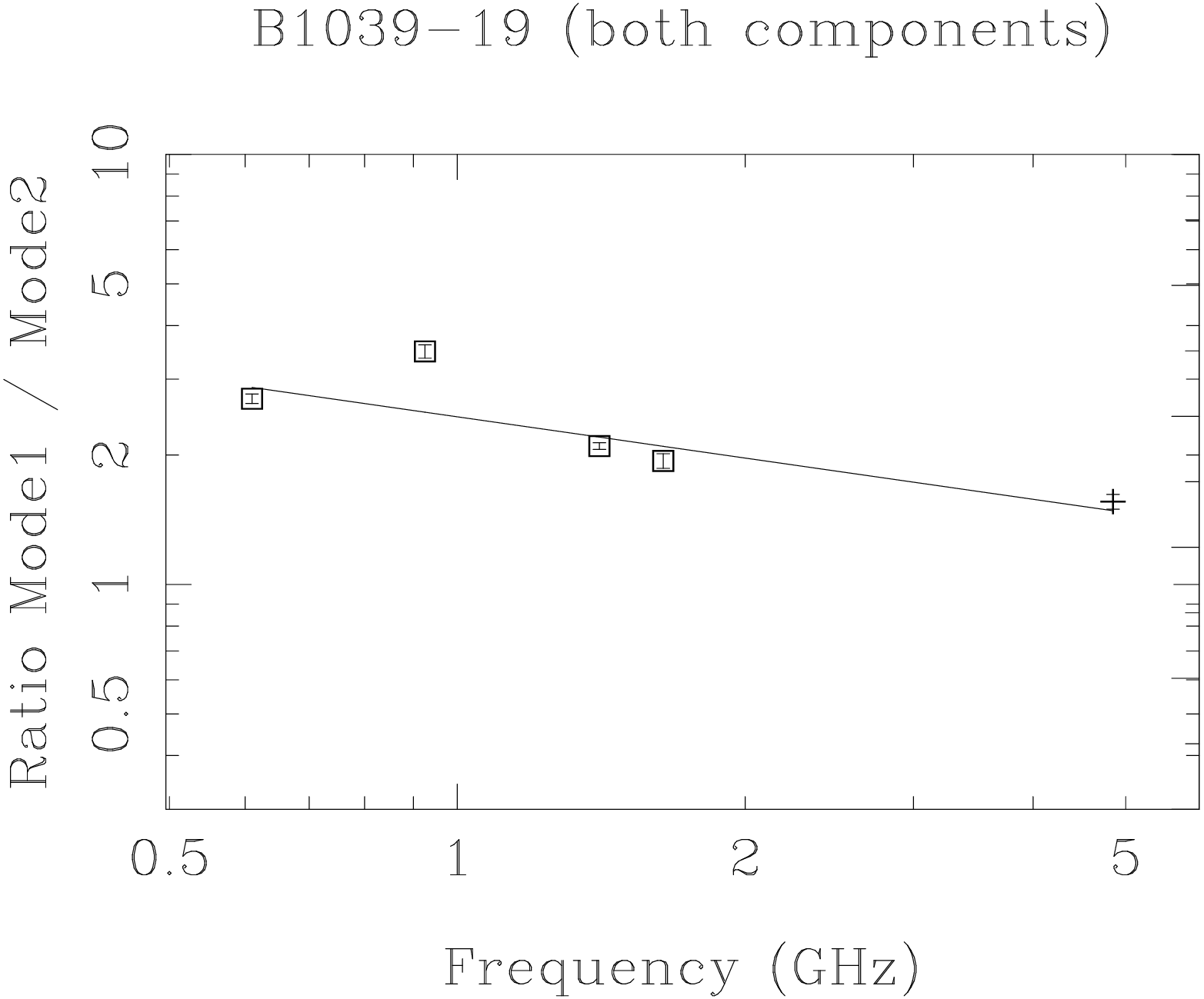}
\\
  \includegraphics[width=0.26\textwidth]{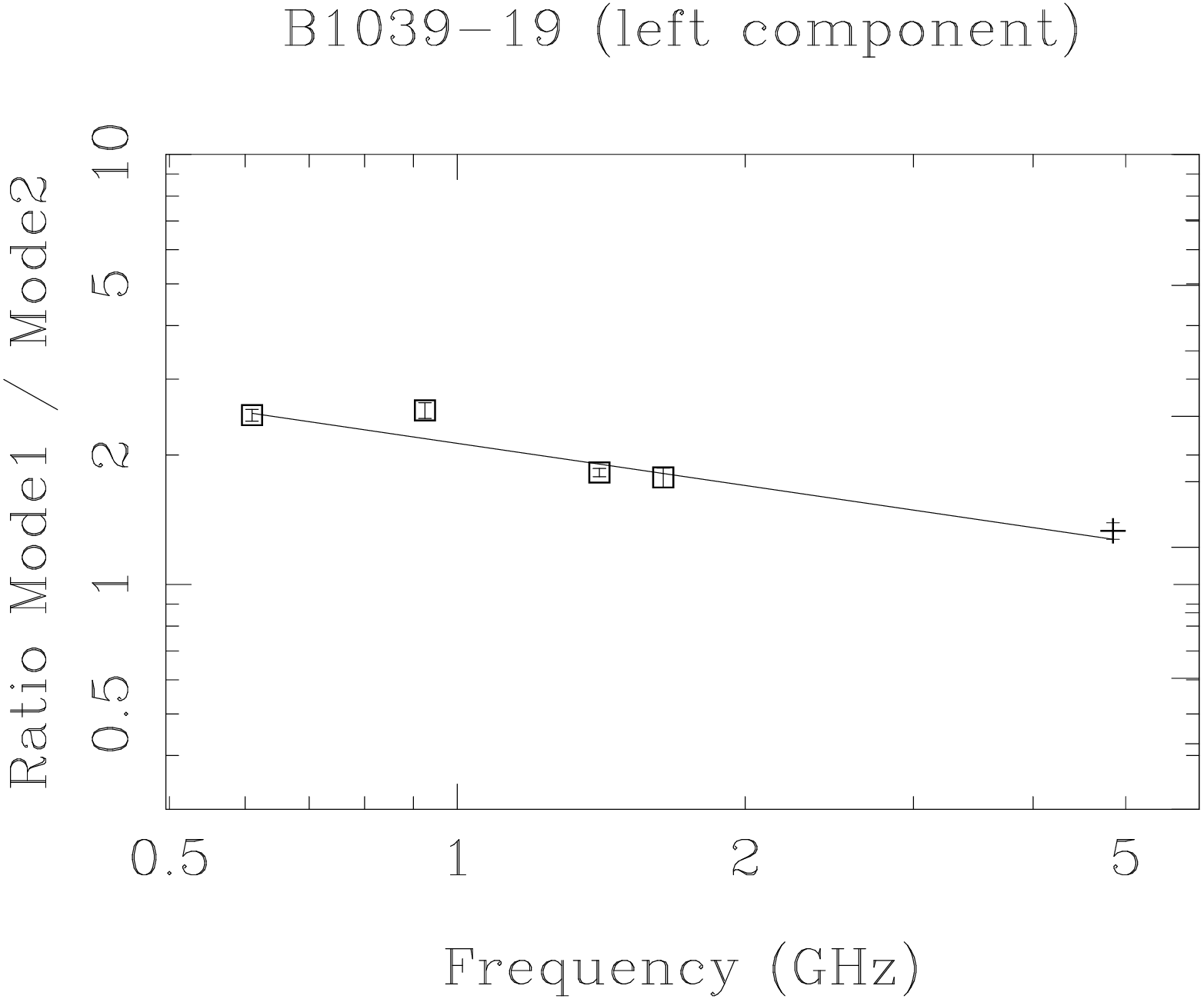}
& \hspace{1cm}
  \includegraphics[width=0.26\textwidth]{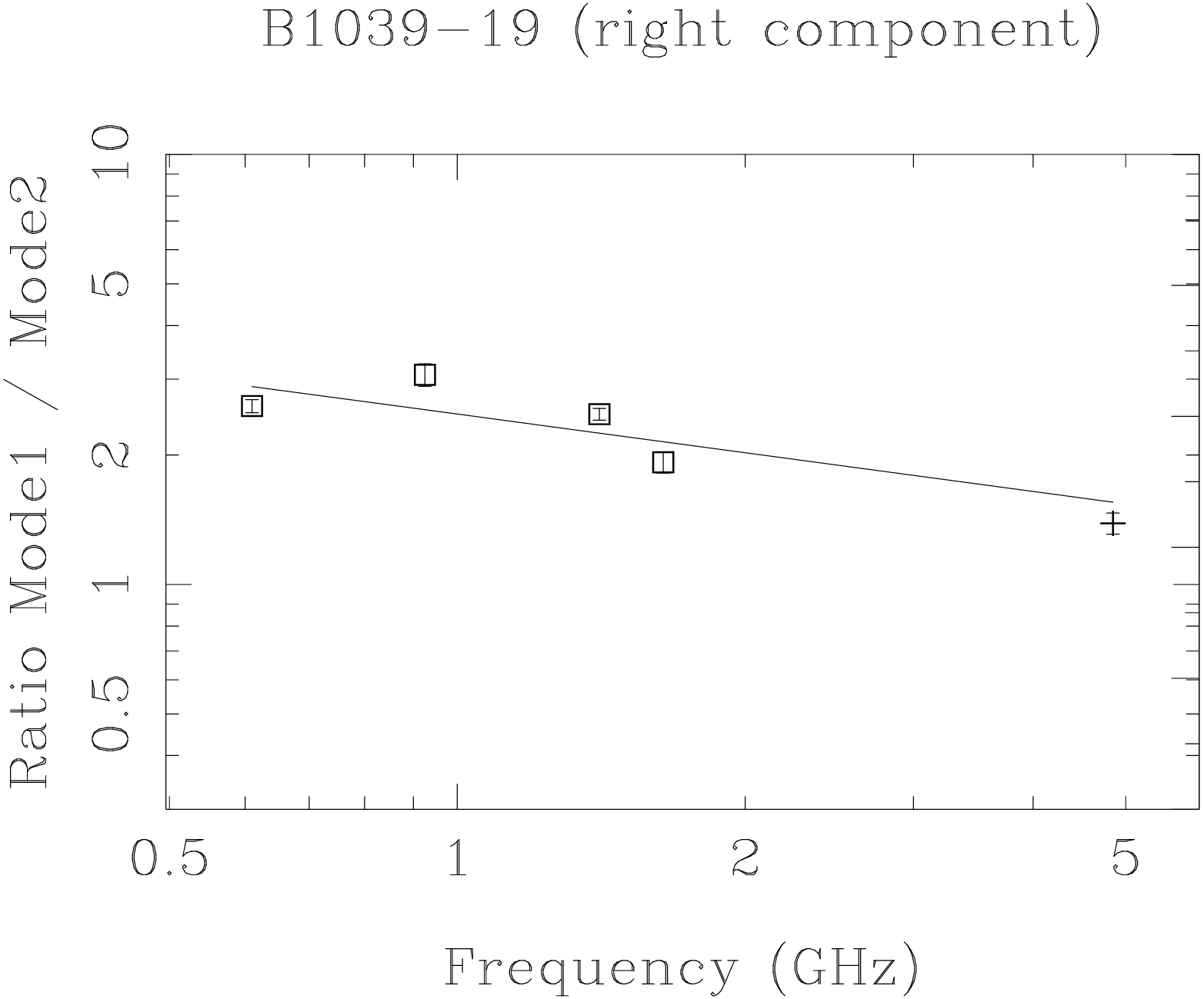}
& \hspace{1cm}
  \includegraphics[width=0.26\textwidth]{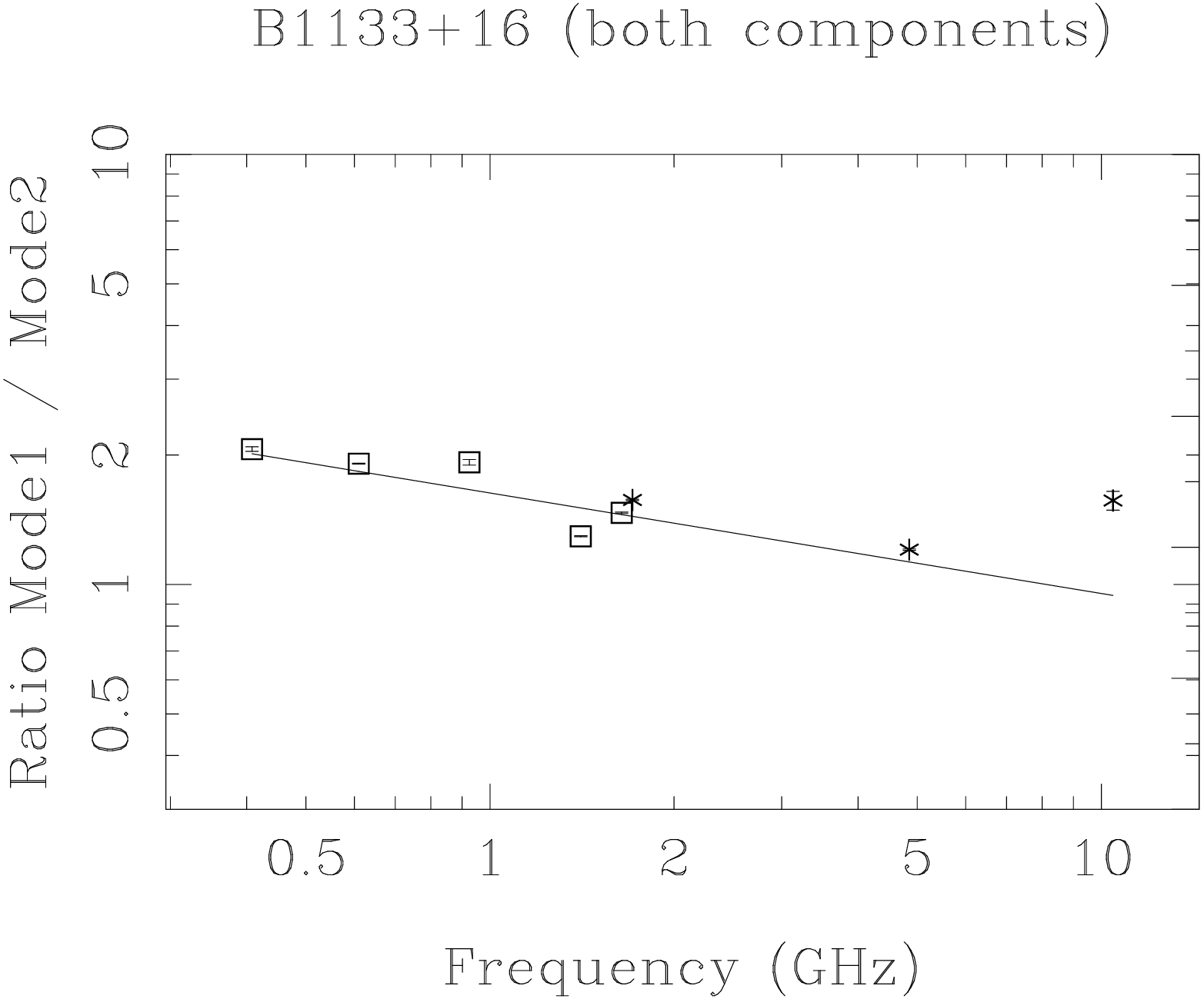}
\\
  \includegraphics[width=0.26\textwidth]{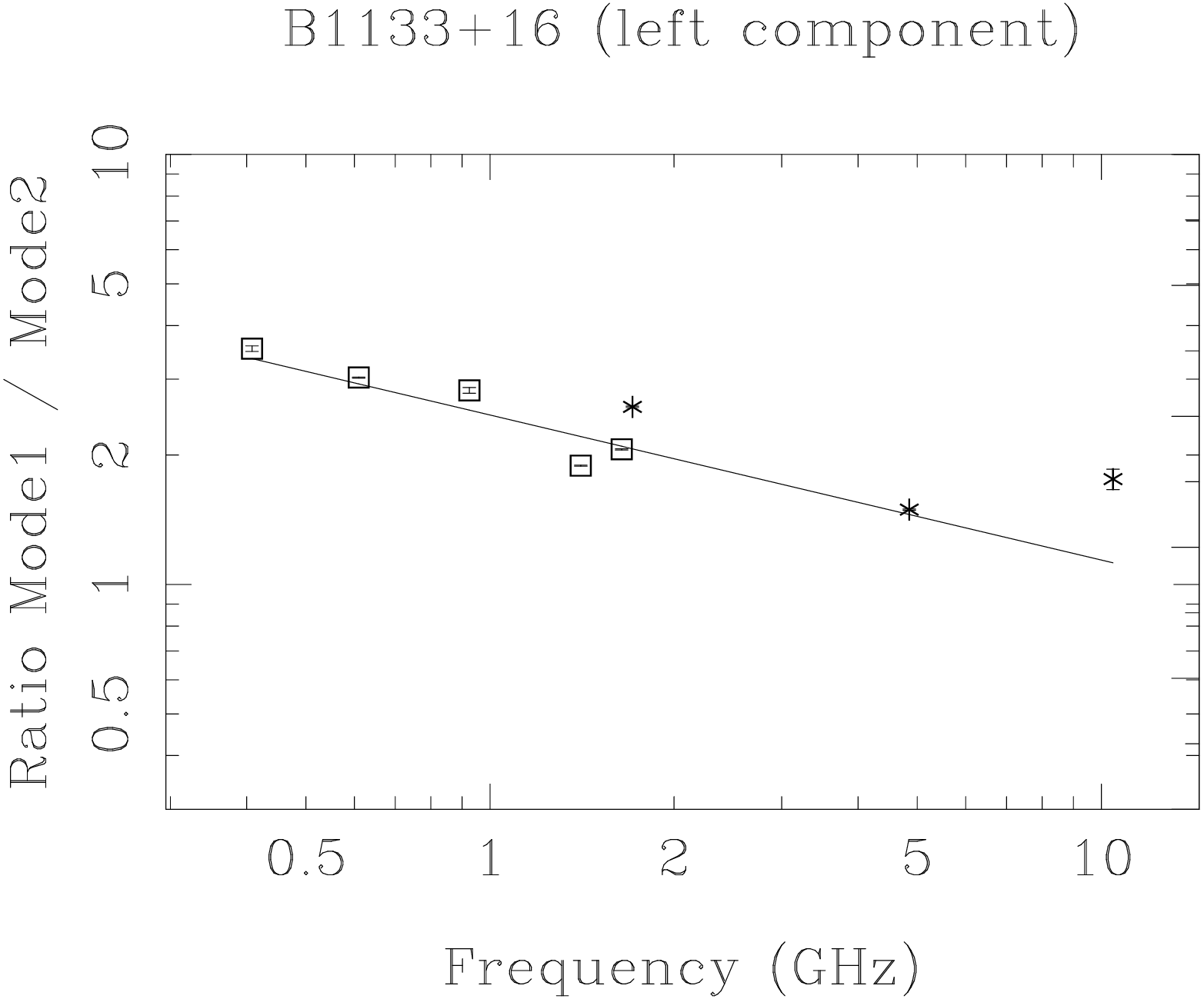}
& \hspace{1cm}
  \includegraphics[width=0.26\textwidth]{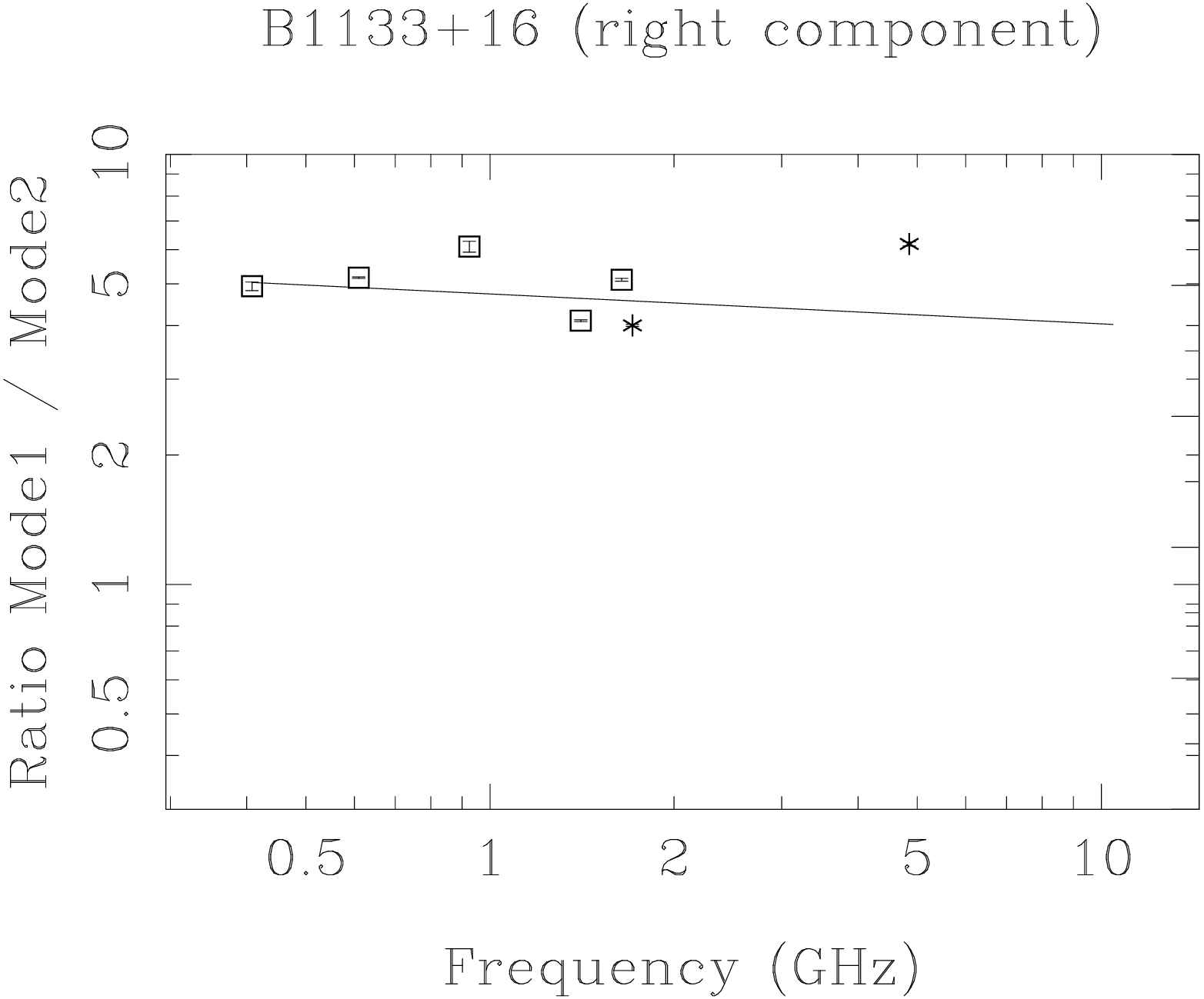}
& \hspace{1cm}
  \includegraphics[width=0.26\textwidth]{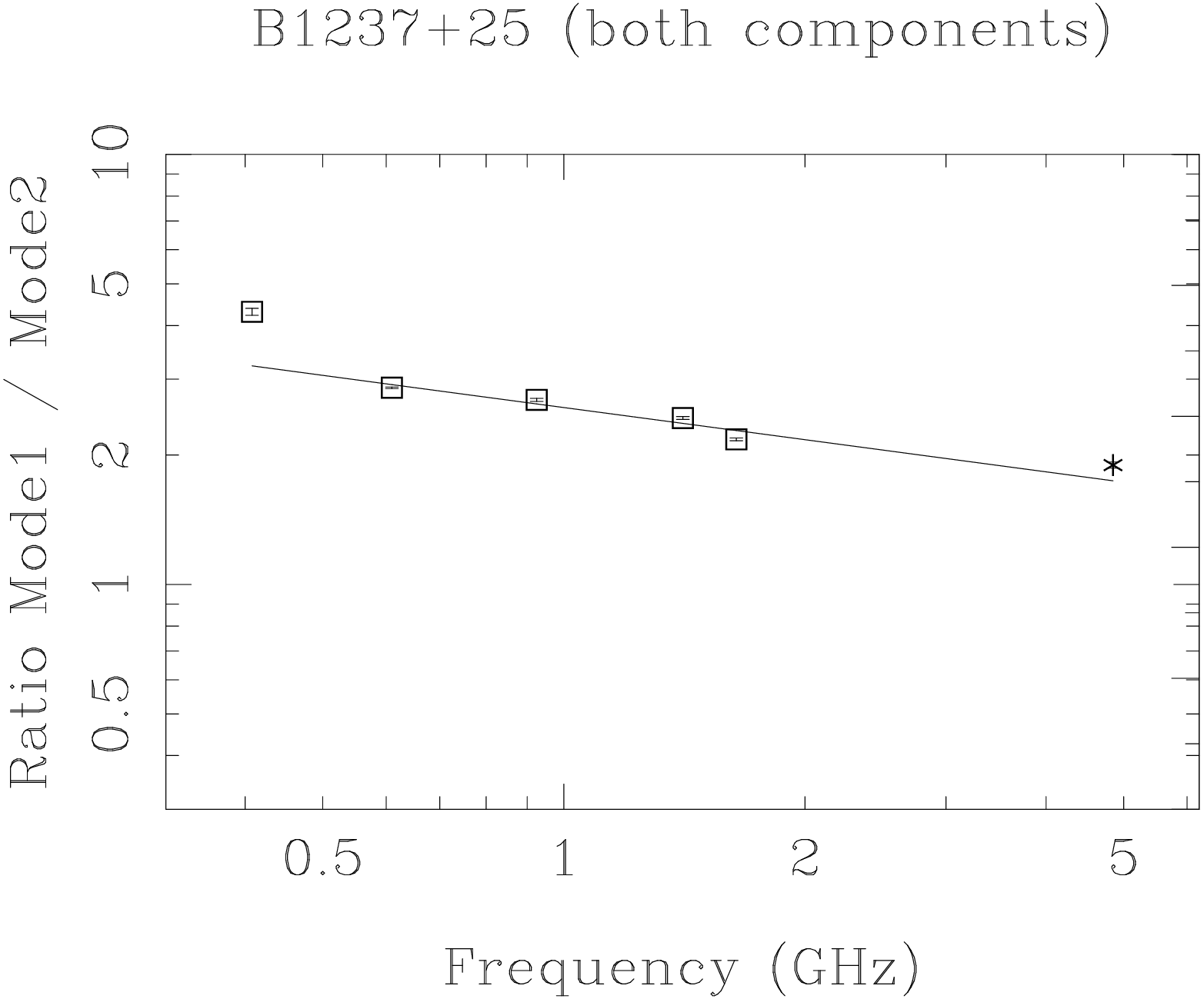}
\\
  \includegraphics[width=0.26\textwidth]{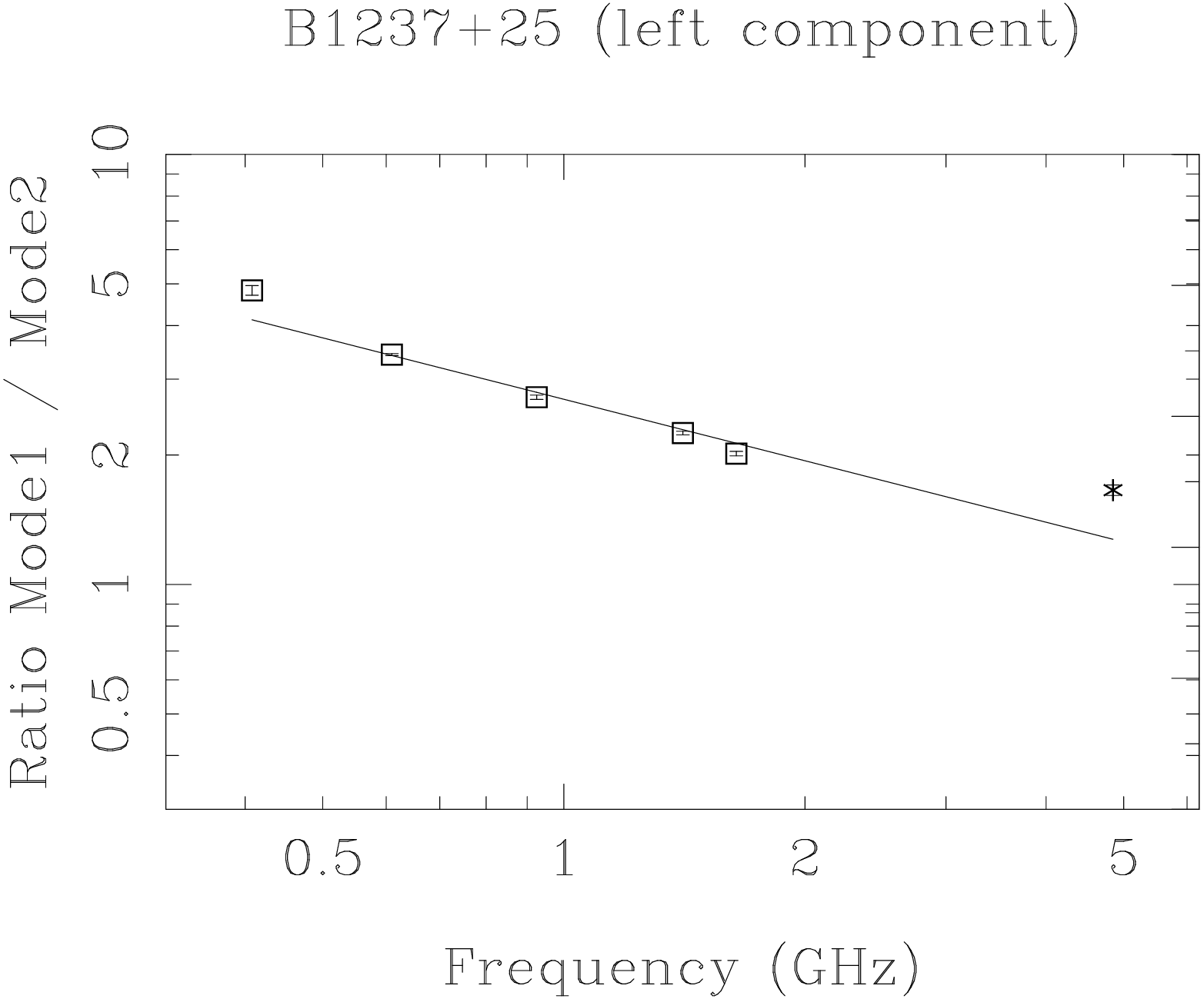}
& \hspace{1cm}
  \includegraphics[width=0.26\textwidth]{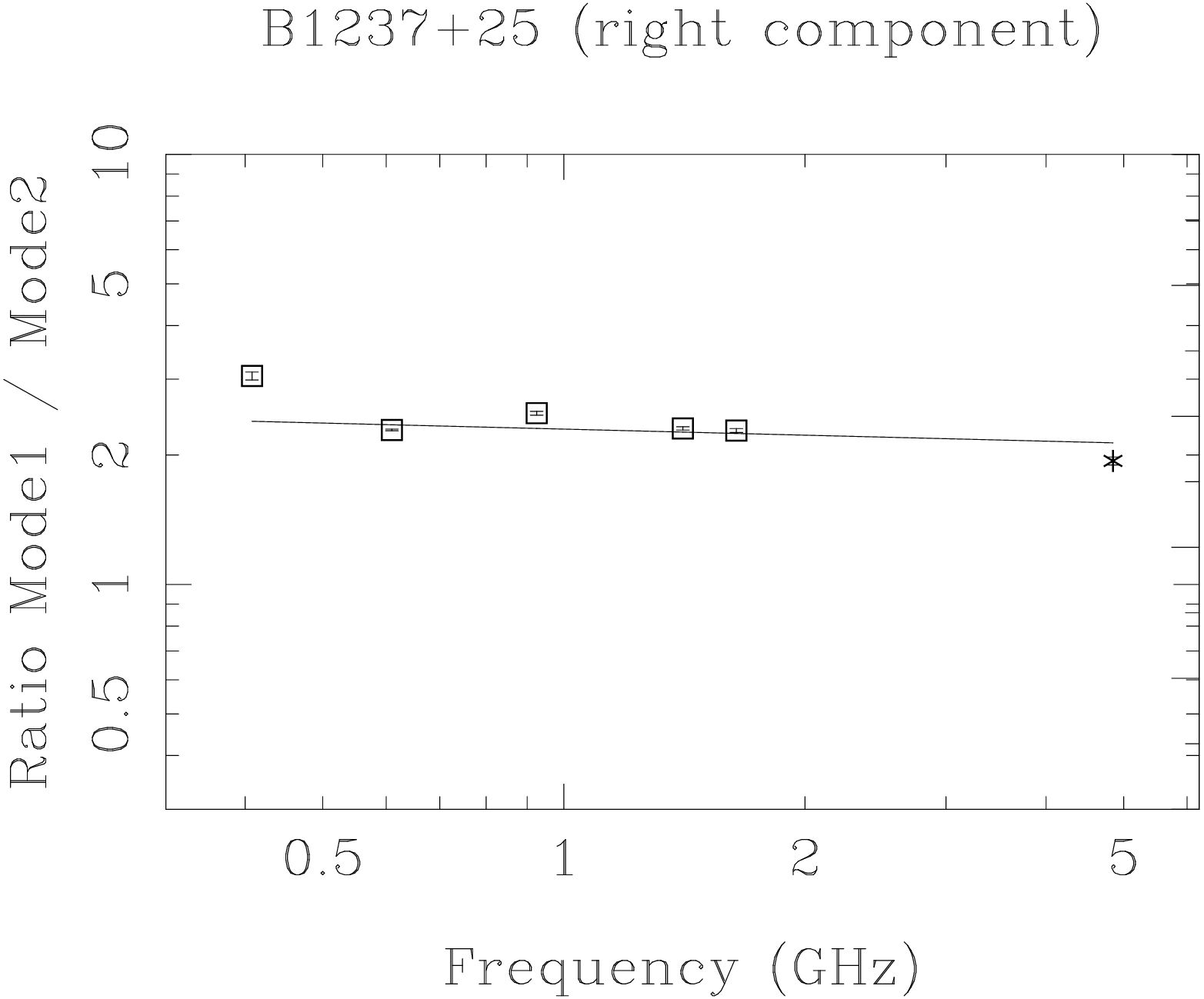}
& \hspace{1cm}
  \includegraphics[width=0.26\textwidth]{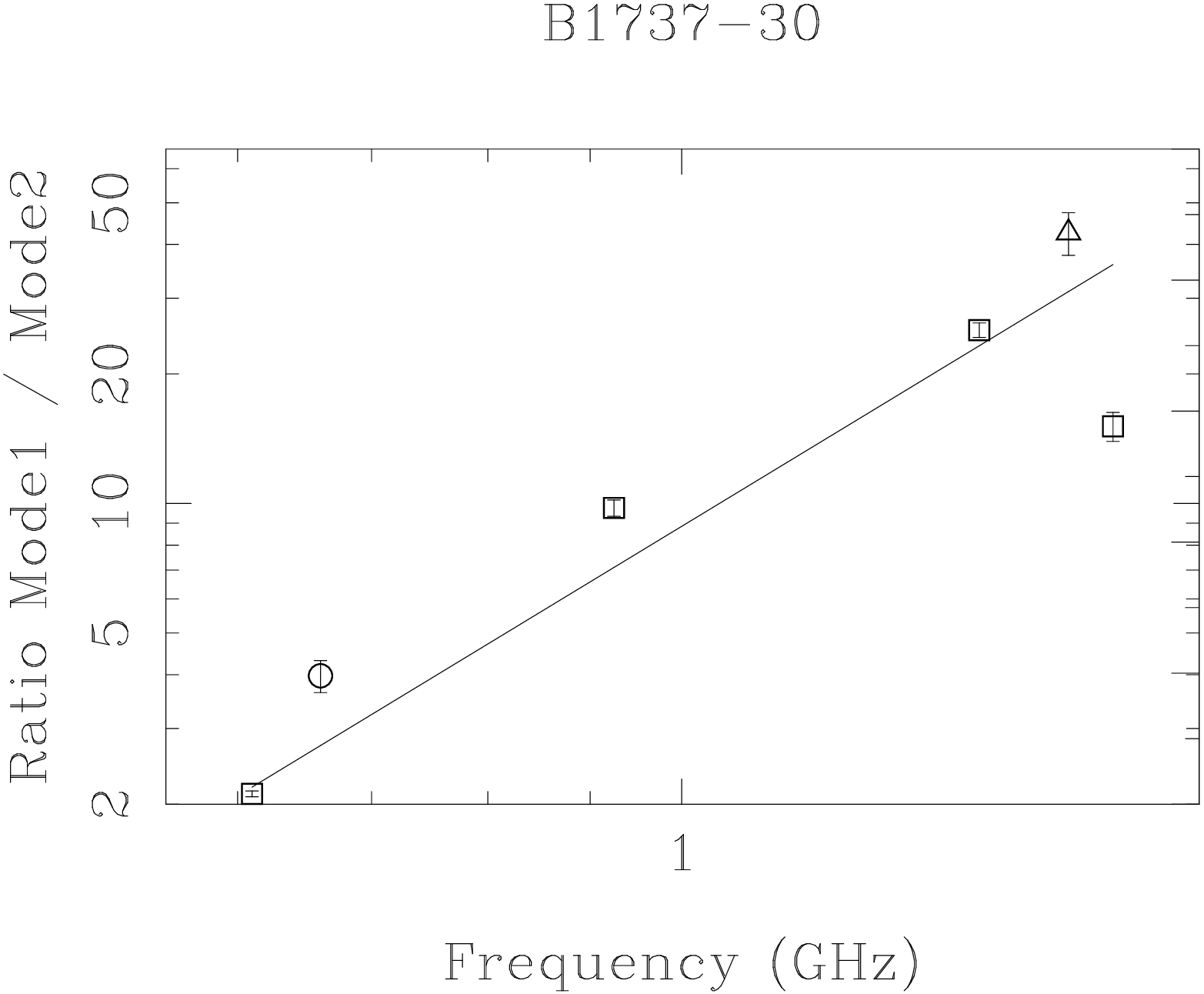}
\\
  \end{tabular}
 \caption{See Fig.~\ref{fig:ratio_of_modes_1} for
 explanation. Note that for PSR B1737$-$30 the vertical axis is
 shifted with respect to the other plots. This does not affect the
 slope.}
 \label{fig:ratio_of_modes_2}
\end{figure*}

\begin{figure*}
 \begin{tabular}{ccc}
  \includegraphics[width=0.26\textwidth]{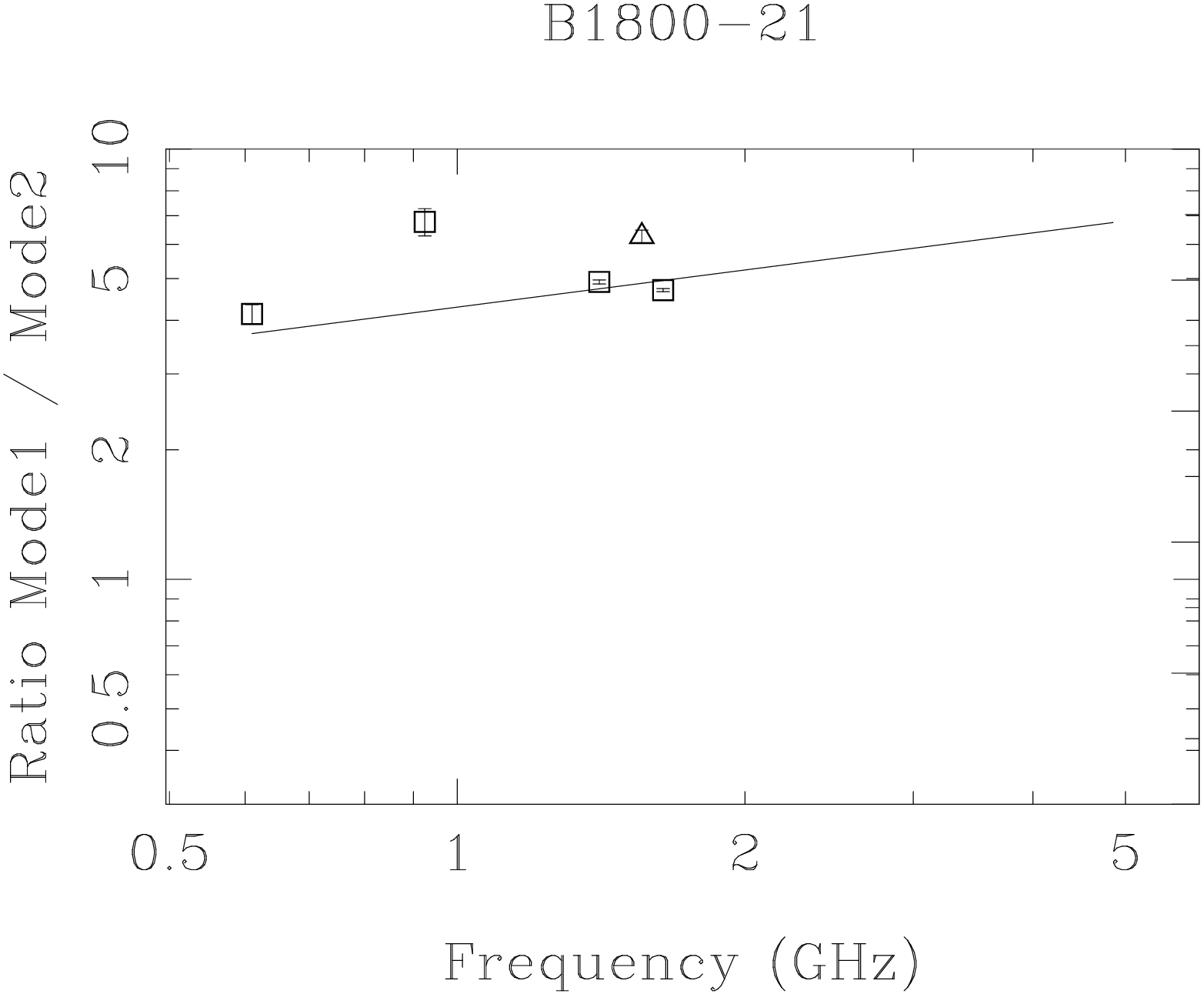}
& \hspace{1cm}
  \includegraphics[width=0.26\textwidth]{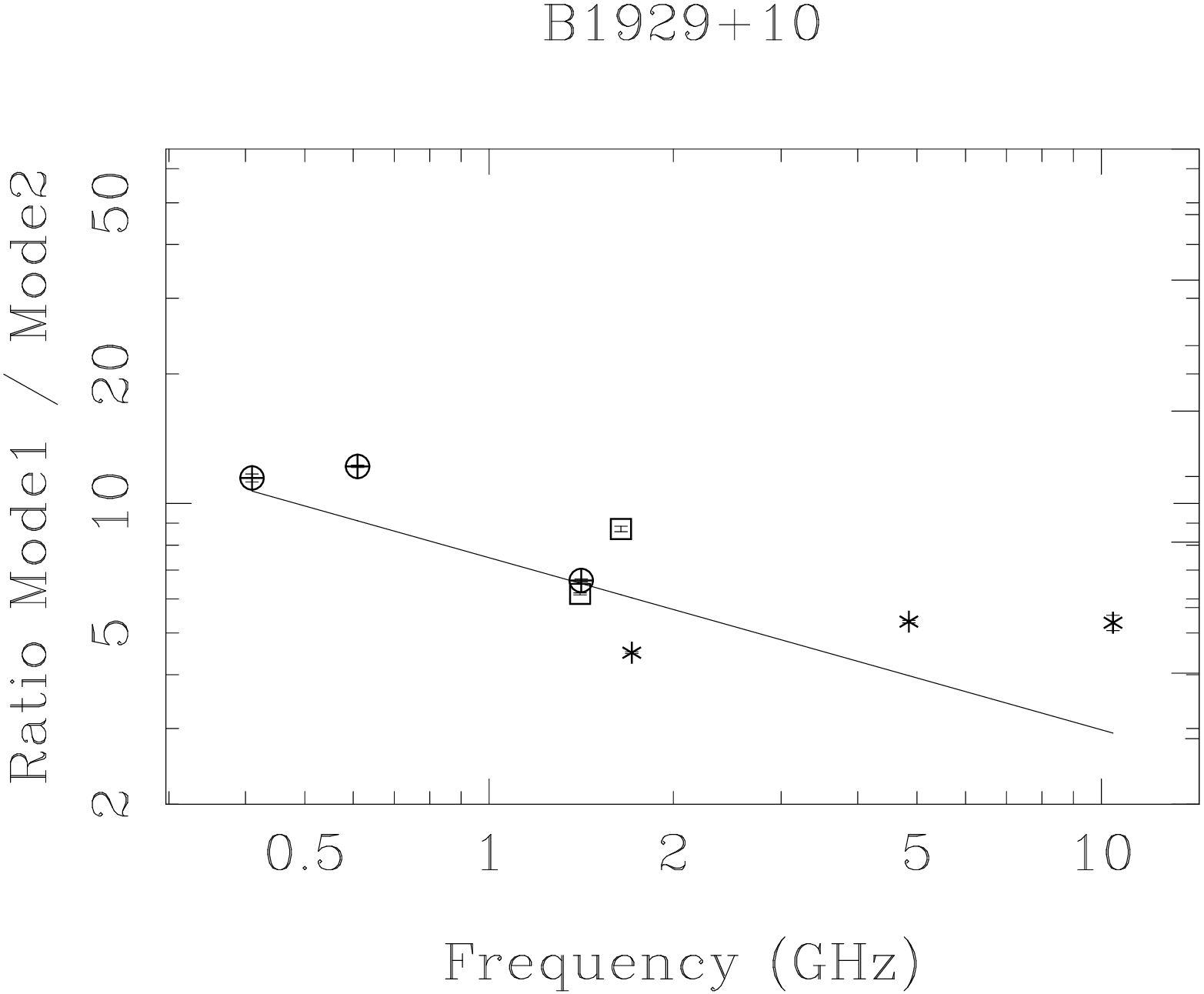}
& \hspace{1cm}
  \includegraphics[width=0.26\textwidth]{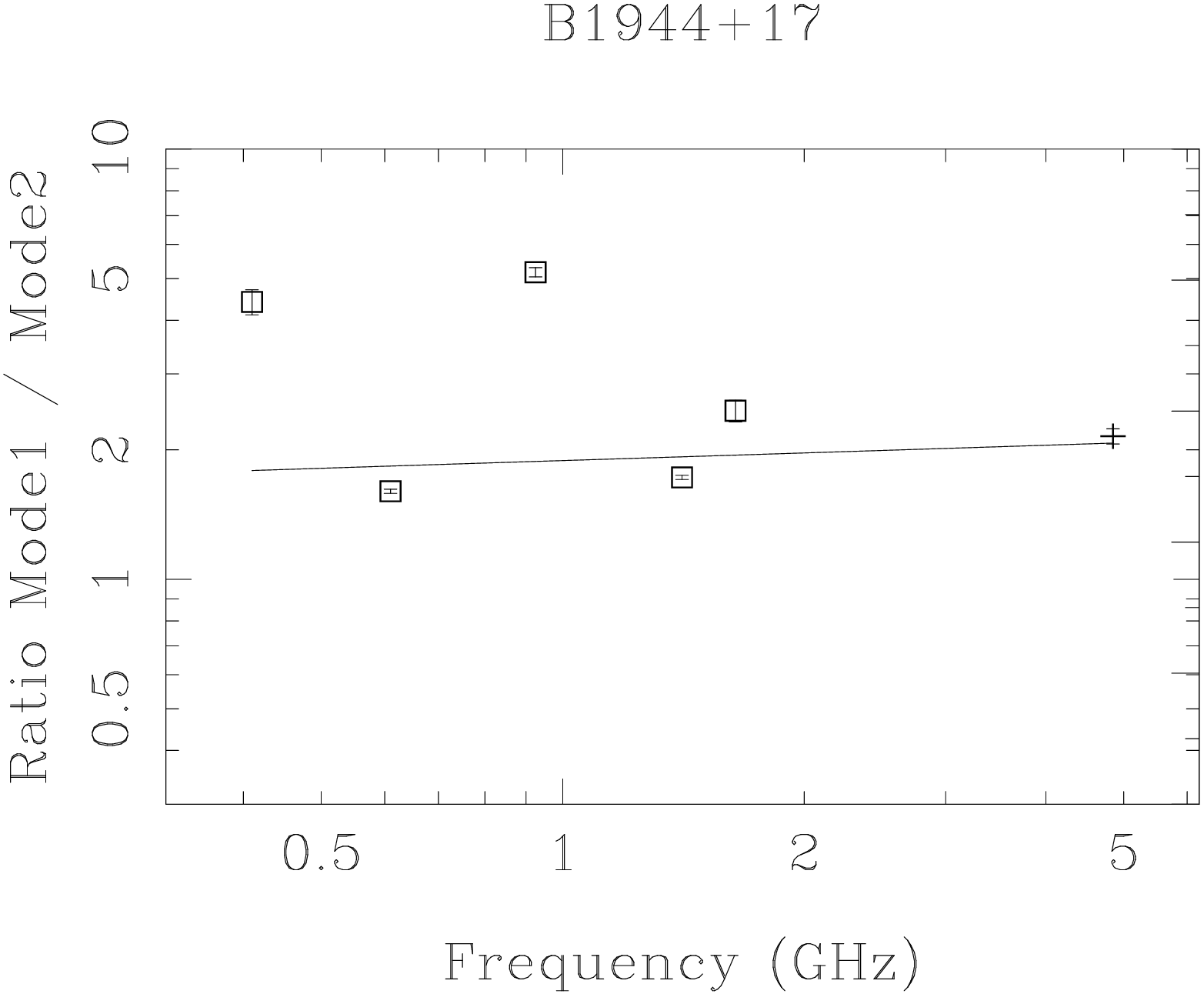}
\\
  \includegraphics[width=0.26\textwidth]{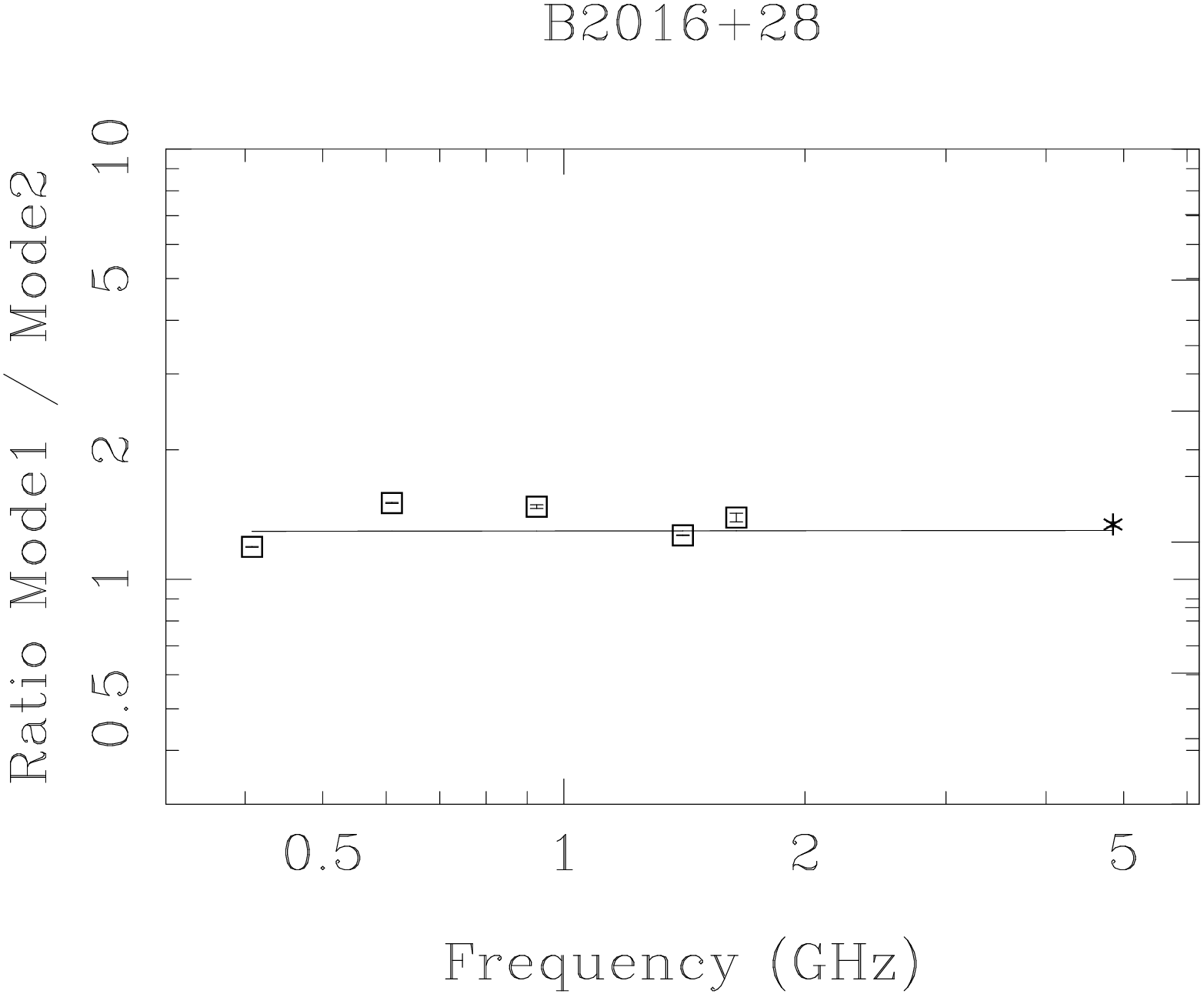}
& \hspace{1cm}
  \includegraphics[width=0.26\textwidth]{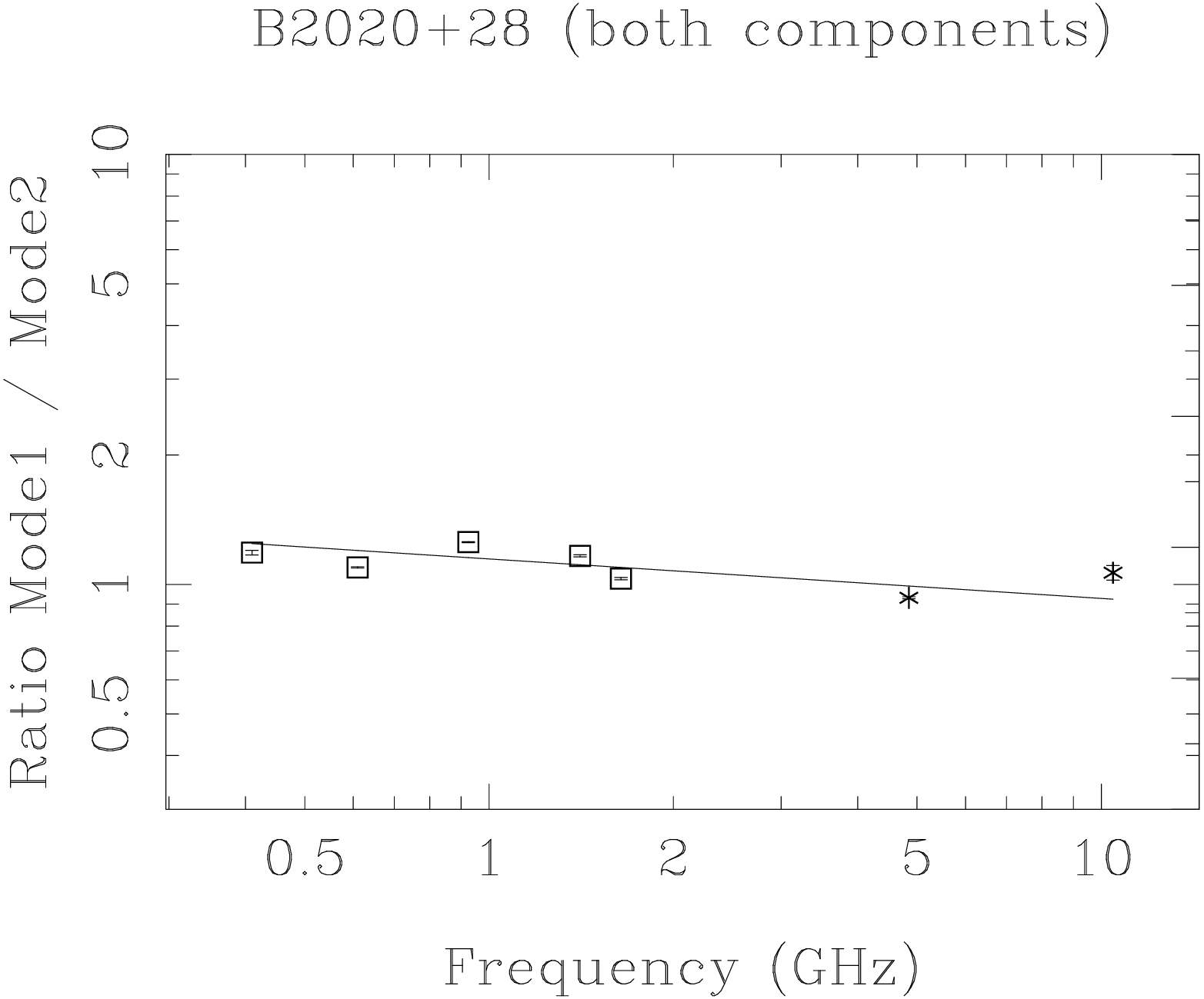}
& \hspace{1cm}
  \includegraphics[width=0.26\textwidth]{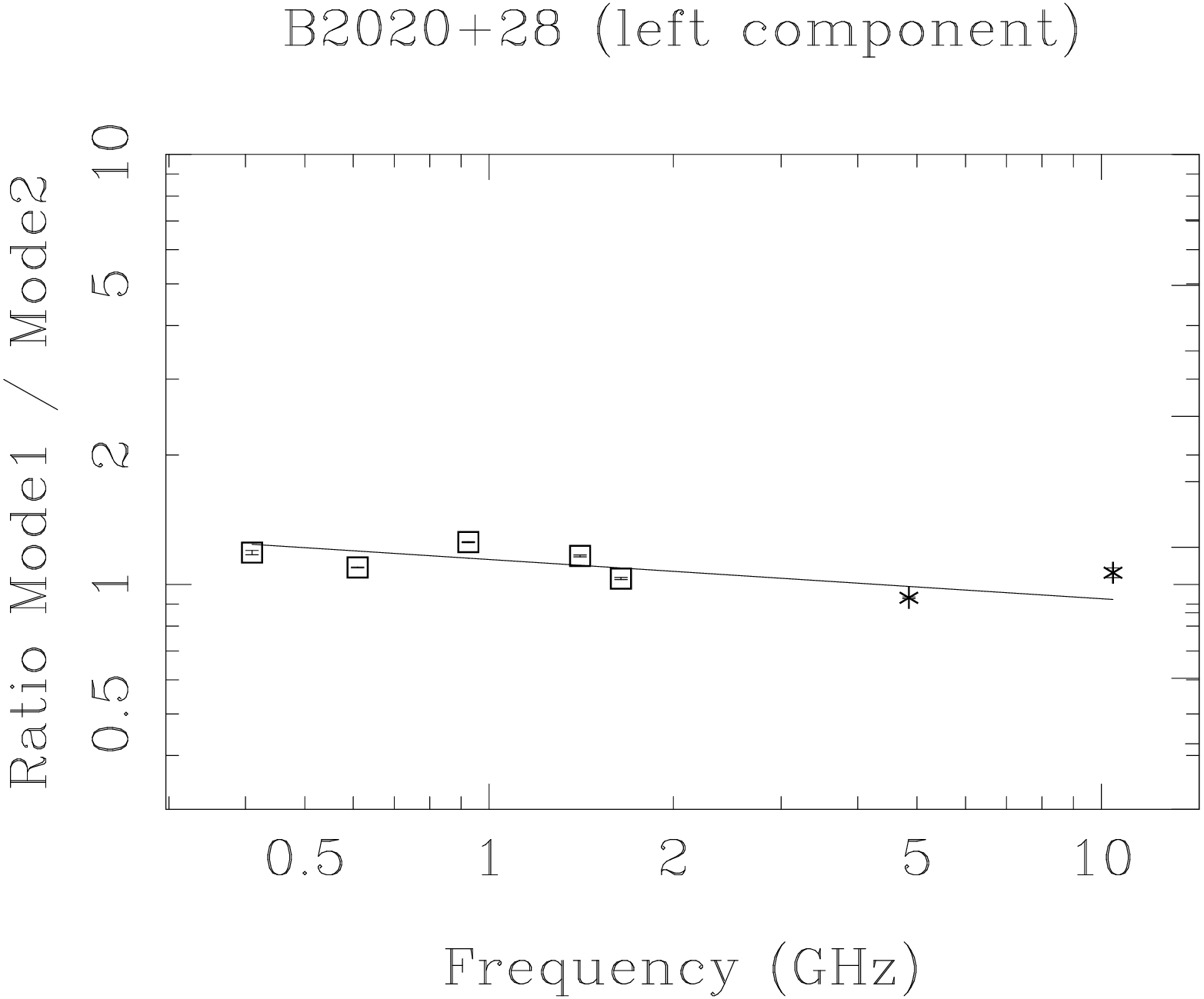}
\\
  \includegraphics[width=0.26\textwidth]{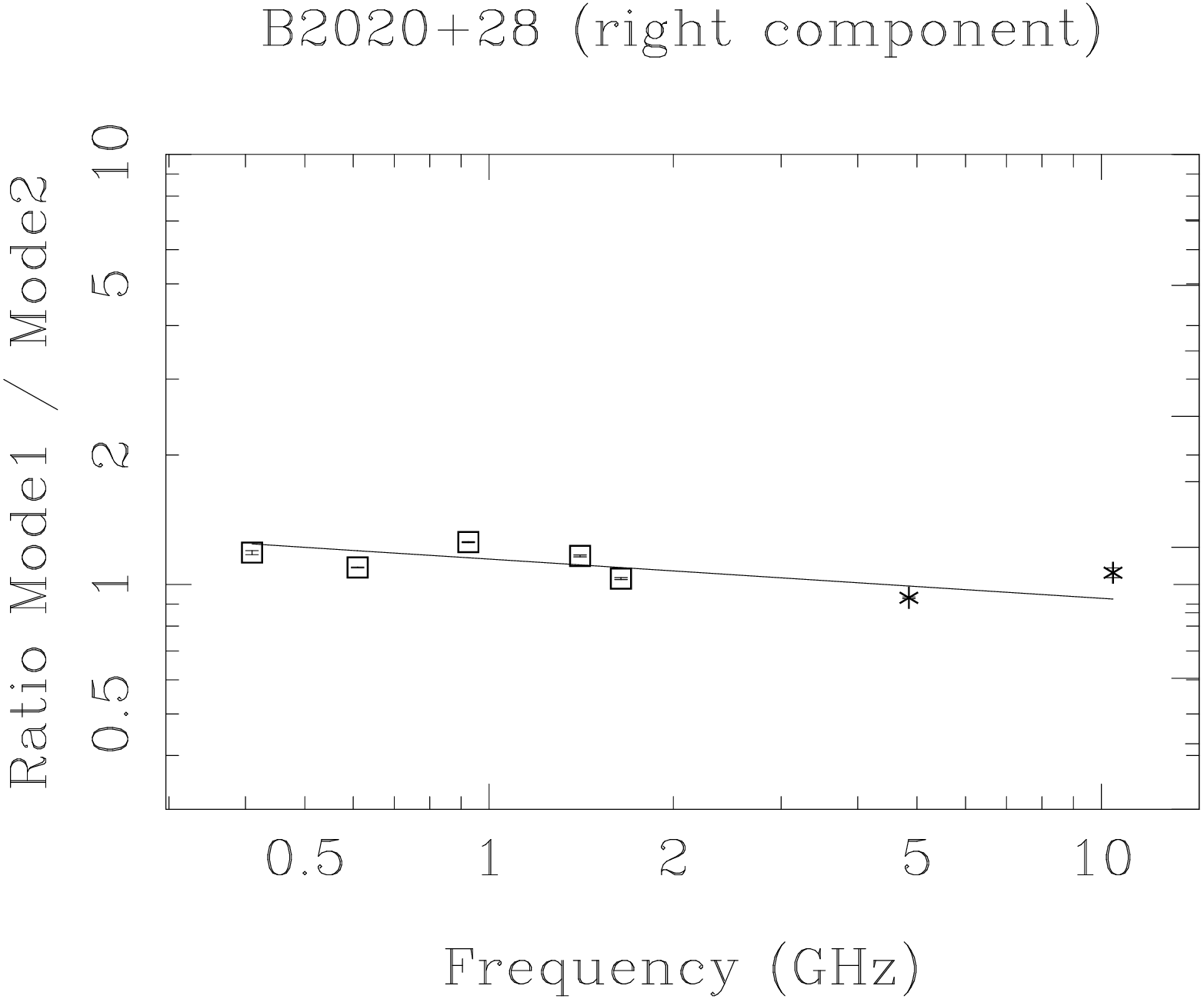}
  \end{tabular}
  \begin{tabular}{cl}
   \hline
  $\Box$        & \citet{Gould98}                    \\
  $-$              & \citet{Hoensbroech98}                  \\
  $\times$         & \citet{Hoensbroech97}              \\
  $\bigcirc$       & \citet{Guojun95}             \\
  $\bigtriangleup$ & \citet{Wu93}                \\
  $\cdot$          & \citet{Seiradakis95}             \\
   \hline
  \end{tabular}
 \caption{See Fig.~\ref{fig:ratio_of_modes_1} for
 explanation. Note that for PSR B1929+10 the vertical axis is
 shifted with respect to the other plots. This does not affect the
 slope.}
\label{fig:ratio_of_modes_3}
\end{figure*}

Because most data does not contain any information on the observed
pulsar flux, the spectral index of the modes could only be determined
relatively. Nonetheless, Figs.~\ref{fig:ratio_of_modes_1}
through~\ref{fig:ratio_of_modes_3} show that in many cases there is a
clear trend in the ratio of the average intensity of the two
modes. The suggested power laws are listed in
Table~\ref{tab:characteristics}. These trends imply that the spectral
indices of the two polarisation modes often differ from each
other. Possible explanations are that the conversion of ordinary modes
into extraordinary modes is frequency dependent (or in the case of
radius to frequency mapping, height dependent), the two modes are
damped differently, or one of the modes is moving out of the field of
view.

\subsection*{Association of modes with components}
Assuming that the different modes follow different paths in the
magnetosphere, each mode can be responsible for several components in
the total intensity profile. In many pulsars a mode can, at some
frequencies, indeed be associated with one or more components of the
average intensity (see Table~\ref{tab:characteristics}).
Figures~\ref{fig:0144} through~\ref{fig:2020} show that this can occur
at any frequency.  This association of polarisation modes with
components suggests that the method used to separate the two
polarisation modes is successful.

\subsection*{Jump in the PA}
In several pulsars, such as B0950+08 (see
Table~\ref{tab:characteristics}), the PA jumps (sometimes gradual)
60$\deg$ through 90$\deg$, while at the same longitudes the
ellipticity angle does not change sign and the intensity plots of the two
modes do not allow for a smooth change in dominance, suggesting that
there is no mode-change. Yet this behaviour does seem to be associated
with mode-changing, because the ellipticity angle usually shows a similar
(but smaller) jump and the modes are of equal strength during such a
jump. A partial (less than 90$\deg$) or gradual jump in the PA cannot
be explained with two completely orthogonal modes, since the PA is
completely determined by the strongest mode and can therefore only
jump by 90$\deg$. In the view of the ordinary and extraordinary modes
being 100\% polarised and responsible for all the polarising
characteristics of pulsar emission, the partial or gradual jump in the
PA can be explained when the two polarisation modes are not always
completely orthogonal \citep{Edwards04}. An assumption that the modes
are not 100\% polarised would not explain these jumps.

\subsection*{Correlation with type}
It is interesting that only one of the observed features, mode
drifting, is correlated with the pulsar's type. And even this one case
involves only two pulsars. Also, the many pulsars we find containing
one component, dominated by one mode at low frequency, and increasing
in intensity with increasing frequency, are of all types. This suggests
that the mechanism behind the frequency behavior of the polarisation
of pulsars is similar in both the core as in the (multiple) conal
components.

\section{Conclusions}
\label{sec:Conclusions}
We have determined the average pulse profiles of two polarisation
modes for \NPulsars pulsars at multiple frequencies, assuming that the
modes are completely orthogonal and 100\% polarised. We find many
cases where each polarisation mode can be wholly associated with one
component in the average intensity and cases where a 90$\deg$ phase
jump in the PA is associated with the two polarisation modes changing
dominance. This suggests that the two polarisation modes are
successfully determined in these cases. In other cases where there is
a (gradual) jump in the PA of less than 90$\deg$ it would appear that
one of the assumptions enabling the determination of the modes is not
always valid. The spectra of the ratio of the integrated intensity of
the polarisation modes show that in many cases there is a trend for
one of the modes to become stronger with increasing frequency. This
can be both the mode which we have defined as strongest, as well as
the weakest mode. We also find that the average profiles of the modes
often differ from each other at different frequencies. In particular,
we find that when a component is dominated by one mode at low
frequency it tends to increase in intensity with increasing frequency
with respect to the rest of the profile. Furthermore, out of \NPulsars
pulsars there are 2 pulsars that clearly show one polarisation mode
changing its longitudinal position with frequency. From our results we
cannot determine whether the longitudinal transitions of one
polarisation mode into the other at a fixed frequency are the result
of the X- and O-modes beams being separated due to refraction
\citep{Barnard86}, or due to switching between significant and
insignificant conversion of O-mode into X-mode
\citep{Petrova01}. Both explanations can also account for some
of the complex changes that occur in the average profiles of the
polarisation modes over frequency, since plasma waves with different
frequency will traverse different paths through the pulsar
magnetosphere. In the model of \citet{Petrova01} complex changes in
the average profile of the polarisation modes over frequency implies
that there are differences between the plasma number densities in
polarisation limiting regions corresponding to emission at different
frequencies. Following the technique as described in \citet{Petrova03}
and assuming radius to frequency mapping, calculation of the plasma
density distributions from polarisation profiles at different
frequencies will then yield changes in the plasma density both as a
function of longitude as well as altitude.

\begin{acknowledgements}
We would like to thank an anonymous referee for his/her useful
comments. This research has made extensive use of the database of
published pulse profiles maintained by the European Pulsar Network,
available at: \texttt{http://www.mpifr-bonn.mpg.de/div/pulsar/data/}
\end{acknowledgements}

\bibliographystyle{apj} 
\bibliography{opr}

\pagestyle{empty}

\clearpage

\clearpage

\clearpage

\clearpage

\end{document}